\documentclass{article}

\usepackage{graphicx}
\usepackage[margin=2.2cm]{geometry}
\usepackage{subcaption}
\usepackage{amsmath, amssymb, amsthm}
\usepackage{bbm}
\usepackage{mathtools, extarrows}
\usepackage{booktabs}
\usepackage{titling}
\usepackage{float}
\usepackage{braket}
\usepackage[table]{xcolor}
\usepackage{longtable}
\usepackage{multirow}
\usepackage{enumitem}

\definecolor{yesgreen}{HTML}{1B7F3B}
\definecolor{nored}{HTML}{B3261E}
\definecolor{nagrey}{HTML}{8A8A8A}
\definecolor{partorange}{HTML}{C2690B}
\definecolor{linkblue}{HTML}{1A4FA3}

\usepackage{hyperref}

\hypersetup{
  colorlinks=true,
  linkcolor=linkblue,
  citecolor=yesgreen,
  urlcolor=linkblue,
}

\newcommand{\cYes}{\textcolor{yesgreen}{Yes}}
\newcommand{\cNo}{\textcolor{nored}{No}}
\newcommand{\cNA}{\textcolor{nagrey}{N/A}}
\usepackage[
  backend=biber,
  style=numeric-comp,
    sortcites=true,
  sorting=none,
  maxbibnames=10,
  minbibnames=10,
  backref=true
]{biblatex}

\DefineBibliographyStrings{english}{%
  backrefpage  = {mentioned on page},
  backrefpages = {mentioned on pages},
}
\usepackage{authblk}
\usepackage{tikz}

\usetikzlibrary{arrows.meta}

\newcommand{\quilt}[1]{\overset{\boxplus}{#1}}

\newcommand{\atomsymbol}{%
  \tikz[
    baseline=-0.45ex,
    scale=0.7,
    transform shape,
    line width=0.35pt,
    line cap=round
  ]{%
    \draw (0,0) ellipse (.48em and .15em);
    \draw[rotate=60] (0,0) ellipse (.48em and .15em);
    \draw[rotate=-60] (0,0) ellipse (.48em and .15em);

    \fill (-.045em,-.026em) circle (.026em);
    \fill ( .045em,-.026em) circle (.026em);
    \fill (0em,.052em) circle (.026em);
  }%
}

\newcommand{\atom}[1]{\overset{\atomsymbol}{#1}}

\newcommand{\nucleussymbol}{%
  \tikz[
    baseline=-0.45ex,
    scale=0.7,
    transform shape,
    line width=0.35pt,
    line cap=round
  ]{%
    \draw (0,0) circle (.27em);

    \fill (-.105em,-.060em) circle (.065em);
    \fill ( .105em,-.060em) circle (.065em);
    \fill (0em,.120em) circle (.065em);
  }%
}

\newcommand{\nucleus}[1]{\overset{\nucleussymbol}{#1}}
\newcommand{\ft}{\mathbb{F}_2}
\newcommand{\Pauli}{\mathsf{P}}
\newcommand{\Stitch}{\operatorname{Stitch}}
\newcommand{\Stab}{\operatorname{Stab}}

\newcommand{\wt}{\operatorname{wt}}

\newcommand{\dotG}{%
  \mathrel{%
    \tikz[baseline=(c.base)]{
      \node[inner sep=0pt] (c) {$G$};
      \fill ([xshift=0.04em, yshift=0.9]c.center) circle (1pt);
    }%
  }%
}

\newcommand{\dotd}{%
  \mathrel{%
    \tikz[baseline=(c.base)]{
      \node[inner sep=0pt] (c) {$d$};
      \fill ([xshift=-0.03em, yshift=-1.3]c.center) circle (0.6pt);
    }%
  }%
}

\theoremstyle{plain}
\newtheorem{theorem}{Theorem}[section]
\newtheorem{lemma}[theorem]{Lemma}
\newtheorem{proposition}[theorem]{Proposition}
\newtheorem{corollary}[theorem]{Corollary}

\newtheoremstyle{definitionstyle}{8pt}{8pt}{\normalfont}{}{\bfseries}{.}{0.5em}{}

\theoremstyle{definitionstyle}
\newtheorem{definition}[theorem]{Definition}
\AddToHook{env/definition/begin}{\pushQED{\qed}}
\AddToHook{env/definition/end}{\popQED}

\theoremstyle{remark}
\newtheorem{remark}[theorem]{Remark}

\title{\huge Stitching Quantum Codes into Quilts
\vspace{0.3cm}\\ \Large{Universal Logic with no Ancillary Qubits}}
\author{Stasiu Wolanski \thanks{\texttt{stasiu.wolanski@riverlane.com}}\\
\vspace{0.4cm}
\textit{Riverlane, St Andrew's House, Cambridge, UK} \vspace{0.2cm}\\
\textit{
Dept. of Physics and Astronomy, University College London, UK}}
\date{November 2026}

\begin{document}
\maketitle

\begin{abstract}
Proposals abound for performing fault-tolerant logical Pauli-product measurements between quantum codes beyond the surface code. We introduce a new construction, the code quilt, whereby a target joint logical operator of two or more codeblocks is directly measured without the use of an ancillary system. We construct explicit circuits to perform a logical measurement using the quilt construction, which also require no physical qubits beyond the ancillas already in place to perform idle syndrome extraction on the codeblocks in question. We give a detailed structural analysis of the construction, which is based on a subsystem code, and provide several sets of sufficient conditions under which the relevant code-capacity distance is preserved. We give two variants of the protocol. The first, where the gauge generators of the subsystem code are measured directly, is provably not fault tolerant in the large-distance limit but admits a particularly practical circuit implementation in which the physical qubit degree increases by at most two. The second has more favourable asymptotic properties, although we do not prove its fault tolerance in the general case. We then perform an extensive empirical analysis of the technique, in particular of the first variant, as applied to bivariate bicycle and surface codes, finding code-capacity-distance preserving quilts for measuring 25,000 randomly selected logical operators. We optimise to find sparse and decongested detector bases for each and, for a randomly selected subset, find circuits that we estimate to be distance preserving. Finally, we perform decoding experiments on a subset of these circuits and find that their logical error probability is comparable to that of idle memory for operations involving the 144-qubit bivariate bicycle code and distance-12 rotated surface code at physical error rates between $1.5\times 10^{-3}$ and $3\times 10^{-3}$.
\end{abstract}

\newpage

\section*{Introduction: summary of construction, and comparison to other work}

High-rate QLDPC codes \cite{tillich2014quantum,bravyi2024high,panteleev2021degenerate,okada2026highgirth} promise to reduce the overhead of fault-tolerant quantum computing. In order to be useful, we must have a prescription for performing logical operations with these codes. In `Tour de gross', Yoder et al.\ provided the first concrete proposal for a universal computer using bivariate bicycle codes~\cite{yoder2025tourdegross}, which combined code automorphisms~\cite{sayginel2024automorphisms} with a technique for performing intra- and inter-block Pauli-product measurements known as generalised surgery \cite{cohen2022lowoverhead,cross2024improvedsurgery}. Whilst these primitives are universal (when combined with a suitable means of generating resource states for non-Clifford logic), this proposal incurred a heavy ($\approx20\times$) time overhead, and required the addition of ancillary `logical processing units' to facilitate these measurements. These ancillary systems have a size of around half that of the codeblocks themselves. Building on this and earlier work, many techniques have been proposed that improve upon this proposal. Using generalised bicycle codes in place of bivariate bicycle codes can reduce the time overhead by providing access to all of the logical operators of the code using only a single shift automorphism~\cite{webster2026pinnacle}. The techniques themselves have been developed into fast~\cite{cowtan2025hypergraph,zheng2025highrate,chang2026constanttime,berent2026lifted} and parallelisable~\cite{cowtan2025parallel,gu2026qgpu} variants, also in an attempt to reduce time overhead.
Progress has also been made in reducing the space overhead associated with generalised surgery by reducing the size of the ancilla system \cite{yuan2026parsimonious}. Magic state preparation using QLDPC codes has also been explored~\cite{liu2026magicinjection,gidney2024cultivation}.

In this work, we present two novel methods for performing Pauli-product measurements between QLDPC codeblocks.
Based on the same underlying subsystem code construction, they are designed to be as simple and practical as possible, whilst providing empirical fault-tolerance in practical settings, at the cost of analytically certifiable universal applicability 
(there is nonetheless substantial analysis of the construction described in the main text and detailed in the appendix).

Most generalised surgery constructions draw on two basic design principles (speaking in very broad terms).
First, one typically designs a protocol for measuring a general logical operator of a \textit{single code}, such that a product measurement across two or more codeblocks can be thought of as a measurement of a single logical on a larger tensor-product code.
Second, one starts from a design ansatz, generally a graph~\cite{cross2024improvedsurgery}, hypergraph~\cite{cowtan2025hypergraph} or CSS complex~\cite{ide2025homological}, whose purpose is to ensure that the resulting stabilisers of some deformed code commute whilst containing the target logical operator. Starting from this ansatz, we then attempt to prove fault-tolerance by showing distance preservation and LDPC-ness, possibly adding additional ancillary structure in the process.

The method presented in this paper deviates from both of these principles.
The fundamental measurement primitive is taken to be a Pauli-product measurement over two or more codeblocks, and the deformed code is explicitly a subsystem code comprising non-commuting gauge operators.
The resulting subsystem code, which we call a \textit{code quilt} (or simply a quilt) has LDPC gauge generators by construction, and further the form and locality of the original stabilisers of the two codeblocks is preserved, although some are downgraded to gauge operators. This subsystem code can be viewed as a generalised version of the subsystem lattice surgery construction described in~\cite{nautrup2017interface}.
We prove sufficient conditions for code-capacity distance preservation and for combining smaller quilts into parallel and higher-weight logical measurements. Sparse inherited generators are constructed under additional conditions on the local check presentations.
Although these results give insight into the structure of code quilts, the specific code quilts provided to measure logical operators given as part of the numerical experiments in this work are found by applying a simulated annealing \cite{kirkpatrick1983annealing} optimisation over several heuristics.
For bivariate bicycle and surface codes, representative searches provide evidence of code-capacity distance preservation beyond these sufficient conditions. Selected implementations are also assessed through circuit-distance searches and decoding experiments.
This suggests that future work may be able to analytically establish much looser sufficient conditions for preserving fault-tolerance properties.

Based on this subsystem code, we propose two concrete protocols for performing a logical measurement. We perform circuit-distance estimation and numerical simulation of the first protocol, and give a full description of circuits for both.
\begin{enumerate}
  \item \textit{Subsystem protocol.} The primary analytical and numerical focus of this work; we construct a subsystem code that measures one or more desired logical operators across two or more codeblocks.
  The gauge generators retain the sparse structure of the constituent codes.
   Certain subspaces of the stabilisers in its centre admit bounds on the support under additional conditions on the interface between the codes, and the sparsity and congestion of the other stabilisers are examined empirically. The detectors corresponding to repeated measurements of the target logical operator can grow in weight with the distances of the codes involved. 
   Their circuit-level performance is assessed empirically here.
  \item \textit{Stabiliser protocol.} We gauge-fix the subsystem code into a stabiliser code, and measure its generators directly. This preserves the code-capacity distance bounds of the subsystem code. The generators are products of original checks. 
\end{enumerate}
 
 The analytical results primarily concern code-capacity distance and extend to the gauge-fixed stabiliser protocol. The subsystem protocol also has a particularly simple physical implementation. Circuit-level distance is studied for selected implementations using fault searches and decoding experiments.
 We give explicit circuit constructions for each, and give simulation results for the subsystem protocol.
Neither protocol requires an ancillary system of data qubits. 
More remarkably, neither requires any measurement qubits in addition to those already in place to perform idle syndrome extraction, so these protocols have no spatial overhead at all.
Logical measurements are performed by means of $d$ syndrome extraction rounds.
Using the subsystem protocol, the length of the syndrome extraction rounds during a measurement is generally between one and three timesteps longer than for a memory circuit (depending on the type of the measured operator and the number of codeblocks involved in the measurement) so the time overhead of implementing measurements is minimal, although it is not generally single-shot.

In performing the investigations described herein, we developed techniques that may be of use beyond their application here: a unified formalism for reasoning about code-capacity distance in the context of generalised surgery; a technique called `detector error model folding' that losslessly reduces the complexity of estimating circuit distance; efficient open-source distance finders for both code-capacity and circuit-level problems; a statistical method for estimating the sensitivity of a general distance-finding algorithm; and various tricks to help with decoding large, non-CSS circuits.
\section*{Availability of source code and data, and use of AI}

All of the data presented in the figures of this paper, and all of the source
code required to reproduce them, are available on GitHub and Zenodo~\cite{quantumquilts, wolanski2026quiltsdata}.

Various AI tools were used extensively throughout all stages of this research project, especially to write the software required to perform the numerical investigations described herein. It is intended that the source code and data, especially when paired with AI tools to parse them, will allow the reader to resolve any ambiguity in the text as to the specifics of the constructions described or the experiments performed, as well as to reproduce and build upon the results presented.

\section*{Acknowledgements}

I am deeply grateful to the following people for valuable conversations that contributed to the results presented in this work: Ophelia Crawford, Annie Ray, Zhiyang `Sunny' He, Andrew Patterson, Setiawan, Daniel Sabsovich, Joschka Roffe, Earl Campbell, Mackenzie Shaw, Mark Walters, and particularly my supervisor, Dan Browne.
This work was supported by Riverlane and the Engineering and Physical Sciences Research Council grant
number EP/Y035046/1.

\newpage
\tableofcontents
\newpage

\section{Manuscript structure and summary of main results}\label{sec:results-summary}

\begin{figure}[H]
    \centering
    \includegraphics[width=\linewidth]{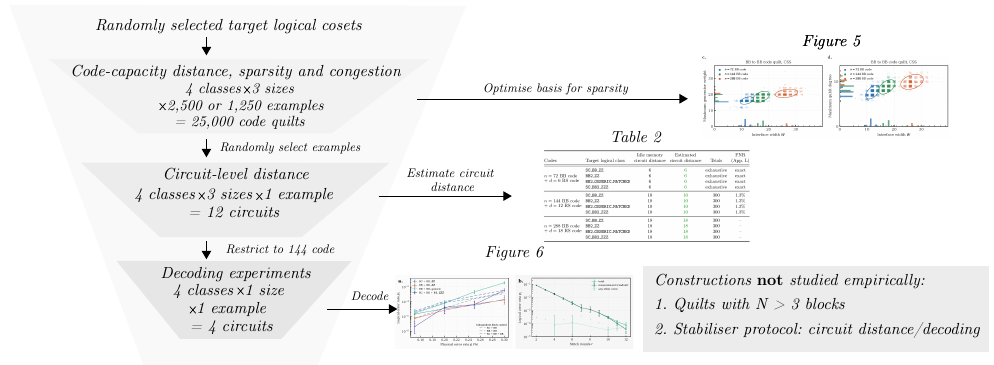}
\end{figure}

The structure of the paper is as follows. We begin with some preliminaries, including notation and the definitions of partially initialised codes and operational distance, which may be of independent interest.
Section~\ref{sec:twoblock} introduces the two measurement protocols, which are based on a single subsystem code construction, through surface-code examples. Section~\ref{sec:quilts} formalises that construction and extends it to measurements between several codes and to simultaneous measurements. The rest of the paper details the steps required to create circuits using the construction, and presents analytical and empirical results about their performance. Detailed algebra and proofs, as well as computational details, are given in the appendices. In addition, Section~\ref{sec:pullback} very briefly gives an interpretation of a simple case of code quilts using the standard language of homological measurement.
We conclude with some discussion of the implications of the work and future directions in Section \ref{sec:discussion}.

In the remainder of this section, we summarise the construction and the results.

\subsection{Constructing a logical Pauli-product measurement circuit}
Using the basic code quilt construction, we develop a pipeline for creating and simulating full circuits to measure arbitrary logical operators between stabiliser codes:
\begin{enumerate}
    \item We construct code quilts that include a given target logical operator as a stabiliser. In order to do this, we optimise for suitable representatives of the target logical coset (Section \ref{sec:repfinding}, Appendix \ref{sec:repsearch}).
    \item These quilts are subjected to a screen for code-capacity distance (Section \ref{sec:distance}, Appendix \ref{sec:tweedietest}).
    \item We optimise a basis for the centre of these codes (Section \ref{sec:sparsity}, Appendix \ref{sec:sparsityempirical}).
    \item We construct syndrome extraction circuits for them (Section \ref{sec:circuits}), and subject these circuits to circuit-level distance searches (Section \ref{sec:distance}, Appendices \ref{sec:circuit-distance-methods} and \ref{sec:tweedietest}).
    \item We perform various transformations on the resulting detector error model to make decoding easier (Section \ref{sec:circuitsims}, Appendix \ref{sec:simulationdetails}).
    \item We decode the circuits to find their logical error probabilities, comparing the performance to that of idle memory in the same codes using the same noise model and decoder (Section \ref{sec:circuitsims}, Appendix \ref{sec:simulationdetails}).
    
\end{enumerate}

\subsection{Summary of results}
Through a combination of analytical work and empirical studies using the above pipeline, we obtain various results that explore the structure and performance of code quilts, which are summarised here:
\begin{enumerate}
    \item \textbf{Analytical code-capacity distance and sparsity/congestion:} Section~\ref{sec:distance} gives sufficient conditions under which the measurement preserves protection against errors on the data qubits. These results cover the measured outcome and the quantum information retained after measurement. They apply, for example, when the measured operators have minimum weight, or when the codes and measured operators have the same structure on each block. The proofs are given in Appendices~\ref{sec:distancetheorems} and~\ref{sec:gauge-fixing}, using the operational-distance framework of Appendix~\ref{sec:operationaldistance}. Using the structure of the code quilt developed in Appendix~\ref{sec:structuretheorems}, and in particular its various classes of stabilising operators, we summarise in Section~\ref{sec:sparsity} and detail in Appendix~\ref{sec:sparsityanalytical} analytical bounds on the sparsity and congestion of certain classes of stabilisers in the code quilt.
    \item \textbf{Empirical code-capacity distance and sparsity/congestion:} Section~\ref{sec:repfinding} describes code quilts chosen to perform 25,000 target logical measurements between bivariate bicycle and surface codes randomly selected in advance. 
    The targets are selected over four classes of logical operator (for instance, one class consists of joint non-CSS operators over two bivariate bicycle codes), with each class studied over three distances/sizes: the 72-qubit, 144-qubit, and 288-qubit bivariate bicycle codes and, in some cases, matched-distance rotated surface codes.  (The impatient reader can look ahead to Figure \ref{fig:classes} for illustrations of the four classes.) Each choice satisfies the construction's algebraic conditions and is tested by randomised searches for errors below the desired distance, and in some cases, by exact exhaustive search, although the representatives are not required to satisfy the sufficient conditions for distance preservation above. This supplies evidence for distance preservation beyond the analytical conditions. The search procedure is described in Section~\ref{sec:repfinding}. Both these code-capacity distance estimates and, to a lesser extent, the circuit-level distance estimates, are subject to a thorough numerical analysis to estimate the sensitivity of the distance-finding tools we employ, detailed in Appendix \ref{sec:tweedietest}. We also numerically optimise a basis for the stabilisers associated with the quilts, including both those operators whose weight is analytically bounded by the structure theorems and those to which the theorems do not apply. In Section~\ref{sec:sparsity} and Appendix~\ref{sec:sparsityempirical}, we provide plots showing empirically the relationship between the optimised sparsity and congestion of these stabilisers and the sizes of the bivariate bicycle and surface codes involved in the construction. We find slow growth of operator weight and congestion with the code distance: tripling the code distance from $6$ to $18$ increases the median maximum generator weight and the median maximum qubit degree by between 25\% and 50\%. 
    \item \textbf{Circuit-distance evidence:} Section~\ref{sec:circuits} introduces circuits for the two measurement protocols. Section~\ref{sec:distance} estimates the circuit distances of the subsystem protocol circuits derived from one randomly selected code quilt from each class of logical measurement at each of the three code sizes. We reduce this search to a shorter circuit description using a map that preserves the spacelike logical effects, and bound the distance of the remaining timelike error direction analytically using the circuit structure. Details are given in Appendix~\ref{sec:circuit-distance-methods}. Not every code quilt that is distance preserving at the code-capacity level gives rise to a circuit that preserves circuit distance; in every case we found an alternative representative in whose circuit no distance-reducing fault was found. The reliability of that conclusion is quantified at $n=144$ in Appendix~\ref{sec:tweedietest}.
    \item \textbf{Decoded logical performance.} Section~\ref{sec:circuitsims} considers measurements involving the 144-qubit bivariate bicycle code. Taking the four circuits (one in each of the four classes) that we estimated in the previous stage to preserve circuit-level distance, we estimate their logical failure probabilities under circuit noise and compare them with idle memory experiments of the same duration.
    
\end{enumerate}

\section{Preliminaries}

\subsection{Notation}

Throughout, we identify Pauli strings on $n$ qubits with their phase-free representation as members of the binary symplectic vector space $\ft^{2n}$, equipped with the usual symplectic product. 
We then prefer the notation and language of linear algebra throughout to that of group theory, so that we may speak of \textit{subspaces} $A \le \ft^{2n}$ and understand each to be the image of some subgroup $\mathcal{A} \subseteq \mathcal{P}_n$ of the usual Pauli group under the phase-removing quotient map $\mathcal{P}_n \rightarrow \ft^{2n}$.
We relax our language somewhat, and, where the meaning is clear, identify the Pauli subgroups with the subspaces representing them.
We may then:
\begin{itemize}
    \item Write an orthogonal complement $A^\perp$ to mean the image of the Pauli subgroup that consists of operators that commute with every operator in $\mathcal{A}$.
    \item Employ the usual linear algebraic operations $+, \oplus, \setminus, /, \cap$... on subspaces (generally written with capital letters $A$, $B$).
    \item Generate new subspaces from matrices $M \in \ft^{m \times 2n}$ as $\langle M \rangle$ or from lists of binary vectors.
    \item Quotient one subspace by another $Q = A/B$, and choose arbitrary complementary subspaces $C$  so that $C\cong Q$ and $A = B + C = B \oplus C$.
\end{itemize}

\subsection{Subsystem and stabiliser codes}
Subsystem codes are defined as subgroups of the Pauli group that are not necessarily abelian; correspondingly they are represented here by subspaces $G \le \ft^{2n}$ that are not necessarily isotropic.
The stabiliser group of a subsystem code is the centre of the gauge group, represented by $Z(G) = G \cap G^\perp$.
A stabiliser group $S$ is a subsystem code that equals its centre, or equivalently is represented by a subspace $S \le \ft^{2n}$ that is isotropic $S \cap S^\perp = S$.
The (non-gauge) logical qubits of a subsystem code are manipulated using the bare logical operators $G^\perp$, whose actions are classified into cosets $\mathcal{L} = G^\perp / Z(G)$.

\subsection{Operational distance and partially initialised codes}

Defining distance for logical measurement procedures, even at the code-capacity level where neither gate nor measurement errors are considered, is quite subtle.
This is because \textit{logical classical readout errors} (where the result obtained does not correspond to the state collapse that actually occurred) and \textit{logical quantum memory errors} must both be accounted for, yet cannot be described together as the logical operators of any one stabiliser or subsystem code (as they do not form a symplectic subspace). Further explanation, and additional details about the following definitions, are given in Appendix \ref{sec:operationaldistance}.

In summary, in order to simplify the presentation, we define the \textit{partially initialised code} $(G,C)$ as a subsystem code $G$ with a distinguished subspace of \textit{trusted stabilisers} $C \le Z(G)$.
$C$ corresponds to those stabilisers whose value is known and available for error detection and correction, as opposed to the logical operators we intend to measure.
We can then encapsulate protection against both failure modes in a single quantity, the \textit{operational distance}:
        \begin{equation}
            d_{\mathrm{op}}
            =
            \min\{|E|:E\in C^\perp\setminus G\}.
        \end{equation}

This quantity is upper bounded by the usual definitions of bare and dressed distance of a subsystem code, whilst also accounting for the possibility of logical classical readout errors. Results relating to the code-capacity distance are calculated and presented in terms of the operational distance throughout. The partially initialised code is equivalent to the hybrid quantum-classical subsystem codes of \cite{nemec2022encoding}, and it may prove a useful formulation, of interest independently of the specific constructions in the present work.

\section{Two-block logical measurements: simple code quilts} \label{sec:twoblock}

\begin{figure}

    \centering
    \includegraphics[width=\linewidth]{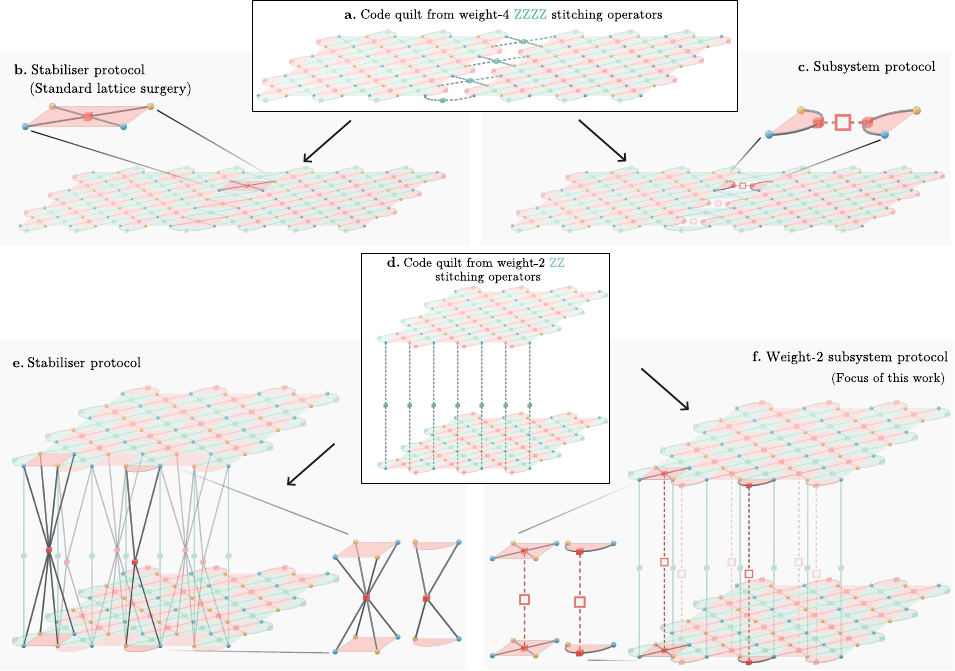}
    \caption{\textit{Four ways of performing a logical $\bar{Z}\bar{Z}$ measurement between two distance-7 rotated surface codes.} The top three diagrams show the two possible protocols as applied to the code quilt defined by the decomposition of the logical operator into the shown weight-4 $ZZZZ$ stitching operators, and the bottom three show the same for the code quilt defined by the decomposition into the shown weight-2 $ZZ$ stitching operators.
    The top diagram is shown with the two codes side-to-side and reflected, in order to match the standard visual presentation of surface code lattice surgery. The bottom diagram positions the codes atop one another; we will use this presentation in the remainder of figures in this work.
    \textbf{a.} Decomposing the chosen weight-14 logical representative into three weight-4 $Z$-type stitching operators and a single weight-2 stitching operator, defining a code quilt.
    \textbf{b.} Fixing the stitching gauges and measuring the generators of the resulting stabiliser code reproduces standard surface code lattice surgery.
    \textbf{c.} Alternatively, we can measure the generators of the quilt subsystem code directly. This uses lower-weight measurements.
    \textbf{d.} Decomposing the chosen weight-14 logical representative into 7 weight-2 $Z$-type stitching operators, defining a second code quilt.
    \textbf{e.} Fixing the stitching gauges gives a stabiliser code with generators of weight up to 8.
    \textbf{f.} The subsystem protocol: we measure the anticommuting generators of the quilt subsystem code directly, and multiply their outcomes to derive the measurements of the central stabilisers.
    Its code-capacity distance is preserved; circuit-level performance also depends on the extraction schedule.
    We study this protocol numerically for finite bivariate bicycle and surface codes, comparing logical error probabilities with memory experiments.}
    \label{fig:surfacecode}
\end{figure}
Here we motivate our general construction by describing four procedures for measuring a joint logical operator across two rotated surface-code patches.
A more general description follows in Section \ref{sec:quilts}, with a full structural analysis given in Appendix \ref{sec:structuretheorems}.
The four procedures come from first describing two subsystem code constructions called code quilts, and then applying to each both the stabiliser and the subsystem protocol.
The constructions are illustrated in Figure~\ref{fig:surfacecode}.
Suppose we have two logical qubits each encoded in a $d\times d$ rotated surface-code patch and wish to measure the logical $\bar{Z^1}\bar{Z^2}$ operator of the joint system (here and throughout, superscripts on logical operators denote the codeblock upon which the logical operator acts).
Let the joint stabiliser group of the two codeblocks be $G_0 = G_0^1 + G_0^2$.
Fix a representative of this logical consisting of the product of the weight-$d$ operators $\bar{Z^1}$ and $\bar{Z^2}$ each running along an edge of the two patches. 
(In panels $\textbf{a}$, $\textbf{b}$, $\textbf{c.}$ of Figure \ref{fig:surfacecode} the two patches are placed side-by-side, with one reflected, so that the logical operators face each other, to match the usual presentation of surface-code lattice surgery. Panels $\textbf{d}, \textbf{e}, \textbf{f.}$ use the convention adopted for the remainder of the paper, where the codeblocks are stacked vertically atop one another.)
We cannot measure this weight-$2d$ operator directly as the measurement would be impracticably high-weight, requiring a long syndrome extraction cycle and risking an abundance of damaging hook errors.
Instead, we decompose the measured operator into a product of disjoint lower-weight operators.
Two of many possible decompositions are: 
\begin{enumerate}
    \item A set of $\lfloor d/2 \rfloor$ weight-4 $ZZZZ$ operators and, for odd $d$, a single weight-2 $ZZ$ operator. The weight-4 operators each act on an edge of an $X$ plaquette incident on the logical operator in block one and the corresponding edge in block two, and the weight-2 operator completes the two operators. Alternatively,
    \item a set of $d$ weight-2 physical $ZZ$ operators each of which acts on one qubit in block 1 and the corresponding qubit in block 2.
\end{enumerate}
We will call these operators, which span two codeblocks, \textit{stitching operators}, and denote the group generated by them as $S$.
The first decomposition is presented because it reproduces standard surface-code lattice surgery, and this section is the only place that it is used. The analytical sparsity results, circuit constructions and numerical data in this work concern decompositions of the second kind, in which each stitching operator acts on exactly one qubit per codeblock.
It is immediate that these operators are each physically local, and immediate that their product is the $\bar{Z^1}\bar{Z^2}$ operator we wish to measure.
We appear to face an equally immediate problem: the stitching operators of both decompositions anticommute with some of the $X$-type stabiliser generators of the two original codeblocks.
This does not immediately preclude fault tolerance.
We proceed by treating the whole system of noncommuting operators as a subsystem code that we call a \textit{code quilt}, defined by 
\begin{equation}
    \quilt{G} = G_0 + S.
\end{equation}
The code quilt $\quilt{G}$ has $k^1 + k^2 - 1 = 1$ logical qubit, with the $\bar{Z}^1\bar{Z}^2$ logical operator demoted to a stabiliser in the centre of the group $Z(\quilt{G})$.
We can then either:
\begin{enumerate}
    \item Deform to the subsystem code by measuring its physical generators directly, and inferring the value of the central stabilisers $Z(\quilt{G})$; we will refer to this as \textit{the subsystem protocol}.
    \item Directly measure the stitching operators $S$ as well as a generating set for the operators in $\quilt{G}$ that commute with them. This is equivalent to deforming to a stabiliser code $\quilt{S} = \quilt{G}\cap S^\perp$ (this code results from $\quilt{G}$ by gauge-fixing). We will refer to this as \textit{the stabiliser protocol}.

\end{enumerate}
Deforming from two or more idle codeblocks to either the quilt subsystem code $\quilt{G}$ or the quilt stabiliser code $\quilt{S}$ implements the logical measurement we desire.
We defer the full treatment of these codes as partially initialised codes to Appendix \ref{sec:structuretheorems}. Intuitively, those stabilisers whose eigenvalues are inherited from the source codes as well as those obtained by gauge fixing in the stabiliser protocol form the trusted subspace of stabilisers.
In Section~\ref{sec:distance} we give various analytical sets of sufficient conditions for the stitched code to have the required operational distance, all of which are satisfied by this example.

The subsystem protocol applied to surface codes resembles the procedure described in \cite{nautrup2017interface}.
The reader may anticipate how the procedures presented in this section, particularly the subsystem protocol using weight-2 stitching operators, can be generalised to perform logical measurements over more than two codeblocks.
In the following section, we show how to perform joint and parallel measurements between any set of stabiliser codes, including those that are not as obviously topological as the surface code.

However, the construction fails on a single codeblock---a version on a single codeblock whereby individual qubits in the support of a logical operator are measured out would be trivially distance-reducing.
This invites a distinction between these protocols and other generalised surgery procedures that define measurement on a single codeblock and then either define higher-weight measurements as operating on a single direct-sum code or define adapters to combine the measurements.

\section{Higher-weight and parallel logical measurements: general code quilts}\label{sec:quilts}

\begin{figure}
    \centering
    \includegraphics[width=\linewidth]{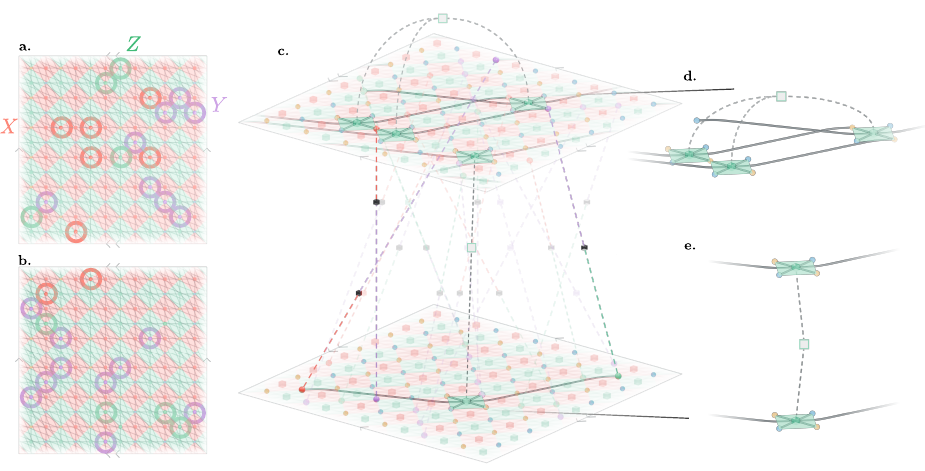}
    \caption{\textit{A general two-block code quilt consisting of two 72-qubit bivariate bicycle codes.} \href{https://stasiu51.github.io/Shatter/\#id=fig2-bb72-bb2generic-1087-viz\&bookmark=1}{Click to visualise this circuit in Shatter.}
    A quilt subsystem produced using our optimiser for measuring a randomly selected choice of $4^{12}\times 4^{12}/4$ parity-matched generally non-CSS logical operators.
    \textbf{a.} and \textbf{b.} The weight-18 logical representatives optimised to measure a randomly selected target logical (stored as \texttt{BB2\_GENERIC\_MATCHED-1087} in the accompanying data).
    \textbf{c.} The quilt stitch, showing with dashed lines the weight-2 stitching operators into which the optimised logical representative is decomposed.
    Also shown are one of the intra-block product stabiliser generators of the quilt on the upper block, consisting of the product of three $Z$-type checks, as well as one of the inter-block product stabilisers, consisting of one $Z$-type stabiliser on each of the two blocks. (These are elements of $H_{c}$ and $H_{ac}$ respectively, as described in Section \ref{sec:sparsity}.)
    Note that the $YX$ stitching operator that anticommutes with the stabiliser highlighted on the bottom block also anticommutes with its partner on the top codeblock via wrapping around the top torus.
    \textbf{d.} and \textbf{e.} Cut-away views of the intra- and inter-block product stabilisers.}
    \label{fig:bb-stitch}
\end{figure}

A natural generalisation of the stitching procedure is to joint logical measurements between more than two codeblocks.
In \cite{litinski2019game}, this is achieved by measuring long `snakes' between more than two patches of surface code.
In other generalised surgery protocols, this may be achieved by treating multiple codeblocks as a direct-sum code \cite{cohen2022lowoverhead}, or by developing adapters between extractor modules \cite{swaroop2024adapters,he2025extractors}.
Here, we take a direct approach that acknowledges the distinct identity of the multiple codeblocks but does not build any bridging ancilla system between them.
Our framework also accommodates a limited form of parallel logical measurement alongside higher-weight logical measurement.
This is described using a small classical linear code $B$ called the \textit{incidence code}, whose definition we motivate here with some examples. We also generalise the construction to codes other than the surface code.

For example, we may wish to measure some product $\bar{Z^1} \bar{Z^2_5} \bar{Y^3_3}$ supported on a surface code and two bivariate bicycle codes (a logical operator $Z^i_j$ refers to the $j$-th $Z$-type logical operator of the $i$-th codeblock).
We may also wish to reduce the time overhead of our logical circuit by performing some logical measurements \textit{in parallel}, so that one codeblock is involved in more than one logical measurement at once.
We concern ourselves with parallel measurements exemplified by the pair of simultaneous measurements $\bar{Z^1} \bar{Z^2_5}$, $\bar{Z^2_5} \bar{Y^3_3}$. 
We enforce that each measured product agrees as to which operator(s) is(are) to be measured on each codeblock---in this case, $\bar{Z^1}$, $\bar{Z^2_5}$, and $\bar{Y^3_3}$ respectively.

We can model general setups on $N$ codeblocks like this by specifying
\begin{enumerate}
    \item on each block, a measured subspace of logical operators $\mathcal{M}^i$; and
    \item a classical binary linear code called the incidence code, $B \le \ft^N$ on $N$ bits, with distance at least 2.
\end{enumerate}
Note that in the analytical analysis of operator weight throughout this work, we consider $N$ to be a small constant.
Joint and parallel measurements are then differentiated by the codewords of $B$: the examples above are described by the same per-block operators $\mathcal{M}^1 = \langle\bar{Z^1}\rangle$, $\mathcal{M}^2 = \langle\bar{Z^2_5}\rangle$, and $\mathcal{M}^3 = \langle\bar{Y^3_3}\rangle$, but differ in their incidence codes.
The weight-3 logical measurement of $\bar{Z^1} \bar{Z^2_5} \bar{Y^3_3}$ is described by $B = \langle 111 \rangle$, the 3-bit repetition code. 
The two parallel measurements $\bar{Z^1} \bar{Z^2_5}$, $\bar{Z^2_5} \bar{Y^3_3}$ are instead described by $B = \langle 110, 011 \rangle$, the 3-bit even-parity code.

We now give a slightly simplified definition of a \textit{code quilt} in which the measured products involve one nontrivial logical operator on each block.
A code quilt is naturally a subsystem code---more precisely a partially initialised code, in the sense of Appendix~\ref{sec:operationaldistance}, since the value of the measured product is not known in advance. We refer to it simply as a subsystem code throughout the main text. The subsystem protocol measures the gauge operators of this subsystem code $\quilt{G}$ directly, and the stabiliser protocol fixes its commuting stitching gauges and measures the resulting stabiliser code $\quilt{S}$.
Definition~\ref{def:codequilt} in Appendix~\ref{sec:structuretheorems} is more general, allowing several independent logical operators on each block, at the cost of requiring more mathematical machinery to analyse.

\begin{definition}[Code quilt, simplified]\label{def:codequilt-simplified}
    A \textit{code quilt} can be specified by the following data:
    \begin{enumerate}
        \item $N \ge 2$ disjoint codeblocks each supporting a stabiliser code $\mathcal{C}^i$ with parameters $n^i$, $k^i$, $d^i$, and stabiliser subspace $H^i \le \Pauli_{n^i}$;
        \item for each codeblock $\mathcal{C}^i$, a set of $W$ disjoint operators acting on the qubits of the codeblock, called \textit{local stitching operators}, written as the rows of a full-rank matrix $S^i \in \mathbb{F}_2^{W\times 2 n^i}$, whose product is a nontrivial logical operator $l_i$ of $\mathcal{C}^i$, where no proper subset of $l_i$ multiplies to a (trivial or nontrivial) logical operator;
        \item the \textit{incidence code}: a classical linear code $B\le \mathbb{F}_2^N$ of distance $d_B\ge 2$.
    \end{enumerate}
    For $p\in\Pauli_{n^i}$, write $p_\uparrow^{(i)}$ for the Pauli on the full $N$-block system with $p$ on block $i$ and no support on any other block. For $x\in\ft^W$ and $T\in\ft^{W\times N}$, define
    \begin{equation}
        \Stitch_i:\ft^W\rightarrow \Pauli_{n^i},
        \qquad
        \Stitch_i(x)=xS^i,
        \qquad
        \Stitch(T)=\prod_{i=1}^N \bigl(\Stitch_i(T_{\cdot i})\bigr)_\uparrow^{(i)}.
    \end{equation}

    These data first define the underlying subsystem code of the quilt, whose gauge group is generated by the local stabilisers and the stitching operators:
    \begin{equation}\label{eq:quiltdef-simplified}
        \quilt{G}
        =
        \left\langle
        H^i_\uparrow\ (1\le i\le N),\quad
        \Stitch(T)\ (T\in \ft^W\otimes B)
        \right\rangle.
    \end{equation}
\end{definition}

The stitching operators are therefore represented by $W\times N$ matrices $T$. Each row is a codeword of $B$, specifying the blocks on which a corresponding local stitching operator acts; each column selects the local stitching operators on one block.

In the subsystem protocol, the generators of the original codes $\mathcal{C}^i$ continue to be measured, along with the stitching operators $S^i$, and the stabilisers of the code quilt, including the logical result, are inferred from their values.
The stabiliser protocol uses the same structure, but fixes the gauges associated with the stitching operators to form the stabiliser code $\quilt{S}$.
According to the following theorem, this can only improve the code-capacity distance of the quilt. The downside is that in the stabiliser protocol we must measure the central stabilisers directly, rather than inferring their values from the product of stabiliser generators of the original code.

\begin{theorem}[Stabiliser protocol: code-capacity distance]\label{theorem:gauge-fixing-main}
    Fix the commuting stitching gauges of a code quilt, with their fixed values included among the error-detecting checks. The resulting partially initialised stabiliser code has operational distance at least that of the quilt.
\end{theorem}
\begin{proof}[Proof sketch]
    A harmful error that commutes with the fixed gauges cannot be a harmless gauge operator of the original quilt: completeness of the gauge fixing would make it a stabiliser of the fixed code. Thus every operational error of the fixed code is an operational error of the original quilt, and therefore any error that causes a failure of the stabiliser protocol would also cause one in the subsystem protocol. The precise statement and proof are given in Appendix~\ref{sec:gauge-fixing}.
\end{proof}

\section{Circuit implementations of the two protocols}\label{sec:circuits}
In this section we give explicit circuits to implement the subsystem and stabiliser protocols, given a code quilt.
For these circuits, and for the empirical analysis in the remainder of the paper, we restrict to the case where each local stitching operator is weight-1, equivalently, each stitching operator is weight-$N$ with $N$ the number of codeblocks involved in the measurement. For example, the second of the two code quilts described in Section~\ref{sec:twoblock} for the surface code is of this kind.

We assume that the codeblocks that are involved in the logical measurement procedure are equipped with a standard syndrome extraction circuit, which proceeds by resetting a series of ancilla qubits, performing a series of entangling gates between data and ancilla qubits, and then measuring the ancilla qubits (this coincides with the definition of a `normal syndrome extraction circuit' in~\cite{shaw2026optimising}).

The two protocols are illustrated in panels \textbf{b.} and \textbf{c.} of Figure~\ref{fig:circuits}.
 Panel \textbf{a.} first establishes the visual notation used in the subsequent panels by showing two rounds of an alternating syndrome extraction circuit for a single $X$-type stabiliser of an idle bivariate bicycle code.
Detecting regions of the circuit are shown as coloured outlines around qubits; sometimes these outlines also extend between qubits for emphasis. The state of the detectors is shown in each frame before the effect of the gates depicted in that frame has been applied to them (note this is the opposite convention to Stim's slice diagrams).
In panel \textbf{a.}, the first round consists of an $X$-type reset of a single ancilla qubit in the left-hand frame. 
The reset is followed by 6 CNOT gates in the middle-top frame that exchange the support of a pink $X$-type detector with that of a red $X$-type detector, leading to the red detector being localised onto the ancilla qubit. Measurement of the ancilla in the right hand frame then `closes off' the red detecting region and resetting it creates a new, weight-one detecting region of a single ancilla. In the second round, the same CNOTs are performed in reverse order, leading back to the starting configuration whereupon the pink detector is closed off by measurement of the ancilla in the left-hand frame.

\subsection{Subsystem protocol circuit}
\begin{figure}
    \centering
    \includegraphics[width=0.96\linewidth]{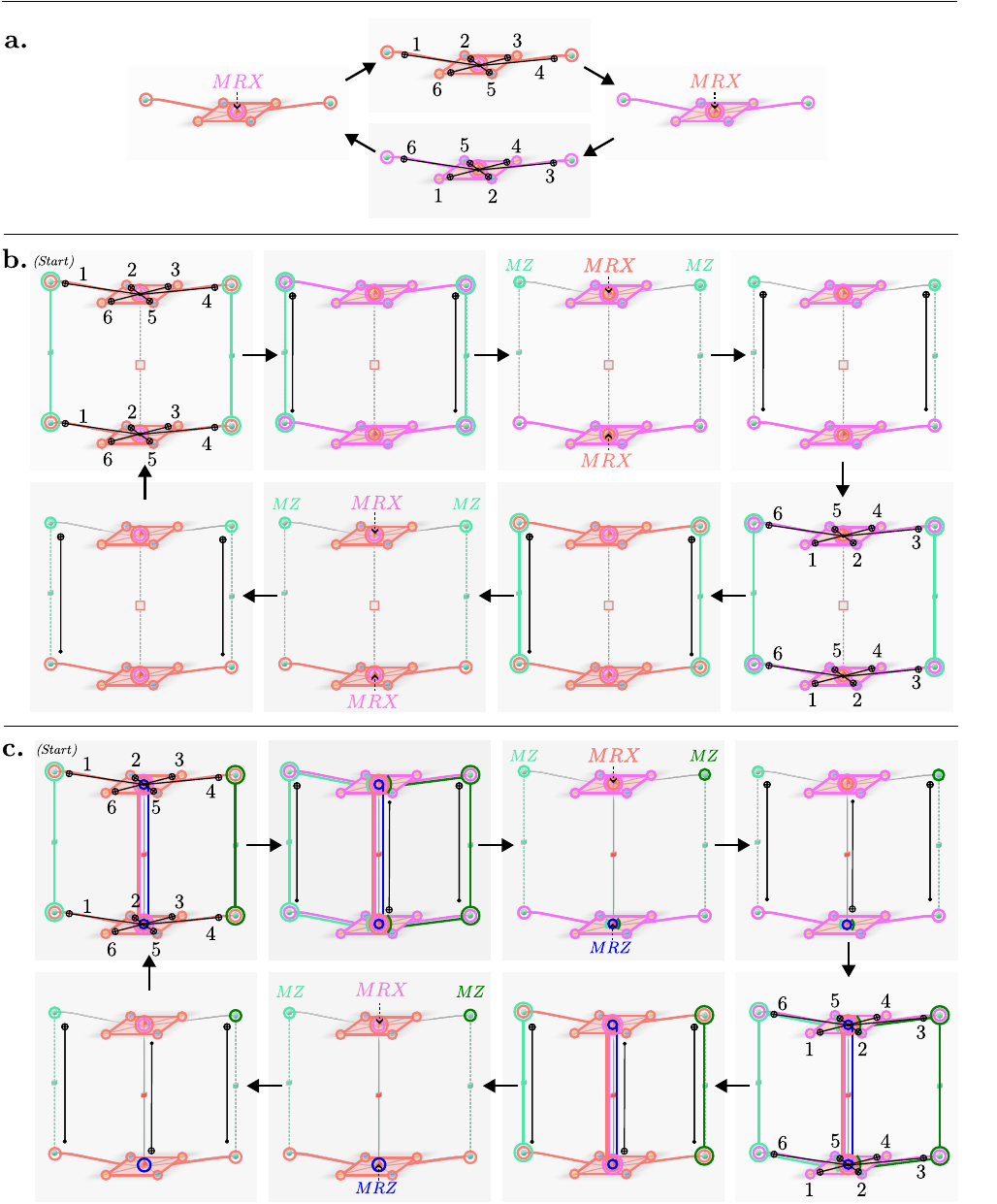}
    \caption{\textit{Quilt code syndrome extraction circuits in the subsystem and stabiliser protocols.} 
    Coloured overlays indicate detectors; red and pink are $X$-type; bright green, dark green, and blue are $Z$-type.
    \textbf{a.} Two rounds of an alternating syndrome extraction circuit for a bivariate bicycle code. Every other round, the CNOT gates are performed in reverse order.
    \textbf{b.} The subsystem protocol, the main focus of this work, illustrated for a code quilt consisting of two codeblocks: the stitching operators are measured at the same time as the original stabiliser generators thanks to two additional `morphing' CNOT rounds.
    \textbf{c.} The stabiliser protocol, showing the direct measurement of a cross-block product stabiliser using a temporary GHZ state created between the ancilla qubits of the two blocks. In this case, the stabiliser protocol has the same circuit depth as the subsystem protocol; in general the product stabilisers will be higher weight and require more rounds to construct GHZ states to measure them.
    }
    \label{fig:circuits}
\end{figure}

The subsystem protocol is illustrated for the case of a $ZZ$-type measurement across $N=2$ codeblocks of a bivariate bicycle code in panel \textbf{b.} of Figure~\ref{fig:circuits}. Note that we start our description in the second-from-left frame in the upper row of the panel.
\begin{enumerate}
    \item The usual gates of the syndrome extraction circuits for the constituent codes are performed.
    Every other round, the order of the CNOTs of the local SEC is reversed, following the alternating-circuit approach of~\cite{shaw2026optimising}.
    \item Single-qubit gates are applied to transform the stitching operators into (WLOG) $Z$-type operators. (In the figure example, the operators are already $Z$ type, so no single-qubit gates are performed.)
    \item A tower of `morphing' CNOTs~\cite{mcewen2023relaxing,shaw2026optimising} is performed to contract the support of each stitching operator onto a single qubit on one of the codeblocks. This and the previous step can often be interleaved with idle spots at the start or end of step 1, reducing time overhead. (In the figure example, there are $N=2$ codeblocks so the morphing towers in fact consist of a single layer of CNOTs.)
    \item The ancilla qubits and data qubit that are single-qubit roots of the CNOT towers are measured in the appropriate bases, and the syndrome ancillas are then reset for the next round.
    \item The tower of CNOTs and the local basis changes are undone.
\end{enumerate}

\subsection{Stabiliser protocol circuit}

The stabiliser protocol fixes the commuting stitching gauges and directly measures the product stabilisers of the code quilt. Together these generate the gauge-fixed stabiliser code of Appendix~\ref{sec:gauge-fixing}.
We can measure these product stabilisers using a constant number of additional CNOT layers, by constructing and contracting a GHZ state over the ancilla qubits associated with the original stabilisers that constitute each product stabiliser. In panel \textbf{c.}, we show how a stabiliser that is the product of one $X$-type original stabiliser on each of $N=2$ codeblocks is measured by constructing a GHZ state across two ancilla qubits, with a single $Z$-type flag detector defined on the bottom ancilla qubit. The construction is similar in spirit to the Hadamard-measurement step in the original magic state cultivation protocol~\cite{gidney2024cultivation}. Note that, in contrast to the subsystem protocol, in which the stitching operators multiply to a single high-weight detector shown in bright green, the two $Z$-type stitching operators each give rise to distinct detectors (bright green and dark green, respectively).

The construction of the GHZ state can happen in parallel with the `morphing tower' CNOT operations described in the subsystem protocol circuit.
To measure stabilisers that are products of both $X$- and $Z$-type original-code stabilisers, single-qubit gates are performed between the GHZ construction/contraction and the local syndrome extraction circuit. Stabilisers that are products of more than two original-code stabilisers are measured by producing a larger GHZ state using a tree of CNOTs, and correspondingly give rise to flag detectors on all but one of the ancilla qubits involved.

\section{Finding fault-tolerant measurements between bivariate bicycle and surface codes}\label{sec:repfinding}
In this section, we narrow our focus to certain specific classes of logical measurements between bivariate bicycle and surface codes.  Over the rest of this paper, we describe the generation and optimisation of quilts to measure a wide range of randomly selected target logical operators over these classes.
The quilts we generate allow us to empirically demonstrate the existence of distance-preserving quilts, apparently universally, for the whole range of logical operators in these classes. Although we give various analytical conditions which guarantee distance preservation in Section \ref{sec:distance}, most notably the existence of a representative of the target logical of minimum weight, the quilts we develop throughout show that we can commonly find distance-preserving quilts even where these conditions are not satisfied.
They also provide data for an empirical study of the sparsity and congestion of the detectors that result from the construction.
We consider four different patterns of logical measurement, listed below, consisting of various combinations of measurements on copies of bivariate bicycle codes with surface codes of matching distance:
\begin{figure}
    \centering
    \includegraphics[width=\linewidth]{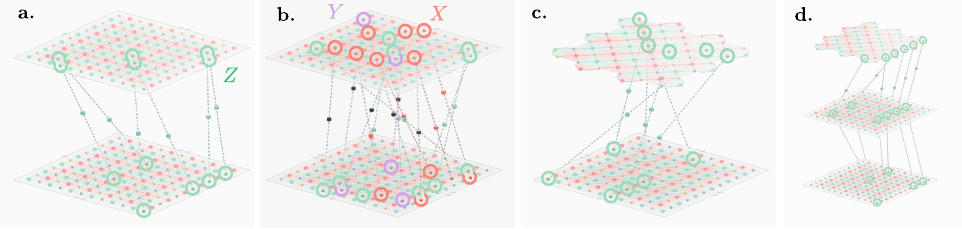}
    \caption{\textit{The four classes of code quilt generated from representatives of logical operators in the four classes of target logical cosets that we empirically study in this work.} 
    Stitching operators are shown with dashed lines with endpoints coloured to indicate Pauli type; product stabilisers are not shown.
    \textbf{a.} (\texttt{BB2\_ZZ}) a code quilt for measuring a  joint $Z$-type logical operator between two bivariate bicycle codes (here the 72-qubit bivariate bicycle code). 
    \textbf{b.} (\texttt{BB2\_GENERIC\_MATCHED}) a code quilt for measuring a general non-CSS logical operator between two bivariate bicycle codes.
    \textbf{c.} (\texttt{SC\_BB\_ZZ}) a code quilt for measuring a joint $Z$-type logical operator between a bivariate bicycle code and a surface code (here the distance-6 rotated surface code).
    \textbf{d.} (\texttt{SC\_BB2\_ZZZ}) a code quilt for measuring a joint $Z$-type logical operator between two bivariate bicycle codes and a surface code.
    }
    \label{fig:classes}
\end{figure}
\begin{enumerate}
    \item \texttt{BB2\_ZZ}: the product of two randomly selected $Z$-type logical operators, one on each of two bivariate bicycle codes. Accounting for the symmetry between the blocks, there are $2^{12}\times 2^{12}/2$ logical cosets of the joint code in this class.
    \item \texttt{BB2\_GENERIC\_MATCHED}: the product of two randomly selected logical operators with no restriction on Pauli type, one on each of two bivariate bicycle codes, with matched Hamming-weight parity $|L_1| = |L_2|$ (mod 2), as explained below. Accounting for this parity condition as well as symmetry, there are $4^{12}\times 4^{12}/4$ total logical cosets in this class (half of all possible two-block logical measurements).
    \item \texttt{SC\_BB\_ZZ}: a randomly selected $Z$-type logical operator of the bivariate bicycle code, multiplied with the $Z$-type logical operator of the rotated surface code. This class contains $2^{12} = 4{,}096$ logical cosets.
    \item \texttt{SC\_BB2\_ZZZ}: two randomly selected $Z$-type logical operators of the bivariate bicycle code, multiplied together with the $Z$-type logical operator of the rotated surface code, giving a measured product across three codeblocks. This class contains $2^{12}\times 2^{12}/2$ logical cosets.
\end{enumerate}

The latter two classes are chosen to model the process of teleporting a non-Clifford gate into one or more bivariate bicycle codes using a resource state stored in a surface code \cite{litinski2019game, gidney2024cultivation}. 

A code quilt in which every stitching operator has weight $N$ requires the projections of the measured logical operator onto its codeblocks to have equal weight. Every stabiliser of a bivariate bicycle code has even weight in each sector separately, so the weight parity $|L|$ (mod 2) of a logical coset is well defined. 
A pure $X$- or pure $Z$-type logical operator is automatically of even weight, so any two logical cosets automatically have the same weight parity, and in practice representatives can be found of equal weight.
However, the weight of a general non-CSS operator of the BB codes may be odd or even, owing to the overlap between its $X$ and $Z$ parts. Requiring that the weight parities of the target logical operators on the two blocks agree leads to the factor of two reduction in membership of the \texttt{BB2\_GENERIC\_MATCHED} class vs the entire two-block logical space.

We define 12 classes of target logical coset by applying each of the four patterns to the 72-qubit, 144-qubit (`gross') and 288-qubit (`two-gross') bivariate bicycle codes, in each case in combination with the rotated surface code of matching distance.
From each class, we randomly selected 2,500 distinct target logical cosets at $d=6$ and at $d=12$, and 1,250 at $d=18$: 10,000, 10,000 and 5,000 targets respectively, or 25,000 in total. Note that for the \texttt{SC\_BB\_ZZ} classes, the chosen targets cover $61\%$ of the class at $d=6$ and $d=12$, and $31\%$ at $d=18$. 

We then attempted to find explicit logical representatives for each of these target logical cosets and decompositions into weight-2 stitching operators, so that each representative and decomposition defines a code quilt construction that satisfies the various conditions in Definition~\ref{def:codequilt-simplified} and is distance preserving at the code-capacity level.

To find the representatives and decompositions, we ran a heuristic optimisation algorithm based on simulated annealing (source code here~\cite{quantumquilts}), requiring each representative to satisfy the following desiderata:
    \begin{enumerate}
        \item The logical representative must be in the correct target logical coset.
        \item It must be irreducible; that is, it contains no complete stabilisers or other nontrivial logical operators in any proper subset of its support.
        \item The projections of the logical operator onto the respective codeblocks (e.g. the representatives of $L_1$ and $L_2$ in the first class) must have equal weight so the overall logical operator admits a decomposition into weight-$N$ stitching operators. 
        \item It must satisfy the SkipTree connectivity condition, which bounds the weight and the qubit degree of a basis for a certain subspace ($H_{ac}$, see later) of the stabilisers of the quilt, as explained in Section \ref{sec:sparsity} and Appendix~\ref{sec:sparsityanalytical}.
        \item The code quilt resulting from the representative and decomposition must have an operational distance equal to that of the source codes.
    \end{enumerate}
    Examples of the generated code quilts for each of the four classes involving the 72-qubit bivariate bicycle code and the distance-6 surface code are shown in Figure~\ref{fig:classes}.
    The first four desiderata are cheap to check computationally, whereas the distance condition (5) is not.
    To estimate the code-capacity operational distance, we primarily used a modification of \texttt{QDistRnd}~\cite{pryadko2022qdistrnd} optimised for non-CSS codes and computational efficiency.
    To bias the annealer toward representatives that give rise to quilts of the required distance, it is guided by the following heuristics:
    \begin{enumerate}
        \item \textit{$X$--$Z$ overlap.} For general non-CSS targets, the number of sites at which a representative's $X$ and $Z$ supports coincide is pushed above a threshold, which prevents the representative from degenerating towards a CSS operator on either block.
        \item \textit{Low weight.} Among representatives that satisfy the above, lower total weight is preferred, which keeps the interface width $W$ close to the code distance and tends to give sparser inherited stabilisers.
    \end{enumerate}
    The decomposition of the logical representative into weight-2 stitching operators reduces to choosing (a) bijection(s) between the projections of the representatives onto the respective codeblocks. We first try the deterministic pairing given by the SkipTree ordering described in Appendix~\ref{sec:sparsityanalytical}; if the resulting quilt is not found to be distance preserving, several random pairings are tried. Often, where a representative was found not to give rise to a distance-preserving code quilt under one decomposition, choosing a different random decomposition restored the distance. 
    It may interest the reader that we derived a heuristic for optimising the decomposition, based on anyonic charge, designed to optimise the likelihood of distance preservation by forcing a pathological logical operator to `use all of the stitching operators' between the codeblocks. This is detailed in Appendix~\ref{sec:chargepairing}. 
    This heuristic was found to be almost-always satisfied by randomising the decomposition, and no more effective than randomisation, so it was not enforced over the final corpus of representatives. However, it may find application in future work.

    The collection provided with this paper contains 25,000 representatives, together with decompositions of them, that satisfy the deterministic interface conditions and are estimated using \texttt{QDistRnd} to be code-capacity distance preserving. Our distance estimates are obtained by running \texttt{QDistRnd} in an attempt to find dressed logical operators of weight less than the target distance; we declare a quilt to be distance preserving when 200,000 trials have failed to find one. How much confidence this result deserves is quantified in Appendix~\ref{sec:tweedietest}.
    The resulting library of representatives is available at~\cite{quantumquilts}.
    This evidence motivates further analytical sufficient conditions for code-capacity distance preservation.
    We did not attempt to construct circuits and check their circuit-level distance for each of the 25,000 chosen representatives, instead analysing representative examples from each class, as detailed in the following section.

    \subsection{Circuit depth of syndrome extraction for the four classes of logical measurement}
    We take the length of a syndrome extraction circuit to be the number of timesteps required to do all the unitary gates, measure, and reset the qubits, assuming all those instructions take the same unit time. This is clearly a dubious assumption that depends strongly on which physical platform implements the circuits.
    The alternating technique that we use (Section~\ref{sec:circuits}), though very convenient for the purposes of circuit-distance estimation, comes at the cost of preventing us from interleaving the circuits by offsetting the syndrome extraction cycles of certain qubits relative to others, as could be done in the original, non-alternating schedule \cite{bravyi2024high}. In the control case of idle memory, which we also alternate in our decoding comparisons, this increases the length of the cycle by one, so our control cases for the bivariate bicycle code each take 9 timesteps per round including measure and reset instructions. In comparison, our merged codes, using the subsystem protocol, have syndrome extraction rounds that take between 10 and 12 timesteps. The \texttt{SC\_BB\_ZZ} and \texttt{BB2\_ZZ} circuits each take 10 timesteps, due to the insertion of morphing CNOTs, which can be partially interleaved with the existing CNOTs. 
   The circuits for \texttt{BB2\_GENERIC\_MATCHED} and \texttt{SC\_BB2\_ZZZ} quilts each take 12 timesteps, in the former case due to the single-qubit Clifford required to measure non-CSS stitching operators, and in the latter case due to the weight-3 stitching operators. The circuits are therefore between 1 and 3 timesteps longer than alternating idle memory, and would be between 2 and 4 timesteps longer than idle memory in the standard schedule.

\section{Distance results}\label{sec:distance}
In this section we collect analytical and empirical results related to the code-capacity and circuit-level distances of code quilts.

\subsection{Analytical distance results}

We first give the main analytical results of the paper, whose proofs are given in full in the Appendix along with additional structural analysis of the code quilts. The quilts studied empirically in the remainder of this work do not generally satisfy any of the sufficient conditions required for the following theorems to apply, and yet are commonly found to be distance preserving anyway. This suggests further analytical work, perhaps limited to the structure of specific codes, may give rise to broader sufficient conditions for distance preservation than the ones given here. 

\begin{theorem}[Weight-$D$ quilt distance]\label{theorem:weightDquilt-main}
    Let the constituent codes of a quilt have distance $D$. If corresponding local stitching operators have the same weights $|S^i_a|=w_a$ on every block and $\sum_{a=1}^W w_a=D$, then the quilt has (code capacity) operational distance at least $D$.
\end{theorem}
\begin{proof}[Proof sketch]
    Call a subset of the local stitching operators on one block \textit{closed} when its product commutes with every stabiliser of that block. Under the hypothesis, the only closed subsets are the empty one and the whole set: any other would multiply to an operator of weight strictly below $D$ which, together with its complement, would supply a nontrivial logical operator of $\mathcal C^i$ below the code distance.

    Now take a silent error $E$---one which flips no trusted detectors of the partially initialised code---of weight below $D$. On each block, $E$ fails to commute with some set of the local stitching operators, and multiplying $E^i$ either by that set or by its complement closes it off into an operator of the original code. The two completions differ by the measured logical operator, so at least one of them is nontrivial and costs at least $D$. Since the local stitching operators are disjoint with weights summing to $D$, the two completions cost $D$ between them, and $|E^i|$ is at least the cost of whichever one was not used. Summing over the blocks either reaches $D$, or else every completion is an original-code stabiliser and $E$ was a harmless gauge operator all along. Theorem~\ref{theorem:weightDquilt} gives the full argument, in which the admissible patterns of anticommutation are organised by a tensor subspace $Y^\perp\otimes B^\perp$.
\end{proof}

A \textit{homogeneous quilt} has identical constituent codes and local stitching operators, up to permutations and single-qubit gates, as per Definition~\ref{def:homogeneousquilt}.

To describe its distance, consider a single block and add all its local stitching operators to the gauge group. This defines the block's \emph{perforated code}. Its error-detecting checks are the original stabiliser products that commute with every local stitching operator, forming the space $H_c^i$. An undetected error is harmless if it is a product of original stabilisers and local stitching operators. The perforated operational distance $\dotd_i$ is the least weight of an error that commutes with $H_c^i$ and lies outside this enlarged gauge group. Thus it measures the protection remaining on one block when the local stitching operators are treated as gauge degrees of freedom. Definition~\ref{def:perforatedcode} gives the formal construction. For a homogeneous quilt these distances coincide, and we write their common value as $\dotd$.

\begin{theorem}[Homogeneous quilt distance]\label{theorem:homogeneousquiltdistance-main}
    Let the constituent codes of a homogeneous quilt have distance $D$, and let the incidence code have distance $d_B$. If $d_B\dotd\ge D$, then the quilt has operational distance at least $D$.
\end{theorem}
\begin{proof}[Proof sketch]
    Because every block carries the same code and the same local stitching operators, the content of one block can be carried onto any other without changing any weight. An error spread across several blocks can therefore be collected onto a single block, where its logical effects add and where it is no heavier than the error we started with. Were the blockwise logical effects to differ, that collected error would be a nontrivial logical operator of one constituent code of weight below $D$. So a silent error of weight below $D$ must act with the same logical effect on every block.

    The perforated code is defined such that an error realising this common logical effect on one block, with the local stitching operators treated as gauge degrees of freedom, has weight at least $\dotd$. An operationally nontrivial pattern of blockwise effects obeys the incidence code (so that it must have support on a subset of the codeblocks corresponding to a codeword of $B$) and so occupies at least $d_B$ blocks, giving a total weight of at least $d_B\dotd$. The full argument is Theorem~\ref{theorem:homogeneousquiltdistance}.
\end{proof}

Theorem~\ref{theorem:homogeneousquiltdistance-main} means that, given logical representatives of a target logical coset that produce a perforated code with distance $d/n$, any measurement of logical operators of at least $n$ codeblocks of the same code, where each logical operator is related to every other by permutations and single-qubit gates, is automatically distance preserving. Its cross-block stabilisers ($H_{ac}$, Section~\ref{sec:sparsity}) are also manifestly sparse as they are generated by products of at most one original code stabiliser on each block, where we assume, as we do throughout this work, that $N$ is a small constant. By leveraging automorphisms, we found code quilts satisfying $\dotd \ge d/2$ for $65.8\%$ of the $Z$-type logical cosets of the 72-qubit bivariate bicycle code, $99.0\%$ of those of the 144-qubit code and $100\%$ of those of the 288-qubit code; for $100\%$ of those of the symmetric 162-qubit bivariate bicycle code described in~\cite{eberhardt2024logical}; and for $100\%$ of those of the $[[30,8,4]]$, $[[62,10,6]]$, $[[126,12,10]]$, $[[254,14,16]]$ and $[[510,16,24]]$ generalised bicycle codes of~\cite{kovalev2013quantum,panteleev2021degenerate,lin2025singleshot} as deployed in~\cite{webster2026pinnacle}. The representatives found this way, one per automorphism orbit together with the evidence for each, are provided as a separate bank and summarised in Appendix~\ref{sec:homogeneousbank}.

Each of the remaining results addresses the case in which a single measured product runs across every codeblock.

\begin{definition}[Atomic and nuclear quilts, simplified]\label{def:atomicnuclear-main}
    An \textit{atomic quilt} $\atom{\mathcal C}$ is a quilt whose incidence code is the $N$-bit repetition code $B=\langle\mathbbm 1_N\rangle$.

    Two codeblocks have the same \textit{local code/stitch type} when their codes and local stitching operators are represented by the same pair $(H,S)$, that is, when they agree up to the weight-preserving identifications used for homogeneity. Write $m_t$ for the number of blocks of type $t$. 
    
    When at least two types occur, the \textit{nuclear quilt} $\nucleus{\mathcal C}$ is the atomic quilt carried by one representative block of each type. When only one type occurs the atomic quilt is homogeneous, and Theorem~\ref{theorem:homogeneousquiltdistance-main} applies directly. Definition~\ref{def:atomicnuclear} gives the general statement.
\end{definition}

For example, consider a code quilt consisting of $N=9$ codeblocks, of which 3 are 144-qubit `gross' bivariate bicycle codes, 3 are 288-qubit `two-gross' bivariate bicycle codes, and 3 are the distance-12 rotated surface code, and where the local stitching operators multiply to the joint logical operator
\begin{equation}
\bar{L} = \bar{X}_1^1\bar{X}_1^2\bar{X}_1^3\ \bar{Z}_2^4\bar{Z}_2^5\bar{Z}_2^6\ \bar{X}^7\bar{X}^8\bar{X}^9,
\end{equation}
so three unique logical operators are measured three times apiece. We say that within each group of three codeblocks, the codeblocks have the same stitching type.
If the incidence code is $B = \langle 111\ 111\ 111 \rangle$, then the quilt $\atom{\mathcal{C}}$ is atomic.
Its nuclear quilt $\nucleus{\mathcal C}$  would consist of one gross code, one two-gross code, and one distance-12 rotated surface code, and local stitching operators that multiply to 
\begin{equation}
\bar{L} = \bar{X}_1^1 \bar{Z}_2^2 \bar{X}^3,
\end{equation}
with incidence code $B = \langle 111\rangle$. As with all nuclear quilts, it is atomic, and is its own nuclear quilt. 

As with homogeneous quilts, this definition allows for weight-preserving identifications such as automorphisms. For instance, a code quilt like the $N=9$ example above but where the local stitching operators instead multiply to 
\begin{equation}
\bar{L} = \bar{X}_1^1\phi^2(\bar{X}_1^2)\phi^3(\bar{X}_1^3)\ \bar{Z}_2^4\phi^5(\bar{Z}_2^5)\phi^6(\bar{Z}_2^6)\ \bar{X}^7\phi^8(\bar{X}^8)\phi^9(\bar{X}^9),
\end{equation}
where $\phi^i$ are qubit permutations that preserve the stabiliser group and hence map logical operators to other logical operators, would have the same $N=3$ nuclear quilt as above. Note neither atomicity nor homogeneity implies the other. All the codes we consider in the empirical studies in this work are atomic and nuclear, but are not necessarily homogeneous.

A \textit{subquilt} is obtained by keeping a subset $I$ of the codeblocks of a quilt together with their local stitching operators and restricting the incidence code to those coordinates; this is again a quilt whenever the restricted code retains distance at least $2$. A family of subquilts \textit{generates} the quilt when the constraints they impose, lifted back to all $N$ blocks, intersect to exactly the incidence code and measured logical subspace of the whole quilt, so that consistency checked piecewise is consistency checked globally. Definitions~\ref{def:subquilt} and~\ref{def:generatingsubquilts} make this precise.

\begin{theorem}[Generator-subquilt distance]\label{theorem:generatorsubquilt-main}
    If a quilt admits a generating set of subquilts, each of operational distance at least $D$, then the quilt has operational distance at least $D$.
\end{theorem}
\begin{proof}[Proof sketch]
    A silent error of weight below $D$ restricts on each generator to an error compatible with that subquilt and of no greater weight, so the assumed distance makes each restriction a harmless gauge operator there. Its logical label therefore lies in every lifted measured subspace, and the generating property collapses that intersection to the measured subspace of the full quilt, making the error harmless globally. The full argument is Theorem~\ref{theorem:generatorsubquilt}.
\end{proof}

For an atomic quilt the incidence constraint is generated by weight-two checks, so pairs of blocks suffice at the incidence level and the generating family can be taken small.

\begin{corollary}[Atomic quilt generators]\label{corollary:atomicgenerators-main}
    Let $\atom{\mathcal C}$ be an atomic quilt in which at least two local code/stitch types occur. Suppose the nuclear quilt $\nucleus{\mathcal C}$ has operational distance at least $D$, and that for each type $t$ of multiplicity $m_t>1$ the copies of $t$ are joined by a connected family of homogeneous atomic subquilts of operational distance at least $D$. Then
    \begin{equation}
        \atom d_{\mathrm{op}}\ge D.
    \end{equation}
\end{corollary}
\begin{proof}[Proof sketch]
    The nuclear quilt ties one representative of each type together and the within-type generators propagate that agreement to every remaining block, so the two families together generate the atomic incidence code and measured subspace. Theorem~\ref{theorem:generatorsubquilt-main} then applies. See Corollary~\ref{corollary:atomicgenerators}.
\end{proof}

Because a homogeneous atomic subquilt on $m_t$ blocks has incidence distance $m_t$, Theorem~\ref{theorem:homogeneousquiltdistance-main} discharges the within-type generators against a single number per type, giving the certificate we actually evaluate.

\begin{corollary}[Atomic quilt perforated-distance certificate]\label{corollary:atomicperfcertificate-main}
    Let $\atom{\mathcal C}$ be an atomic quilt in which at least two local code/stitch types occur, and write $\dotd_t$ for the perforated operational distance of type $t$. If the nuclear quilt $\nucleus{\mathcal C}$ has operational distance at least $D$, and
    \begin{equation}
        m_t\dotd_t\ge D
        \qquad\text{for every local code/stitch type } t,
    \end{equation}
    then
    \begin{equation}
        \atom d_{\mathrm{op}}\ge D.
    \end{equation}
\end{corollary}
\begin{proof}[Proof sketch]
    Each within-type subquilt is homogeneous on $m_t$ blocks with incidence distance $m_t$, so $m_t\dotd_t\ge D$ certifies it through Theorem~\ref{theorem:homogeneousquiltdistance-main}. Corollary~\ref{corollary:atomicgenerators-main} then gives the result. See Corollary~\ref{corollary:atomicperfcertificate}.
\end{proof}

\subsection{Empirical code-capacity distance of selected code quilts on bivariate bicycle and surface codes}

As described in Section~\ref{sec:repfinding}, through an optimisation algorithm, we were able to find logical representatives for the 25,000 target logical cosets, chosen randomly in advance, that were estimated to give rise to code quilts that preserved the code-capacity operational distances of the input codes.
The code-capacity operational distances of the constructed code quilts were estimated using a specialised implementation of the standard \texttt{QDistRnd} distance-finding algorithm~\cite{pryadko2022qdistrnd}. In Appendix~\ref{sec:tweedietest} we develop a statistical model of the distance-estimation protocol. For the $d=6$ quilts we check it against an exact meet-in-the-middle search: in all four classes the finder and the exact search agreed on every quilt, and the model recovered the true fraction of distance-preserving quilts. Assuming that the same model describes the qualitative behaviour of the $d=12$ and $d=18$ quilts, we derive estimates of the number of quilts in the data provided with this paper that are not in fact distance preserving: overwhelmingly none for three of the four $d=12$ classes; at most a few for one $d=18$ class, and around 5 of 1,250 for another. For the remaining three classes the finder conceded too rarely for the model to be fitted; we suspect that the rates of failure are low for these as well, since rare concession is itself what a population of mostly distance-preserving quilts looks like. The estimates are summarised in Table~\ref{tab:codecap-distance}; the methodology of Appendix \ref{sec:tweedietest} may be of independent interest. The full set of chosen representatives (as well as code to reproduce the full code quilt) is available here~\cite{quantumquilts}.

\begin{table}[htbp]
\centering
\small
\setlength{\tabcolsep}{4pt}
\begin{tabular}{@{}llccccc@{}}
\toprule
Codes & Target logical class
      & \begin{tabular}[b]{@{}c@{}}Representatives\\ found\end{tabular}
      & \begin{tabular}[b]{@{}c@{}}Idle\\ distance\end{tabular}
      & \begin{tabular}[b]{@{}c@{}}Estimated\\ quilt distance\end{tabular}
      & Trials
      & \begin{tabular}[b]{@{}c@{}}FNR\\ (App.~\ref{sec:tweedietest})\end{tabular} \\
\midrule
\midrule
\multirow{4}{*}{\shortstack[l]{$n=72$ BB code\\ $+\ d=6$ RS code}}
  & \texttt{SC\_BB\_ZZ} & 2{,}500 & 6 & \textcolor{green!55!black}{6} & $2\times10^5$ & $<10^{-6}$ \\
  & \texttt{BB2\_ZZ} & 2{,}500 & 6 & \textcolor{green!55!black}{6} & $2\times10^5$ & $<10^{-6}$ \\
  & \texttt{BB2\_GENERIC\_MATCHED} & 2{,}500 & 6 & \textcolor{green!55!black}{6} & $2\times10^5$ & $<10^{-6}$ \\
  & \texttt{SC\_BB2\_ZZZ} & 2{,}500 & 6 & \textcolor{green!55!black}{6} & $2\times10^5$ & $<10^{-6}$ \\
\midrule
\multirow{4}{*}{\shortstack[l]{$n=144$ BB code\\ $+\ d=12$ RS code}}
  & \texttt{SC\_BB\_ZZ} & 2{,}500 & 12 & \textcolor{green!55!black}{12} & $2\times10^5$ & $<2\times10^{-5}$ \\
  & \texttt{BB2\_ZZ} & 2{,}500 & 12 & \textcolor{green!55!black}{12} & $2\times10^5$ & $<10^{-5}$ \\
  & \texttt{BB2\_GENERIC\_MATCHED} & 2{,}500 & 12 & \textcolor{green!55!black}{12} & $2\times10^5$ & $<2\times10^{-5}$ \\
  & \texttt{SC\_BB2\_ZZZ} & 2{,}500 & 12 & \textcolor{green!55!black}{12} & $2\times10^5$ & -- \\
\midrule
\multirow{4}{*}{\shortstack[l]{$n=288$ BB code\\ $+\ d=18$ RS code}}
  & \texttt{SC\_BB\_ZZ} & 1{,}250 & 18 & \textcolor{green!55!black}{18} & $2\times10^5$ & $2\times10^{-4}$ \\
  & \texttt{BB2\_ZZ} & 1{,}250 & 18 & \textcolor{green!55!black}{18} & $2\times10^5$ & $4\times10^{-3}$ \\
  & \texttt{BB2\_GENERIC\_MATCHED} & 1{,}250 & 18 & \textcolor{green!55!black}{18} & $2\times10^5$ & -- \\
  & \texttt{SC\_BB2\_ZZZ} & 1{,}250 & 18 & \textcolor{green!55!black}{18} & $2\times10^5$ & -- \\
\bottomrule
\end{tabular}
\caption{\textit{Code-capacity operational distance of the code quilts in each of the 12 classes.} `Trials' is the number of trials of the distance finder that every quilt survived, and FNR is the estimated fraction of surviving quilts that nonetheless have distance below the estimate, from Table~\ref{tab:codecap} of Appendix~\ref{sec:codecapcalibration}, with a dash where the class conceded a sub-distance logical operator too rarely to be calibrated.}
\label{tab:codecap-distance}
\end{table}

For some classes, finding representatives that give rise to distance-preserving quilts was remarkably easy: for instance, $78\%$ of the quilts first generated by the optimiser for the \texttt{BB2\_GENERIC\_MATCHED} class at $d=12$ were estimated to be distance preserving. Of those that were not, randomly permuting the pairing of the qubits within the stitching operators a few times typically restored the distance.

\subsection{Empirical circuit-level distance of quilt surgery}

Reliably estimating the circuit-level distance of a protocol is substantially more challenging than estimating code-capacity distance. We randomly selected one target logical coset from each of the 12 classes in the representative dataset, and constructed the subsystem protocol circuits detailed in Section~\ref{sec:circuits} as Stim~\cite{gidney2021stim} circuits from code quilts that were estimated to be distance preserving at the code-capacity level. Each circuit consists of $d$ rounds of the merged code quilt, with a single round of the split codes both before and afterwards.

We then applied a generalised version of the circuit `folding' introduced by Shaw and Terhal~\cite{shaw2026optimising} to reduce the resulting detector error model to one corresponding to a single stitching round. We emphasise that this technique, which we treat using non-CSS fault complexes, may be of independent interest. We search the folded model for undetectable faults of weight below the target using a purpose-built distance finder, described in Appendix~\ref{sec:circuitfinder}, and so estimate the circuit distance of errors in all but one of the possible logical cosets. A simple analytical argument then bounds the circuit-level distance of errors in the final (`timelike') logical coset. The details of this analysis are given in Appendix~\ref{sec:circuit-distance-methods}.

The strength of the evidence differs by code size. For the $n=72$ circuits the search is exhaustive, so those distances are exact. For $n=144$ and $n=288$ we accept a circuit when 300 independent trials of the finder find no fault. At $n=144$ we calibrated this rule on several hundred further circuits, as for code capacity (Appendix~\ref{sec:circuitcalibration}): we estimate that an accepted circuit has a $1.3\%$ chance of harbouring an undiscovered fault ($95\%$ interval $0.15$--$7.8\%$). At $n=288$ trials were too costly to calibrate, and we report only that no fault was found in 300 trials. The finder is nevertheless sensitive at this size: each of the six circuits that we rejected for the hardest $n=288$ target gave up a fault within its first few trials.

In each case the best circuit distance we could hope to achieve is that of the underlying syndrome extraction circuit. Not every code quilt that is distance preserving at the code-capacity level gives rise to a circuit that preserves circuit distance. Where the first representative failed we drew another for the same target logical coset. The results are summarised in Table~\ref{tab:circuit-distance}. The hardest case, the measurement between a surface code and a single bivariate bicycle code at $n=288$, rejected six representatives before a seventh survived.

\begin{table}[htbp]
\centering
\small
\setlength{\tabcolsep}{4pt}
\begin{tabular}{@{}llcccc@{}}
\toprule
Codes & Target logical class
      & \begin{tabular}[b]{@{}c@{}}Idle memory\\ circuit distance\end{tabular}
      & \begin{tabular}[b]{@{}c@{}}Estimated\\ circuit distance\end{tabular}
      & Trials
      & \begin{tabular}[b]{@{}c@{}}FNR\\ (App.~\ref{sec:tweedietest})\end{tabular} \\
\midrule
\midrule
\multirow{4}{*}{\shortstack[l]{$n=72$ BB code\\ $+\ d=6$ RS code}}
  & \texttt{SC\_BB\_ZZ} & 6 & \textcolor{green!55!black}{6} & exhaustive & exact \\
  & \texttt{BB2\_ZZ} & 6 & \textcolor{green!55!black}{6} & exhaustive & exact \\
  & \texttt{BB2\_GENERIC\_MATCHED} & 6 & \textcolor{green!55!black}{6} & exhaustive & exact \\
  & \texttt{SC\_BB2\_ZZZ} & 6 & \textcolor{green!55!black}{6} & exhaustive & exact \\
\midrule
\multirow{4}{*}{\shortstack[l]{$n=144$ BB code\\ $+\ d=12$ RS code}}
  & \texttt{SC\_BB\_ZZ} & 10 & \textcolor{green!55!black}{10} & 300 & $1.3\%$ \\
  & \texttt{BB2\_ZZ} & 10 & \textcolor{green!55!black}{10} & 300 & $1.3\%$ \\
  & \texttt{BB2\_GENERIC\_MATCHED} & 10 & \textcolor{green!55!black}{10} & 300 & $1.3\%$ \\
  & \texttt{SC\_BB2\_ZZZ} & 10 & \textcolor{green!55!black}{10} & 300 & -- \\
\midrule
\multirow{4}{*}{\shortstack[l]{$n=288$ BB code\\ $+\ d=18$ RS code}}
  & \texttt{SC\_BB\_ZZ} & 18 & \textcolor{green!55!black}{18} & 300 & -- \\
  & \texttt{BB2\_ZZ} & 18 & \textcolor{green!55!black}{18} & 300 & -- \\
  & \texttt{BB2\_GENERIC\_MATCHED} & 18 & \textcolor{green!55!black}{18} & 300 & -- \\
  & \texttt{SC\_BB2\_ZZZ} & 18 & \textcolor{green!55!black}{18} & 300 & -- \\
\bottomrule
\end{tabular}
\caption{\textit{Expected and estimated circuit distance for code quilts
constructed to measure a single randomly selected target logical coset in each
of the 12 classes.} `Trials' is the number of independent trials of the distance finder that the accepted circuit survived, and FNR is the estimated probability that it nonetheless harbours a fault below the estimated distance (Appendix~\ref{sec:circuitcalibration}); a dash marks circuits for which this could not be calibrated.}
\label{tab:circuit-distance}
\end{table}

\section{Sparsity results}\label{sec:sparsity}

Alongside distance, two properties that matter for the fault tolerance of a protocol are stabiliser sparsity (stabilisers and detectors have low Pauli weight) and low congestion (each qubit occurs in few stabilisers and detectors). Both control the entropic contribution to the logical error probability at finite physical error rate~\cite{beverland2019entropy}, and together with growing distance they suffice to prove fault tolerance in the high distance limit~\cite{kovalev2013faulttolerance, gottesman2014constantoverhead}. In this section, we present a combination of analytical and empirical results that, together with the numerical decoding experiments, show the present constructions produce circuits of acceptable sparsity for practical purposes. 
All the results presented here are restricted to the case where the stitching operators of a code quilt are weight-$N$, that is, the local stitching operators are weight-1 on each block, as in the second decomposition of Section~\ref{sec:twoblock}. We also assume that $N$ is low, meaning we do not consider the growth of operator weights or congestion as the number of codeblocks involved in a quilt grows.

\subsection{Relevant operator classes}

Whilst the full structural analysis is deferred to Appendix~\ref{sec:structuretheorems}, it is useful to briefly describe some different relevant operator classes over the two constructions.

\begin{enumerate}
    \item \textit{Physically measured operators}. The physically measured operators for the subsystem protocol consist of the original code stabilisers and the weight-$N$ stitching operators. Those for the stabiliser protocol are the stabilisers of the gauge-fixed code quilt, including the weight-$N$ stitching operators, which are upgraded from gauge operators to stabilisers by the gauge fixing.
\end{enumerate}

In the subsystem protocol, the physical operators do not all correspond to circuit detectors, so we distinguish the following categories of operators that do give rise to detectors:

\begin{enumerate}[start=2]
    
    \item \textit{Untouched stabilisers.} The untouched stabilisers form a subspace of the stabiliser group of each code generated by original-code stabilisers that commute with all of the stitching operators, so pass unchanged into the code quilt (in both protocols).
    \item \textit{Measurement consistency stabilisers.} These are the stabilisers that correspond to the logical operators we intend to measure using either protocol. They are not typical error-detecting stabilisers as their initial value is unknown. We can however form consistency checks between each of the $d$ stitching rounds; this gives $d-1$ detectors in both protocols.
    \item \textit{Stitching consistency stabilisers.} In the stabiliser protocol only, the stitching operators themselves are stabilisers, and therefore form individual detectors between rounds. In both protocols, the initial value of these stabilisers is nondeterministic; in the subsystem protocol, each subsequent measurement is also nondeterministic so they cannot form individual detectors.
    \item \textit{$H_c$ quilt stabilisers.} These are stabilisers of the code quilt that can be generated entirely using original-code stabilisers confined to a single block, excluding the untouched stabilisers. They are identical between the two protocols; in the subsystem protocol their value is inferred by directly measuring original-code stabilisers, whereas in the stabiliser protocol they are measured directly. For instance, the weight-12 stabiliser that is the product of three stabilisers (with some cancellation between them) in the upper codeblock in Figure \ref{fig:bb-stitch} lies in this space.
    \item \textit{$H_{ac}$ quilt stabilisers.} The stabilisers of the code quilt that must necessarily be formed as a product of original-code stabilisers of multiple blocks. (These do not form a subspace but are rather an arbitrary complement of $H_c$; see Appendix~\ref{sec:structuretheorems}.) As with $H_c$ stabilisers, they are inferred in the subsystem protocol and measured directly in the stabiliser protocol. The weight-12 product stabiliser spanning the two codeblocks in Figure \ref{fig:bb-stitch}, illustrated with a square and some dashed lines, lies in this space.
    \item \textit{$H_{ac} + H_c$}. Together, these form a stabiliser subspace, that we can optimise over to find a low-weight basis.
\end{enumerate}

We summarise our results on sparsity and congestion in the following table. A \cYes\ indicates that either an analytic bound exists on a quantity or we have empirical evidence showing it grows slowly in the cases studied. A \cNo\  indicates either that we cannot prove a bound, or in the sole case of the measurement consistency stabiliser, can show growth with distance. \cNA\  indicates that the operator never forms part of a (useful) basis within the given protocol (subsystem/stabiliser).

\begin{center}
\begin{tabular}{@{}l*{4}{>{\centering\arraybackslash}p{2.05cm}}@{}}
\toprule
 & \multicolumn{2}{c}{Analytical bound}
 & \multicolumn{2}{c}{\begin{tabular}[b]{@{}c@{}}Empirically slow\\ growth in $d$\end{tabular}} \\
\cmidrule(lr){2-3}\cmidrule(lr){4-5}
\ \ \ \ Operator family & Subsystem & Stabiliser & Subsystem & Stabiliser \\
\midrule
\midrule
\textit{Physically measured operators}       & \cYes & \cNo & \cYes & \cYes \\
\midrule
\ \ \ \ Detector families & & & & \\
\textit{Untouched stabilisers}               & \cYes & \cYes & \cYes & \cYes \\
\textit{Measurement consistency stabilisers} & \cNo  & \cNA  & \cNo  & \cNA  \\
\textit{Stitching consistency stabilisers}   & \cNA  & \cYes & \cNA  & \cYes \\
\textit{$H_{ac}$ quilt stabilisers only}          & \cYes & \cYes & \cYes & \cYes \\
\textit{$H_c$ quilt stabilisers only}             & \cNo  & \cNo  & \cNo  & \cNo  \\
\textit{$H_c + H_{ac}$ quilt stabilisers}    & \cNo  & \cNo  & \cYes & \cYes \\
\bottomrule
\end{tabular}
\end{center}

\subsection{Analytical sparsity results} \label{sparsity-analytics}

The physically measured operators of the subsystem protocol are trivially low weight, whereas for the stabiliser protocol they correspond to the stabilisers of the gauge-fixed quilt.
The untouched stabilisers in either protocol are also trivially low weight. The measurement consistency stabilisers are only relevant in the subsystem protocol, and are explicitly \textbf{not} low weight: their weight grows as $|L|$, the weight of the chosen representative of the measured logical operator(s). We rely on circuit-level simulations to establish whether this implies that the subsystem protocol has an acceptable logical error probability at finite sizes. The stitching consistency operators only exist in the stabiliser protocol, and are trivially low weight (weight $N$ each).

The weight of stabilisers in $H_{ac}$ can be bounded where the representative satisfies a modest connectivity condition, detailed in Appendix~\ref{sec:sparsityanalytical}.
When this condition applies, the \texttt{SkipTree} algorithm of~\cite{swaroop2024adapters} can be applied to constructively bound each element of a basis for $H_{ac}$ as the product of at most 3 original-code stabilisers per codeblock, and the number of such generators containing any given qubit by $4\kappa$, where $\kappa$ is the qubit degree of the original code's own checks.  All 25,000 logical representatives in the dataset detailed in Section~\ref{sec:repfinding} satisfy the connectivity requirement.
The \texttt{SkipTree} construction does not apply to $H_c$ or, more generally, to the combined subspace $H_c + H_{ac}$, and nothing bounds either axis there.
Homogeneous quilts (Section~\ref{sec:distance}) admit a particularly sparse and decongested basis of $H_{ac}$, of weight at most $2w$ ($w$ is the maximum weight of a stabiliser generator of the original codes) and qubit degree at most $2\kappa$, without the connectivity condition. As in the general case, this does not bound $H_c$. The representatives supporting them are provided separately (Appendix~\ref{sec:homogeneousbank}). Note that this is not the basis that we use in practice, as explained in Appendix~\ref{sec:sparsityempirical}.

The single asymmetry between weight and congestion is the measurement-consistency detector: its weight grows with $|L|$, but it is one detector, so it adds only one to the degree of each qubit it touches. The empirical results below accordingly report weight and qubit degree side by side.

\begin{figure}
    \centering
    \includegraphics[width=\linewidth]{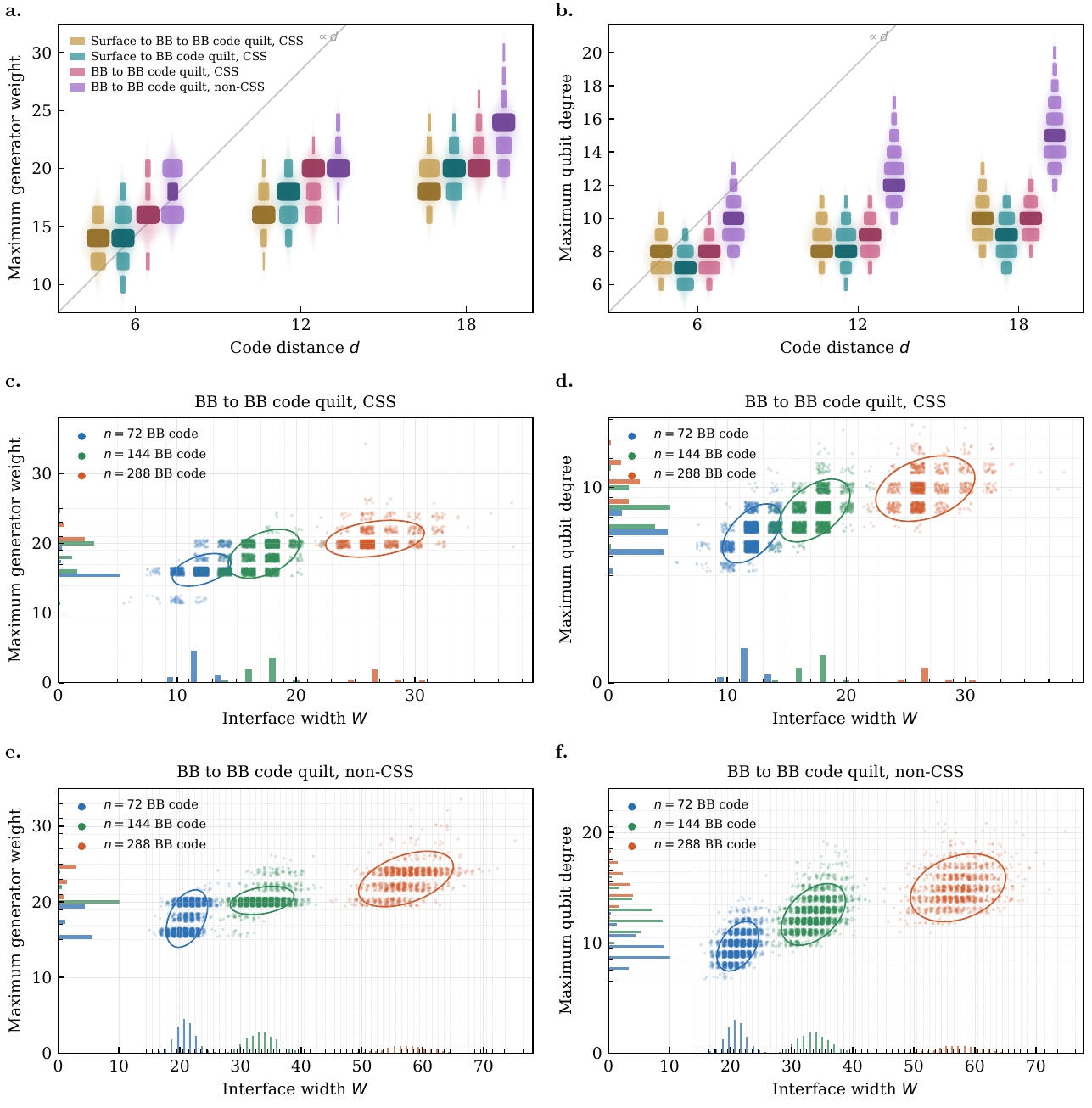}

    \caption{\textit{Sparsity and congestion of the optimised $H_c + H_{ac}$ basis.} \textbf{a,} \textbf{b.} Vertical histograms showing the distribution of maximum operator weights and maximum qubit degree for the $H_{c} + H_{ac}$ stabiliser subspace in the optimised bases, grouped by class of target and shown for each of the three distances under consideration. Note the clearly sub-linear scaling of both quantities. \textbf{c. -- f.} Detailed views into the sparsity and congestion data for two of the four classes \texttt{BB2\_ZZ} and \texttt{BB2\_GENERIC\_MATCHED} (the full data are shown in Appendix \ref{sec:sparsityempirical}). The ellipse drawn for each code size is centred on the mean of that size's points and extends two standard deviations along each principal axis of their sample covariance; no model is fitted. Each dot on each of the four panels represents the result of running our basis optimisation protocol on the $H_c + H_{ac}$ subspace of the stabiliser group of one of the code quilts generated by the chosen representatives described in Section~\ref{sec:repfinding}. The dots in principle lie on an integer lattice, but each dot is placed randomly within its unit square in order to show density. The `interface width' $W$ is the weight of the chosen logical representative of the randomly selected target logical coset. \textbf{Note that, in the subsystem protocol, the maximum weight of any operator physically measured in the circuit in any of the above quilts is 6, and the maximum number of entangling gates with which any qubit must interact in a syndrome extraction round is 8.}
    }

    \label{fig:sparsitycorpus}
\end{figure}

\subsection{Empirical sparsity results}

It remains to empirically analyse the unbounded sector $H_c$. For each of the 25,000 logical representatives chosen by the optimiser in Section~\ref{sec:repfinding}, we optimised a basis for the combined sector $H_c + H_{ac}$, using techniques detailed in Appendix~\ref{sec:sparsityempirical}, aiming to achieve good sparsity and low congestion.
A summary of the results of the sparsity and congestion analyses is shown in Figure~\ref{fig:sparsitycorpus}. It is clear that both operator weight and qubit degree grow slowly with $d$ for the bivariate bicycle and surface codes in question: tripling the code distance from 6 to 18 by moving from the 72-qubit to the 288-qubit BB code increases the median of each quantity by between 25\% and 50\%, the largest increase being in the qubit degree of the worst-behaved class, which is the non-CSS code quilt between two BB codeblocks. A subset of the results is shown in more detail for two of the classes, where the relationships between the weight $W$ of the operator produced by the coset annealing procedure (Section \ref{sec:repfinding}) and the properties of the basis produced by the optimiser are shown.

\section{Circuit-level numerical simulation results}\label{sec:circuitsims}

For the 144-qubit examples we also estimate operational failure probabilities by sampling from the noisy circuits and decoding with Relay-BP~\cite{relaybp2025}, to which we provide a variety of equivalent decoding Tanner graphs (up to an invertible transformation on the detectors), with correspondingly transformed syndromes. The variety in Tanner graphs is achieved by exploiting the freedom in the basis optimisation procedure (Appendix~\ref{sec:sparsityempirical}). Each Tanner graph is also passed through a GARI-style~\cite{maan2026gari} transformation to produce a decoding Tanner graph with fewer edges and fewer four-cycles, which is understood to improve the performance of belief propagation~\cite{raveendran2021trapping}. Relay-BP is run on successive equivalent Tanner graphs until two presentations both allow it to converge to solutions and the solutions agree in logical class (typically, this is achieved using just the first two Tanner graphs). More details about the simulations and decoder used are given in Appendix~\ref{sec:simulationdetails}.

\begin{figure}
    \centering
    \includegraphics[width=\linewidth]{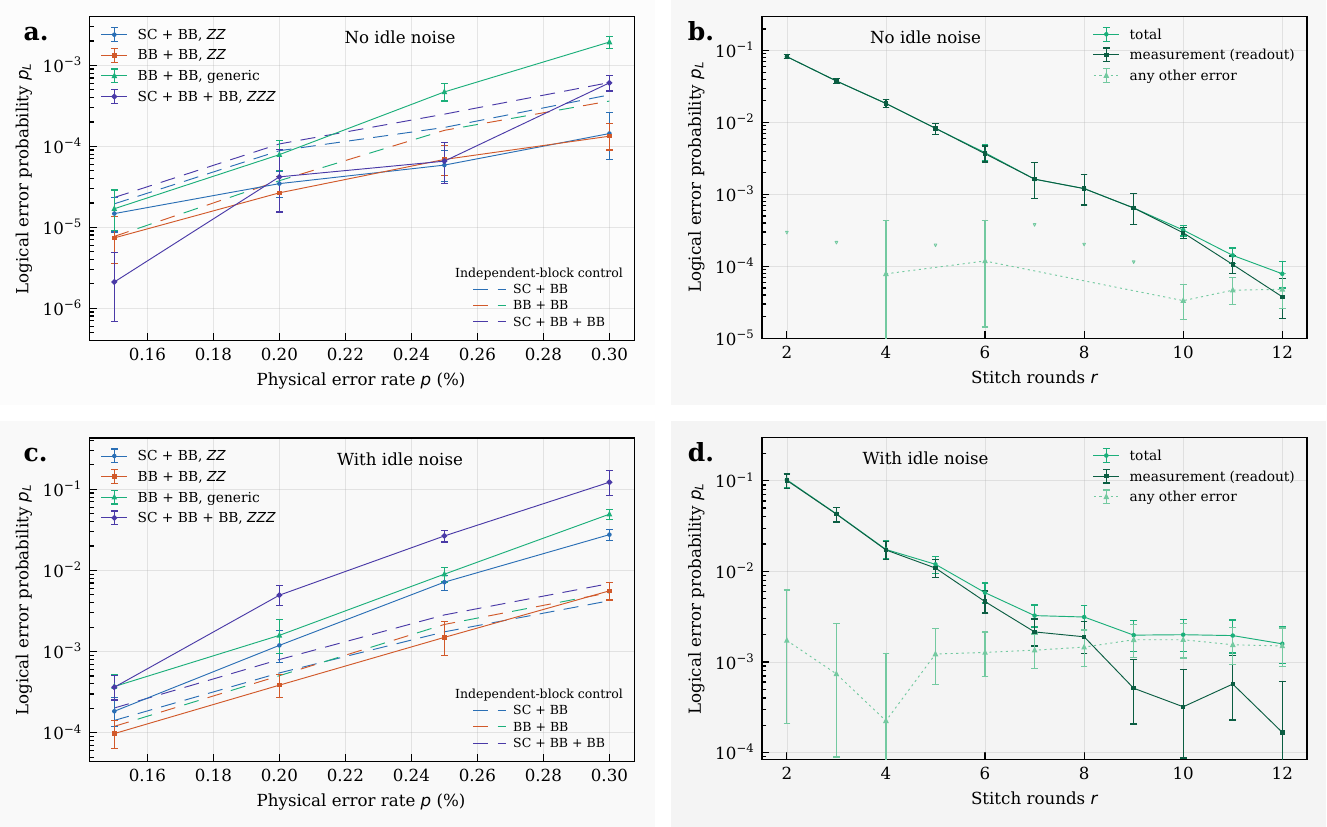}
    \caption{\textit{Results of the circuit-level decoding experiments.} Note the logical error probabilities are not shown per syndrome extraction round, nor per logical qubit, but are failure probabilities for the entire circuit.
    \textbf{a.} and \textbf{c.} The decoding results for all four subsystem protocol circuits involving the 144-qubit (`gross') bivariate bicycle code and the distance-12 surface code, omitting and including idle noise. Shown with dashed lines are matched controls based on simulating an equal number of idle memory blocks of the code for the same number of rounds.
    \textbf{b.} and \textbf{d.} The decoding results for a series of circuits based on the same non-CSS bivariate bicycle code quilt (\texttt{BB2\_GENERIC\_MATCHED}) as in panel (a.) where $p$ is held constant at 0.2\%, and the number of merged stitching rounds varies from $r=2$ to $r=12=d$, omitting and including idle noise. For each $r$ we plot the total failure probability as well as the probability of failures that include a failed classical readout, and the probability of failures that do not include a failed classical readout.}
    \label{fig:decoding-results}
\end{figure}

The comparison with memory experiments uses the same physical error rate and a matched block composition and duration, with results shown in Figure~\ref{fig:decoding-results}.
The failures are broken down into those that include a wrongly reported measurement outcome and those that do not, in Table~\ref{tab:readout-share}.
In the same figure, in the right-hand panels, we show the results of a series of simulations of the \texttt{BB2\_GENERIC\_MATCHED} circuit, where the number of stitching rounds is varied from 2 to $d=12$.
In those plots, we distinguish between logical failures of the classical readout; that is, logical measurement errors, and all other errors in which the readout is performed correctly. We find that the probability of logical measurement errors is suppressed exponentially with the number of stitching rounds, whereas the probability of purely quantum memory errors is roughly constant.
These observations are consistent with the circuit-distance analysis of Section~\ref{sec:distance} and Appendix~\ref{sec:circuit-distance-methods}: the folding technique shows that the distance with respect to the timelike observables is independent of the number of stitching rounds, whereas the distance with respect to the spacelike observables that control measurement error grows with the number of stitching rounds.

Returning to the left-hand panels, we note that the \texttt{BB2\_GENERIC\_MATCHED} case is the only experiment that appears to perform worse, by a factor of 2 or 3, than its control (although the error bars are loose). We attribute this to the significantly higher weight ($W=34$ vs $W\le 18$ for the other three quilts) of the logical-measurement-consistency operator (see Section~\ref{sec:sparsity}) of the non-CSS operator measured by this code quilt as compared to the others.
This drop in performance with operator weight is the sense in which the subsystem protocol is not asymptotically fault tolerant as $d\to\infty$.

Together with the exponential suppression behaviour shown in the right-hand panels, this suggests that at higher error rates it may be profitable to run the stitch for a small number of additional stitching rounds to suppress the probability of measurement failures down to that of the memory failures, which do not suffer from the existence of the high-weight measurement-consistency detectors.

\begin{table}
\centering
\begin{tabular}{@{}lccccc@{}}
\toprule
 & & \multicolumn{4}{c}{Physical error rate $p$}\\
\cmidrule(l){3-6}
Class & $W$ & $0.15\,\%$ & $0.20\,\%$ & $0.25\,\%$ & $0.30\,\%$\\
\midrule
\texttt{BB2\_ZZ} & 18 & 0/10 & 0/18 & 0/23 & 1/30 \\
\texttt{SC\_BB\_ZZ} & 16 & 0/18 & 0/29 & 1/22 & 0/10 \\
\texttt{SC\_BB2\_ZZZ} & 18 & 0/5 & 0/6 & 0/13 & 3/82 \\
\texttt{BB2\_GENERIC\_MATCHED} & 34 & 5/13 & 11/23 & 57/66 & 104/126 \\
\bottomrule
\end{tabular}
\caption{\textit{Logical failures attributable to the classical readout, at $d=12$ stitching rounds.}
Each entry counts the observed failures that flip the `cap' observable against all observed failures,
as (readout failures / total failures). $W$ is the interface width of the representative used.
\texttt{BB2\_GENERIC\_MATCHED} is the only class with a live readout failure mode at this number of rounds.}
\label{tab:readout-share}
\end{table}
\section{CSS code quilts as a pullback of homological measurement complexes}
\label{sec:pullback}

Whilst not the primary language used throughout the present work, it is
possible to interpret the two-block CSS construction as a modification of the
homological measurement framework introduced in Ref.~\cite{ide2025homological} via a pullback (see also the description of code surgery as a pushout in Ref.~\cite{cowtan2024csssurgery}).  We give only a summary of the
interpretation here, and only describe the case where a joint $Z$-type logical
operator is measured between two CSS stabiliser codes.  We do not attempt to
restate, in homological language, the proofs given in the appendices that the
construction measures the desired logical operator, introduces no new
logical qubits, and retains the required distance.
The commuting code described below is the stabiliser protocol's gauge fixing, in which the stitching checks are included as stabilisers.

A CSS stabiliser code is written as a three-term chain complex of vector spaces
over $\ft$,
\begin{equation}
    C_2 \xrightarrow{\partial_2} C_1
    \xrightarrow{\partial_1} C_0.
\end{equation}
Canonically, $C_1\cong\ft^n$ represents the space of Pauli-$X$ operators on
the qubits, $C_2$ represents the space of Pauli-$X$ stabilisers, and $C_0$ is
dual to the space of Pauli-$Z$ stabilisers.  The map $\partial_2$ gives the
$X$ stabilisers as physical operators, while $\partial_1$ records the pattern
of anticommutation between an $X$ operator and the $Z$ stabilisers.  Two
codeblocks are represented by the direct sum of their two chain complexes.

\begin{figure}[H]
    \centering
    \includegraphics[width=0.4\linewidth]{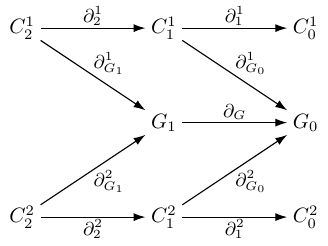}
    \caption{\textit{The two CSS code complexes mapped into the common homological
    measurement complex.}  Superscripts label the codeblock, while subscripts
    label chain degree.  Each of the two squares commutes.}
    \label{fig:pullback-interface}
\end{figure}

Having described the stabiliser codes in this way, homological measurement
introduces an ancilla system, in its simplest form a two-term chain complex
\begin{equation}
    G_1 \xrightarrow{\partial_G} G_0.
\end{equation}
Here $G_0$ indexes the new $Z$ checks whose product contains the desired
logical operator, while $G_1$ indexes additional ancilla qubits introduced to
make those checks commute with the original code.  The maps from the two code
complexes into this common ancilla complex form the commuting diagram shown in
Figure~\ref{fig:pullback-interface}, also known as a mapping cone.

\begin{figure}[H]
    \centering
    \includegraphics[width=0.3\linewidth]{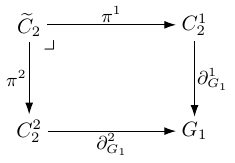}
    \caption{\textit{The pullback defining $\widetilde C_2$.}  Its two projections
    select pairs of original stabilisers with the same image in $G_1$.}
    \label{fig:pullback-square}
\end{figure}

Instead of retaining the auxiliary qubits $G_1$, we pull the two stabiliser
spaces back over their common anticommutation-pattern space:
\begin{equation}
    \widetilde C_2
    :=C^1_2\times_{G_1}C^2_2
    =
    \left\{(h_1,h_2):
    \partial_{G_1}^1(h_1)=\partial_{G_1}^2(h_2)\right\}.
    \label{eq:quilt-pullback}
\end{equation}
The defining pullback square is shown in
Figure~\ref{fig:pullback-square}.  Physically, an element of
$\widetilde C_2$ is a pair of original stabilisers with the same local
anticommutation pattern.  Their product across the two blocks therefore
commutes with every joint stitching check, because over $\ft$ the two
identical patterns cancel.

\begin{figure}[H]
    \centering
    \includegraphics[width=0.4\linewidth]{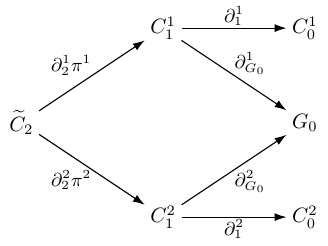}
    \caption{\textit{The CSS quilt after the pullback.}  The image of
    $\widetilde C_2$ supplies the inherited and doubled $X$ stabilisers,
    while $G_0$ supplies the new $Z$-type stitching checks.}
    \label{fig:pullback-result}
\end{figure}

We may therefore remove $C_2^1$, $G_1$, and $C_2^2$ and replace them by the
single space $\widetilde C_2$, which points directly to the data spaces of the
two codes.  The resulting complex is shown in
Figure~\ref{fig:pullback-result}.  It remains a chain complex because the
original rows are chain complexes, the two squares in
Figure~\ref{fig:pullback-interface} commute, and the two interface patterns
agree by the pullback condition.

This description also reproduces the stabiliser decomposition used in the
appendix.  Let $H_c^i=\ker\partial_i^*$ and choose sections
$\sigma_i:Y^\perp\to H^i$.  Then
\begin{equation}
    \widetilde C_2\cong H_c^1\oplus H_c^2\oplus Y^\perp.
\end{equation}
The first two summands give the original stabilisers that already commute with
the stitching operators.  The last sends $y\in Y^\perp$ to the pair
$(\sigma_1(y),\sigma_2(y))$, and hence gives the doubled stabilisers
$\quilt H_{ac}$ of Lemma~\ref{lemma:quiltstabiliserstructure}.  This is the
pullback counterpart of the stabiliser construction used throughout the
appendix.

\section{Discussion, and future work}\label{sec:discussion}

In this work, we introduced code quilts, a technique for performing arbitrary logical Pauli product measurements between arbitrary codes, and examined its applicability to and performance on bivariate bicycle and surface codes.

The code quilt is arguably simpler than other constructions \cite{cohen2022lowoverhead,cross2024improvedsurgery,cowtan2025hypergraph,ide2025homological,swaroop2024adapters,zheng2025highrate} for generalised surgery protocols. 
Theoretical simplicity is desirable in a field where the ultimate goal is to reduce engineering complexity.
The construction makes no assumption about the underlying codes being CSS, which avoids the difficulties that can arise when treating non-CSS codes from a homological algebra point of view \cite{yuan2026noncss}. Nonetheless, as shown in Section \ref{sec:pullback}, in its simplified form it can be interpreted as a pullback on a simple cone complex; perhaps future work could fruitfully formalise this idea further.
Also, there may be more analytical understanding of quilts to be had that may lead to more deterministic methods for constructing code quilts with favourable properties. We have so far found no target logical coset across any combination of bivariate bicycle codes (within the loose parity constraints of Section \ref{sec:repfinding}) and surface codes for which we were unable to find a code-capacity distance-preserving quilt (according to our estimation techniques) to measure it.
The analysis given in Appendices \ref{sec:structuretheorems}, \ref{sec:distancetheorems}, and \ref{sec:sparsityanalytical} assumes nothing about the structure of the codes used to construct the quilt---it may be profitable to examine quilts on specific codes.

The relative simplicity leads to circuits that are naturally commensurate with, and mostly derived from, the existing idle-memory syndrome extraction circuits, and have only slightly increased depth vs these circuits. For the subsystem protocol circuits, we found circuits with no distance-reducing fault for every target we examined, although for one class at the largest size this took seven representatives.

Both the subsystem and stabiliser protocols require each physical qubit to interact with only one or two qubits in addition to the ones they must interact with to perform idle syndrome extraction. However, in contrast to proposals such as \cite{he2025extractors}, the quilt construction given here is only designed to measure a single logical operator. This suggests that to be useful as a primitive for a full architecture for qubits with fixed connectivity, the quilts method would need to be combined with a code with plentiful automorphisms, or be developed further to measure a wider array of operators. For now, perhaps the more obvious application is to neutral-atom or trapped-ion qubits, which can in principle freely rearrange to execute arbitrary quilt measurements in sequence.

This fact, along with the other studies in this work, shows that code surgery can be effectively optimised using numerical tools and heuristics beyond those characteristics we can prove in the asymptotic limit.

Future work should evaluate measurements using code quilts at larger distances, where the lack of asymptotic guarantees may start to show up in decoded performance more so than at distance 12. It may be the case that the large weight of the measurement-consistency operators in the subsystem protocol starts to bite. However, there are also various tricks that could be considered to bring the weight of these detectors down just enough for practical purposes, or more generally to interpolate between an asymptotically fault-tolerant construction and a practical one aimed at finite codes. Whilst the sparsity and congestion results of Section \ref{sec:sparsity} apply to the stabiliser protocol as well as the subsystem protocol, another obvious avenue is to find distances for and simulate the circuits described for the stabiliser protocol.
Preliminary results not reported in this manuscript suggest the spread of circuit-level faults is well-contained by the abundance of flag detectors provided by the protocol.
Despite the relatively short syndrome extraction circuits, the construction still suffers heavily from idle noise. Whilst this is worsened by our simple noise model in which an idling qubit is as dangerous as one undergoing a gate, it seems likely that more bespoke circuit construction, which we have not heavily optimised, could reduce the logical error probability with idling noise substantially.

Finally, the optimisation methods presented here can all be considered proofs-of-concept. Every aspect of the pipeline presented in this work---from a target logical coset, to a selected representative and pairing, to a syndrome extraction circuit, and then decoding---stands to be carefully improved in future work, as well as specialised to particular code choices.
\newpage
\appendix
\section*{Guide to the appendices}
Appendix \ref{sec:operationaldistance} introduces notation for `operational distance', a quantity we find convenient for reasoning about the combined classical and quantum failure modes of a logical measurement operation, as well as a general description of a partially initialised code (PIC), which is a re-packaging of the hybrid quantum-classical codes described in \cite{nemec2022encoding} whose classical information corresponds to the readout of the measurement result in our protocol. 

Appendices \ref{sec:structuretheorems} to \ref{sec:sparsityanalytical} give structural lemmas and theorems describing the various operators that comprise the code quilt, and derive the results relating to code-capacity distance and operator sparsity and congestion described in the main text.

Appendices \ref{sec:overlap} to \ref{sec:simulationdetails} then describe in detail the various empirical aspects of the work, and some readers may wish to skip to this part. We highlight the detector-error-model folding technique of Appendix \ref{sec:circuit-distance-methods} and the `Tweedie test' presented in Appendix \ref{sec:tweedietest} which may be of independent interest.

A symbol glossary containing all the notation used throughout the manuscript is provided in Appendix \ref{sec:symbolglossary} for convenience. 
\section{Operational distance: a natural framing for generalised surgery}\label{sec:operationaldistance}

In this appendix we provide a framework for analysing the code-capacity (or data-level) fault tolerance of protocols that merge and split general subsystem codes with multiple logical qubits in order to measure out a certain subspace of those qubits.
Whilst there exist treatments of related problems~\cite{nautrup2017interface,cross2024improvedsurgery,vuillot2019deformation}, we find the following treatment useful in its explicit distinction between logical `measurement' and `memory' errors, whilst providing a framework to prove robustness against both kinds of errors by analysing the properties of a single subsystem code.
This treatment is convenient for empirical distance finding as well, and may be of independent interest for the analysis of other generalised measurement protocols.
For convenience, a glossary of the symbols used in these appendices is given in Appendix~\ref{sec:symbolglossary}.

At the heart of our treatment is the following definition, which is essentially the construction described by Theorem 4 of~\cite{nemec2022encoding}, although the treatment here is intended to be entirely standalone.

\subsection{Partially initialised codes}

\begin{definition}[Partially initialised code]
    A \textit{partially initialised code} (PIC) is a pair $(G,C)$ consisting of a Pauli subspace $G\leq\ft^{2n}$ on $n$ qubits and a subspace $C\leq Z(G)$ of the centre of $G$.
\end{definition}

For our purposes, $C$ contains the stabilisers that we can use to detect errors: any operator that anticommutes with an operator in $C$ is, by definition, not pathological.
The remainder of the stabiliser group, which we can represent by choosing a complement $M$ so that $Z(G)=C\oplus M$, consists of `stabilisers' whose values \textit{we would like to know}.
Any operator that silently flips the value of an element of $M$ whilst commuting with all of $C$ is pathological, and for fault tolerance we must lower-bound the weight of such errors.
In~\cite{nemec2022encoding}, these are called \textit{classical errors}: this is because at the end of the process their effect is to flip the value of the classically stored bits (perhaps in a control system) that record the measurement outcome.\footnote{The natural language for this reasoning is operator-algebra quantum error correction~\cite{dauphinais2024stabilizer}.}
For this reason, a better name for these errors might be \textit{logical measurement errors}; in this appendix, we use the original language.
These have the unfamiliar property that invisible operators which anticommute with these pathological errors are not necessarily themselves pathological---this is the mathematical complication that precludes their description purely in terms of usual notions of logical operators.
Of course, we must also protect the information that is not measured.
In the language of~\cite{nemec2022encoding}, errors acting on this information are \textit{quantum errors}, and their minimum weight coincides with the standard dressed distance of a subsystem code.
We refer to both classes of pathological errors (and their overlap) as \textit{operational errors}.

\begin{definition}[Distances of a partially initialised code]
    Given a PIC $(G,C)$, define:
    \begin{enumerate}
        \item the \textit{quantum distance}
        \begin{equation}
            d_{\mathrm q}
            =
            \min\{|E|:E\in Z(G)^\perp\setminus G\};
        \end{equation}
        \item the \textit{classical distance}
        \begin{equation}
            d_{\mathrm c}
            =
            \min\{|E|:E\in C^\perp\setminus Z(G)^\perp\};
        \end{equation}
        \item the \textit{operational distance}
        \begin{equation}
            d_{\mathrm{op}}
            =
            \min\{|E|:E\in C^\perp\setminus G\}.
        \end{equation}
    \end{enumerate}
    We take the minimum of the empty set to be $\infty$.
\end{definition}

\begin{lemma}[Operational distance]
    For a partially initialised code $(G,C)$,
    \begin{equation}
        d_{\mathrm{op}}
        =
        \min\{|E|:E\in C^\perp\setminus G\}
        =
        \min(d_{\mathrm q},d_{\mathrm c}).
    \end{equation}
\end{lemma}

\begin{proof}
    Since $C\leq Z(G)$, we have
    \begin{equation}
        G\leq Z(G)^\perp\leq C^\perp.
    \end{equation}
    Therefore
    \begin{equation}
        C^\perp\setminus G
        =
        \bigl(Z(G)^\perp\setminus G\bigr)
        \cup
        \bigl(C^\perp\setminus Z(G)^\perp\bigr),
    \end{equation}
    and the two sets on the right-hand side are disjoint.
    Taking the minimum weight over the two sets gives the result.
    Equivalently, the three error spaces fit into the short exact sequence
    \begin{equation}
        0
        \longrightarrow
        Z(G)^\perp/G
        \longrightarrow
        C^\perp/G
        \longrightarrow
        C^\perp/Z(G)^\perp
        \longrightarrow
        0.
    \end{equation}
\end{proof}

\subsection{Specialisations and admissible merges}

\begin{definition}[Specialisations of the partially initialised code]
    Given a partially initialised code, it is immediate that:
    \begin{enumerate}
        \item if $G$ is isotropic, so $Z(G)=G$, then $(G,C)$ is a \textit{partially initialised stabiliser code};
        \item if $C=Z(G)$, then $d_{\mathrm{op}}=d_{\mathrm q}$ and $d_{\mathrm c}=\infty$, and we recover a usual subsystem code;
        \item if both of the above are true, so $C=Z(G)=G$, then we recover a usual stabiliser code.
    \end{enumerate}
\end{definition}

Before considering fault tolerance, we separate out the requirement that the ideal merge measures the desired logical operators.
Let $G$ be the original gauge group, write $H=Z(G)$, and let $M\leq G^\perp$ be an isotropic subspace of independent logical representatives that we wish to measure.

\begin{definition}[Admissible merge]
    A gauge group $\widehat G\geq G$ is an \textit{admissible merge of $G$ measuring $M$} if, writing $\widehat H=Z(\widehat G)$ and $C=H\cap\widehat H$, we have
    \begin{equation*}
        \text{\rm (1)}\quad \widehat H=C\oplus M;
        \qquad
        \text{\rm (2)}\quad \widehat G\cap H^\perp=G+\widehat H.
    \end{equation*}
    We call the second condition the \textit{clean-splitting condition}.
\end{definition}

The first condition says that the new central operators of the merged code are exactly the logical operators that we intend to measure.
The clean-splitting condition says that any merged gauge operator which is invisible to the restored stabilisers is already harmless: it is a product of an original gauge operator and a merged stabiliser.
Together, these conditions rule out overmeasurement independently of any distance calculation.

\begin{lemma}[Logical count under an admissible merge]\label{lemma:admissiblemergelogicalcount}
    Suppose \(G\) encodes \(k\) logical qubits and \(\widehat G\) is an admissible merge of \(G\) measuring an \(m\)-dimensional logical subspace \(M\). Then \(\widehat G\) encodes
    \begin{equation}
        \widehat k=k-m
    \end{equation}
    logical qubits.
\end{lemma}

\begin{proof}
    Write \(g=\dim G\), \(h=\dim H\), \(\widehat g=\dim\widehat G\), and \(c=\dim C\). Since \(H\le G\le\widehat G\), the commutation pairing between \(H\) and \(\widehat G\) has left kernel
    \begin{equation}
        H\cap\widehat G^\perp
        =
        H\cap\widehat H
        =
        C.
    \end{equation}
    Hence
    \begin{equation}
        \dim(\widehat G\cap H^\perp)
        =
        \widehat g-(h-c).
    \end{equation}
    On the other hand, any element of \(G\cap\widehat H\) commutes with \(G\), so \(G\cap\widehat H=H\cap\widehat H=C\). The clean-splitting condition therefore gives
    \begin{equation}
        \dim(\widehat G\cap H^\perp)
        =
        \dim(G+\widehat H)
        =
        g+m.
    \end{equation}
    Therefore
    \begin{equation}
        \widehat g=g+h-c+m.
    \end{equation}
    Since \(\dim\widehat H=c+m\), the subsystem-code dimension formula gives
    \begin{align}
        \widehat k
        &=n-\frac12\left(\widehat g+\dim\widehat H\right)\\
        &=n-\frac12(g+h)-m\\
        &=k-m.
    \end{align}
\end{proof}

We show in Corollary~\ref{corollary:quiltPIC} that code quilts are admissible merges in this sense.

By an \textit{ideal gauge measurement} for a gauge group \(K\), we mean an ideal sequence of gauge-generator measurements which gauge-fixes into the subsystem code defined by \(K\), and whose classical record determines the eigenvalues of the central operators \(Z(K)\).

\subsection{Fault-tolerant measurement using a PIC (code capacity)}

\begin{lemma}[Code-capacity fault-tolerant measurement using PICs]
    Let $G$ be a subsystem code of dressed distance at least $D$, and let $\widehat G$ be an admissible merge of $G$ measuring $M$.
    Let $C=Z(G)\cap Z(\widehat G)$.
    If the PIC $(\widehat G,C)$ has operational distance at least $D$, then the ideal code-capacity schedule
    \begin{equation}
        \boxed{\textnormal{measure }Z(G)}
        \quad\longrightarrow\quad
        \boxed{\textnormal{ideal }\widehat G\textnormal{ gauge measurement}}
        \quad\longrightarrow\quad
        \boxed{\textnormal{measure }Z(G)}
    \end{equation}
    has operational distance at least $D$ and performs a simultaneous projective measurement of the operators in $M$.
    Here the opening measurement is a noiseless boundary condition, all measurements are otherwise ideal, and arbitrary Pauli data errors may occur in either or both of the two intervening intervals. A fault history is therefore a pair \((E_1,E_2)\), with additive weight \(|E_1|+|E_2|\). It is operationally nontrivial when its complete detector record is trivial but it either changes a reported eigenvalue in \(M\) or leaves a nontrivial dressed logical error after the original code is restored.
\end{lemma}

\begin{proof}
    The admissibility conditions ensure that in the absence of errors the schedule measures precisely $M$ and preserves the remaining logical information.
    Any collection of code-capacity faults occurring within one interval combines into a single Pauli error $E$.
    Since $\widehat G\leq C^\perp$, the following three cases exhaust all possibilities:

    \begin{center}
    \renewcommand{\arraystretch}{1.25}
    \begin{tabular}{p{0.20\linewidth}p{0.34\linewidth}p{0.34\linewidth}}
        \hline
        Error class
        & Between the opening and merged measurements
        & Between the merged and closing measurements
        \\
        \hline
        $E\notin C^\perp$
        & $E$ flips a common stabiliser between the opening and merged measurements, and is detected.
        & $E$ flips a common stabiliser between the merged and closing measurements, and is detected.
        \\
        $E\in C^\perp\setminus\widehat G$
        & $E$ is a nontrivial operational error of $(\widehat G,C)$, and therefore $|E|\geq D$.
        & $E$ is again a nontrivial operational error of $(\widehat G,C)$, and therefore $|E|\geq D$.
        \\
        $E\in\widehat G$
        & $E$ is a merged-code gauge operator. It may change individual gauge-measurement outcomes, but cannot change the values of $C$ or $M$.
        & $E$ is a merged-code gauge operator. Its syndrome under the restored stabilisers is treated by the closing measurement.
        \\
        \hline
    \end{tabular}
    \end{center}

    This classification applies independently to the accumulated errors $E_1$ and $E_2$ in the first and second intervals.
    Because the intermediate measurement is ideal, a common-stabiliser syndrome created by $E_1$ is recorded before $E_2$ occurs, so the two syndromes cannot cancel.
    Consequently, if \(|E_1|+|E_2|<D\) and the complete detector record is trivial, then $E_1,E_2\in\widehat G$.
    Their net effect on the merged state is therefore another merged gauge operator $F\in\widehat G$.

    It remains to check that $F$ cannot become a logical memory error when the code is split.
    Define the restored-stabiliser syndrome map
    \begin{equation}
        s_H:\widehat G\longrightarrow H^*,
        \qquad
        [s_H(F)](h)=h\Omega F^T,
    \end{equation}
    where \(\Omega\) is the binary symplectic form, and fix a section \(q:\operatorname{im}s_H\rightarrow\widehat G\). The closing measurement reports \(s_H(F)\); choose the correction \(Q=q(s_H(F))\).
    Then $F+Q$ commutes with $H$, and hence the clean-splitting condition gives
    \begin{equation}
        F+Q
        \in
        \widehat G\cap H^\perp
        =
        G+\widehat H
        =
        G+M.
    \end{equation}
    The component in $G$ is gauge, while the component in $M$ is a stabiliser of the post-measurement state.
    The residual error is therefore harmless.

    Thus every Pauli error occurring in either interval is detected, harmless, or has weight at least $D$.
    Errors occurring before the opening boundary or after the closing boundary are ordinary memory errors of $G$ and are covered by its dressed distance.
\end{proof}

\section{Code quilt structure}\label{sec:structuretheorems}

Throughout this appendix, for an $\mathbb F_2$-vector space $E$, write
\[
E^* := \operatorname{Hom}_{\mathbb F_2}(E,\mathbb F_2).
\]
For $X\le E$ or $X\le E^*$, write $X^\perp$ for the annihilator of $X$ in the opposite space. When a matrix is used to specify a subspace, its row space is intended. Thus, in this appendix $H^i$, $B$, and $Y$ denote subspaces unless we explicitly refer to a chosen generating matrix.
\subsection{The code quilt}

\begin{definition}[Code quilt]\label{def:codequilt}
    A \textit{code quilt} is specified by the following data:
    \begin{enumerate}
        \item $N \ge 2$ disjoint codeblocks each supporting a stabiliser code $\mathcal{C}^i$ with parameters $n^i$, $k^i$, $d^i$, and stabiliser subspace $H^i \le \Pauli_{n^i}$;
        \item for each codeblock $\mathcal{C}^i$, a set of $W$ disjoint operators acting on that code, called \textit{local stitching operators}, written as the rows of a full-rank matrix $S^i \in \mathbb{F}_2^{W\times 2 n^i}$;
        \item the \textit{incidence code}: a classical linear code $B\le \mathbb{F}_2^N$ of distance $d_B\ge 2$.
    \end{enumerate}
    For $p\in\Pauli_{n^i}$, write $p_\uparrow^{(i)}$ for the Pauli on the full $N$-block system with $p$ on block $i$ and zero binary support on all other blocks. For $x\in\ft^W$ and $T\in\ft^{W\times N}$, define
    \begin{equation}
        \Stitch_i:\ft^W\rightarrow \Pauli_{n^i},
        \qquad
        \Stitch_i(x)=xS^i,
        \qquad
        \Stitch(T)=\prod_{i=1}^N \bigl(\Stitch_i(T_{\cdot i})\bigr)_\uparrow^{(i)}.
    \end{equation}
    Thus $\Stitch_i(x)$ is the product of local stitching operators selected by $x$, and $\Stitch(T)$ is the corresponding cross-block Pauli product. In the binary representation this product is the block-padded sum of the local Pauli vectors. For a subspace $X\le\ft^{W\times N}$, write $\Stitch(X)$ for its image.

    The local stitching operators are required to satisfy the following conditions.
    \begin{enumerate}
        \item The kernels of the anticommutation maps
        \begin{equation}
            \partial_i:\ft^W\rightarrow (H^i)^*,\qquad [\partial_i(x)](h)=h\Omega \Stitch_i(x)^T,
        \end{equation}
        are equal to a common subspace $Y\le\ft^W$:
        \begin{equation}
            \ker\partial_i = Y,\qquad 1 \le i \le N.
        \end{equation}
        \item On each codeblock, the product of all the local stitching operators is a nontrivial logical operator of the code $\mathcal C^i$; this implies that
        \begin{equation}
            \mathbbm{1}_W\in Y.
        \end{equation}
    \end{enumerate}
    These data first define the underlying subsystem code of the quilt, whose gauge group is generated by the local stabilisers and the stitching operators:
    \begin{equation}\label{eq:quiltdefv6}
        \quilt{G}
        =
        \left\langle
        H^i_\uparrow\ (1\le i\le N),\quad
        \Stitch(T)\ (T\in \ft^W\otimes B)
        \right\rangle.
    \end{equation}
    Ultimately, the quilt will be a PIC (Appendix~\ref{sec:operationaldistance}) rather than merely this subsystem code. We first determine the centre of \(\quilt G\), after which Definition~\ref{def:operationalquilt} specifies which part of that centre is inherited from the original codeblocks and can therefore be trusted.
\end{definition}

\begin{definition}[Quilt isomorphism]\label{def:quiltisomorphism}
    Let $\quilt{\mathcal C}_1$ and $\quilt{\mathcal C}_2$ be quilts with the same number of codeblocks, with $n^i_1=n^i_2$ for every $i$. A \textit{quilt isomorphism} from $\quilt{\mathcal C}_1$ to $\quilt{\mathcal C}_2$ is a set of local Clifford operations $\{U^i\}$ on $n^i$ qubits respectively, generated by Hadamard, $S$, and $\mathrm{SWAP}$ gates, that maps corresponding codeblocks and local stitching operators to one another as follows. If $M_{U^i}$ is the induced binary symplectic matrix, write
    \begin{equation}
        U^i(X):=\{pM_{U^i}:p\in X\}
    \end{equation}
    for a Pauli subspace $X$, and apply the same rowwise convention to stitching matrices. We require
    \begin{align}
        U^i(H^i_1) &= H^i_2,\\
        U^i(S^i_1) &= S^i_2,
    \end{align}
    and that the incidence codes agree. If a quilt isomorphism exists between two quilts, we say the two quilts are isomorphic and write $\quilt{\mathcal C}_1\cong\quilt{\mathcal C}_2$.
\end{definition}

\begin{definition}[Stabiliser realisations]\label{def:stab-stitch-realisation}
    Define the dual anticommutation map
    \begin{equation}
        \partial_i^*:H^i\rightarrow (\ft^W)^*,\qquad
        [\partial_i^*(h)](x)=h\Omega \Stitch_i(x)^T.
    \end{equation}
    Since $\ker\partial_i=Y$, the image of $\partial_i^*$ is $Y^\perp$. Choose a section
    \begin{equation}
        \sigma_i:Y^\perp\rightarrow H^i
    \end{equation}
    such that $\partial_i^*(\sigma_i(y))=y$ for all $y\in Y^\perp$.

    Define the local and global stabiliser realisation maps by
    \begin{equation}
        \Stab_i:Y^\perp\rightarrow \Pauli_{n^i},
        \qquad
        \Stab_i(y)=\sigma_i(y),
        \qquad
        \Stab(T^*)=\prod_{i=1}^N \bigl(\Stab_i(T^*_{\cdot i})\bigr)_\uparrow^{(i)}.
    \end{equation}
    Here $T^*\in Y^\perp\otimes(\ft^N)^*$. For a subspace $X\le Y^\perp\otimes(\ft^N)^*$, write $\Stab(X)$ for its image.

    The realisation maps satisfy
    \begin{equation}\label{eq:stab-stitch-pairing}
        \Stab(T^*)\Omega\Stitch(T)^T
        =
        T^*\cdot T
        =
        \sum_{a=1}^W\sum_{i=1}^N T^*_{ai}T_{ai}.
    \end{equation}
    Here $\Stab(T^*)$ and $\Stitch(T)$ are identified with their binary representatives; the value of this pairing is the commutation parity of the two Pauli products.
\end{definition}

\subsection{Gauge and stabiliser group structure}

\begin{lemma}[Quilted code gauge group structure]\label{lemma:quiltgaugestructure}
    Define \(H^i_c:=\ker\partial_i^*\). A quilt's gauge group is generated by three subspaces
    \begin{equation}
        \quilt{G}=\left\langle \quilt H_c,G_{ac},G_s\right\rangle,
    \end{equation}
    where the three spaces are respectively:
    \begin{enumerate}
        \item the commuting original stabilisers,
        \begin{equation}
            \quilt H_c=\left\langle (H^i_c)_\uparrow\ (1\le i\le N)\right\rangle;
        \end{equation}
        \item the chosen anticommuting original-stabiliser representatives,
        \begin{equation}
            G_{ac}=\Stab\left(Y^\perp\otimes(\ft^N)^*\right);
        \end{equation}
        \item the stitching generators,
        \begin{equation}
            G_s=\Stitch\left(\ft^W\otimes B\right).
        \end{equation}
    \end{enumerate}
\end{lemma}

\begin{proof}
    \begin{enumerate}
        \item The subspace $H^i_c=\ker\partial_i^*$ consists of the original stabilisers that commute with all local stitching operators. Lifting these over all codeblocks gives
        \begin{equation}
            \quilt H_c=\left\langle (H^i_c)_\uparrow\ (1\le i\le N)\right\rangle.
        \end{equation}

        \item The chosen section $\sigma_i$ supplies one stabiliser representative for each possible anticommutation pattern in $Y^\perp$. Hence every $h\in H^i$ decomposes as a sum of one commuting stabiliser and one chosen anticommuting representative:
        \begin{equation}
            H^i=H^i_c+\operatorname{im}\sigma_i.
        \end{equation}
        Taking arbitrary products of these chosen representatives over the \(N\) codeblocks gives
        \begin{equation}
            G_{ac}=\Stab\left(Y^\perp\otimes(\ft^N)^*\right).
        \end{equation}

        \item From the definition of the quilt, the stitching generators are exactly
        \begin{equation}
            G_s=\Stitch\left(\ft^W\otimes B\right).
        \end{equation}
    \end{enumerate}
\end{proof}

\begin{lemma}[Quilted code stabiliser group structure]\label{lemma:quiltstabiliserstructure}
    The stabilisers of the quilt $\quilt{\mathcal C}$ are generated by three subspaces
    \begin{equation}
        \quilt H=\left\langle\quilt H_c,\quilt H_{ac},\quilt H_m\right\rangle,
    \end{equation}
    where the three spaces are respectively:
    \begin{enumerate}
        \item the commuting original stabilisers,
        \begin{equation}
            \quilt H_c=\left\langle (H^i_c)_\uparrow\ (1\le i\le N)\right\rangle;
        \end{equation}
        \item the central products of anticommuting original-stabiliser representatives,
        \begin{equation}
            \quilt H_{ac}=\Stab\left(Y^\perp\otimes B^\perp\right);
        \end{equation}
        \item the closed stitching products,
        \begin{equation}
            \quilt H_m=\Stitch\left(Y\otimes B\right).
        \end{equation}
    \end{enumerate}
\end{lemma}

\begin{proof}
    The stabiliser group of a subsystem code is the centre of its gauge group. The commuting parts of the original stabilisers are already in this centre. To find the other two subspaces of central gauge operators, use the concrete matrix pairing in Eq.~\eqref{eq:stab-stitch-pairing}.

    A general element of the remaining gauge group is a sum \(h+s\) with \(h\in G_{ac}\) and \(s\in G_s\). Original stabilisers commute with one another, while the disjoint local stitching operators imply that all stitching generators commute with one another. Consequently, commutation of \(h+s\) with the original stabiliser generators constrains \(s\) alone, and commutation with the stitching generators constrains \(h\) alone. Mixed terms therefore cannot become central through cancellation between these two conditions, so the two subspaces may be determined independently.

    \begin{enumerate}
        \item By definition, the elements of \(\quilt H_c\) commute with all local stitching operators. They are original stabilisers, so they also commute with the original-stabiliser part of the gauge group. Thus \(\quilt H_c\) is central.

        \item The elements of $G_{ac}$ that commute with the stitching operators are exactly those $T^*$ such that
        \begin{equation}
            T^*\cdot T=0\qquad\text{for all }T\in\ft^W\otimes B.
        \end{equation}
        Therefore
        \begin{equation}
            T^*\in \left(Y^\perp\otimes(\ft^N)^*\right)\cap (\ft^W\otimes B)^\perp
            =Y^\perp\otimes B^\perp,
        \end{equation}
        giving
        \begin{equation}
            \quilt H_{ac}=\Stab(Y^\perp\otimes B^\perp).
        \end{equation}

        \item Similarly, the elements of $G_s$ that commute with $G_{ac}$ are given by
        \begin{equation}
            T\in \left(Y^\perp\otimes(\ft^N)^*\right)^\perp\cap (\ft^W\otimes B)
            =Y\otimes B,
        \end{equation}
        giving
        \begin{equation}
            \quilt H_m=\Stitch(Y\otimes B).
        \end{equation}
    \end{enumerate}

\end{proof}

\subsection{The quilt as a partially initialised code}

\begin{definition}[Quilt PIC]\label{def:operationalquilt}
    Let
    \begin{equation}
        G_0
        :=
        \left\langle H^i_\uparrow\ (1\le i\le N)\right\rangle
    \end{equation}
    be the stabiliser group of the unstitched codeblocks. The \textit{inherited centre} of the quilt is
    \begin{equation}
        \quilt C
        :=
        G_0\cap\quilt H
        =
        \left\langle\quilt H_c,\quilt H_{ac}\right\rangle.
    \end{equation}
    The final equality follows from Lemma~\ref{lemma:quiltstabiliserstructure}: the first two summands are original stabilisers, while an element of \(\quilt H_m\) is an original stabiliser precisely when it lies in their span.

    The code quilt of Definition~\ref{def:codequilt} is the PIC
    \begin{equation}
        (\quilt G,\quilt C).
    \end{equation}
    Its operational distance is denoted
    \begin{equation}
        \quilt d_{\mathrm{op}}
        :=
        d_{\mathrm{op}}(\quilt G,\quilt C).
    \end{equation}
\end{definition}

\begin{definition}[Measured logical products]\label{def:measuredlogicalproducts}
    Let $\mathcal L_i:=(H^i)^\perp/H^i$ be the logical quotient space of the $i$th original codeblock. For products of local stitching operators that commute with the original stabilisers, define the local logical-image maps
    \begin{equation}
        \lambda_i:Y\rightarrow \mathcal L_i,
        \qquad
        \lambda_i(y)=[\Stitch_i(y)].
    \end{equation}
    For the quilt incidence code $B$, define
    \begin{equation}
        \lambda_B:Y\otimes B\rightarrow\bigoplus_{i=1}^N\mathcal L_i,
        \qquad
        \lambda_B(T)=\left(\lambda_i(T_{\cdot i})\right)_{i=1}^N.
    \end{equation}
    The logical products measured by the quilt are represented by
    \begin{equation}
        \mathcal M_B:=\operatorname{im}\lambda_B.
    \end{equation}
\end{definition}

\begin{corollary}[Quilts are admissible merges]\label{corollary:quiltPIC}
    The quilt gauge group \(\quilt G\) is an admissible merge of \(G_0\). Its measured logical subspace is represented by \(\mathcal M_B\).
\end{corollary}

\begin{proof}
    By Lemma~\ref{lemma:quiltstabiliserstructure},
    \begin{equation}
        \quilt H
        =
        \left\langle\quilt C,\quilt H_m\right\rangle.
    \end{equation}
    Choose a complement \(\quilt M\le\quilt H_m\) such that
    \begin{equation}
        \quilt H=\quilt C\oplus\quilt M.
    \end{equation}
    Now project the closed stitching products in \(\quilt H_m\) into the logical quotient of the unstitched code,
    \begin{equation}
        G_0^\perp/G_0
        \cong
        \bigoplus_{i=1}^N\mathcal L_i.
    \end{equation}
    The kernel of this projection is \(\quilt H_m\cap G_0\), while its image is \(\mathcal M_B\) by Definition~\ref{def:measuredlogicalproducts}. Since
    \begin{equation}
        \quilt H_m\cap G_0
        =
        \quilt H_m\cap\quilt C
    \end{equation}
    and \(\quilt M\) is a complement of \(\quilt C\) in \(\quilt H\) chosen inside \(\quilt H_m\), the projection restricts to an isomorphism from \(\quilt M\) onto \(\mathcal M_B\). The first admissibility condition therefore holds and the quilt measures precisely the logical products represented by \(\mathcal M_B\).

    It remains to verify clean splitting. Take \(E\in\quilt G\cap G_0^\perp\) and write
    \begin{equation}
        E=h+s,
        \qquad
        h\in G_0,
        \qquad
        s\in G_s.
    \end{equation}
    Since \(E\) and \(h\) commute with \(G_0\), so does \(s\). The stitching operators mutually commute, so \(s\) also commutes with \(G_s\), and hence with every generator of \(\quilt G\). Therefore \(s\in\quilt H\), giving
    \begin{equation}
        E\in G_0+\quilt H.
    \end{equation}
    The reverse inclusion is immediate, so
    \begin{equation}
        \quilt G\cap G_0^\perp=G_0+\quilt H.
    \end{equation}
\end{proof}

\begin{corollary}[Logical count]\label{corollary:quiltlogicalcount}
    The quilt $\quilt{\mathcal C}$ has
    \begin{equation}
        \quilt k=\sum_{i=1}^N k^i-\dim\mathcal M_B
    \end{equation}
    logical qubits. Equivalently, the logical qubits of the original codes survive except for the logical products represented by \(\mathcal M_B\), which are measured out and become part of the stabiliser group.
\end{corollary}

\begin{proof}
    The unstitched stabiliser code \(G_0\) encodes \(\sum_i k^i\) logical qubits. By Corollary~\ref{corollary:quiltPIC}, the quilt is an admissible merge of \(G_0\) measuring the \(\dim\mathcal M_B\) independent logical products represented by \(\mathcal M_B\). The result follows from Lemma~\ref{lemma:admissiblemergelogicalcount}.
\end{proof}

\begin{corollary}[Isomorphic distance preservation]\label{corollary:isodistance}
    Isomorphic quilts have the same dressed distance, operational distance, and number of logical qubits.
\end{corollary}

\begin{proof}
    A quilt isomorphism maps both \(\quilt G\) and its inherited centre \(\quilt C\) to their counterparts in the second quilt. Since local Clifford transformations preserve commutation structure and operator weight~\cite{gottesman1997stabilizer}, the subsystem codes have equal dressed distances and numbers of logical qubits, while their associated PICs have equal operational distances.
\end{proof}

\subsection{Subquilts and generating sets}

\begin{definition}[Subquilt]\label{def:subquilt}
    If \(B\le\ft^N\) is a linear code and \(I\subseteq\{1,\dots,N\}\), write
    \begin{equation}
        B|_I:=\{b|_I:b\in B\}\le\ft^I
    \end{equation}
    for its restriction to the coordinates in \(I\). Given a quilt $\quilt{\mathcal C}$ and a subset of the codeblocks $I\subseteq\{1,\dots,N\}$, if $B|_I$ has distance at least $2$, the \textit{subquilt induced by $I$} is denoted
    \begin{equation}
        \quilt{\mathcal C}|_I.
    \end{equation}
    It consists of the codes $\mathcal C^i$ and local stitching operators $S^i$ for $i\in I$, equipped with incidence code $B|_I$.
\end{definition}

\begin{definition}[Generating set of subquilts]\label{def:generatingsubquilts}
    Let $Q$ be an index set. For each $q\in Q$, choose a codeblock subset $I_q\subseteq\{1,\dots,N\}$ such that
    \begin{equation}
        \quilt{\mathcal C_q}:=\quilt{\mathcal C}|_{I_q},
        \qquad
        B_q:=B|_{I_q}
    \end{equation}
    is a subquilt. Define the lifted consistency code
    \begin{equation}
        \widetilde B_q:=\{v\in\ft^N:v|_{I_q}\in B_q\}.
    \end{equation}
    Let $\mathcal M_q\le\bigoplus_{i\in I_q}\mathcal L_i$ be the measured logical subspace of $\quilt{\mathcal C_q}$, and define its lift
    \begin{equation}
        \widetilde{\mathcal M}_q
        :=
        \left\{\ell\in\bigoplus_{i=1}^N\mathcal L_i:\ell|_{I_q}\in\mathcal M_q\right\}.
    \end{equation}
    We say that the subquilts form a \textit{generating set of subquilts} for $\quilt{\mathcal C}$ if
    \begin{equation}\label{eq:generating-subquilts-incidence}
        B=\bigcap_{q\in Q}\widetilde B_q
        \qquad\text{equivalently}\qquad
        B^\perp=\sum_{q\in Q}\widetilde B_q^\perp,
    \end{equation}
    and
    \begin{equation}\label{eq:generating-subquilts-logical}
        \mathcal M_B=\bigcap_{q\in Q}\widetilde{\mathcal M}_q.
    \end{equation}
\end{definition}

\subsection{Stitch syndromes and compatible errors}

\begin{definition}[Stitch syndrome]\label{def:stitchsyndrome}
    Choose a complementary subspace $Z\le\ft^W$ such that
    \begin{equation}
        \ft^W=Y\oplus Z.
    \end{equation}
    Equivalently, \(Z\) gives a section of the quotient map \(\ft^W\rightarrow\ft^W/Y\).
    For an operator
    \begin{equation}
        E=(E^1,\dots,E^N),\qquad E^i\in \Pauli_{n^i},
    \end{equation}
    which commutes with \(\quilt H_c\), there is a unique vector \(\Gamma^i(E)\in Z\) such that
    \begin{equation}
        E^i+\Stitch_i(\Gamma^i(E))\in(H^i)^\perp.
    \end{equation}
    Indeed, \(\partial_i\) restricts to an isomorphism from \(Z\) onto the functionals in \((H^i)^*\) that vanish on \(H^i_c\).
    The \textit{stitch syndrome} of \(E\) is the matrix with these columns:
    \begin{equation}
        \Gamma:
        \quilt H_c^\perp\longrightarrow Z\otimes\ft^N,
        \qquad
        E\longmapsto
        \Gamma(E):=
        \bigl(\Gamma^1(E),\dots,\Gamma^N(E)\bigr).
    \end{equation}
    Define also the completed operator
    \begin{equation}
        \overline E:=E+\Stitch(\Gamma(E)).
    \end{equation}
    By construction, each local component \(\overline E^i\) commutes with \(H^i\).
\end{definition}

\begin{lemma}[Compatible errors]\label{lemma:stitchsyndrome}
    If an error is compatible with the quilt PIC,
    \begin{equation}
        E\in\quilt C^\perp,
    \end{equation}
    then its stitch syndrome lies in the primal tensor space
    \begin{equation}
        \Gamma(E)\in Z\otimes B.
    \end{equation}
    Its completed operator has an original-code logical label
    \begin{equation}
        \ell(\overline E)
        :=
        \bigl([\overline E^1],\dots,[\overline E^N]\bigr)
        \in
        \bigoplus_{i=1}^N\mathcal L_i,
    \end{equation}
    and
    \begin{equation}\label{eq:gauge-label-criterion}
        E\in\quilt G
        \quad\Longleftrightarrow\quad
        \ell(\overline E)\in\mathcal M_B.
    \end{equation}
\end{lemma}

\begin{proof}
    Since $E$ commutes with all of $\quilt H_{ac}=\Stab(Y^\perp\otimes B^\perp)$, its stitch-syndrome matrix has zero inner product with every matrix in $Y^\perp\otimes B^\perp$. Hence
    \begin{align}
        \Gamma(E)
        &\in (Z\otimes\ft^N)\cap (Y^\perp\otimes B^\perp)^\perp\\
        &=(Z\otimes\ft^N)\cap (Y\otimes\ft^N+\ft^W\otimes B)\\
        &=Z\otimes B,
    \end{align}
    where the last equality follows from the direct-sum decomposition $\ft^W=Y\oplus Z$.

    Since \(\Gamma(E)\in Z\otimes B\), the correction \(\Stitch(\Gamma(E))\) belongs to \(\quilt G\). Suppose first that \(E\in\quilt G\). After this correction, the remaining stitching component is closed on every block and is represented by an element of \(Y\otimes B\). Therefore
    \begin{equation}
        \ell(\overline E)\in\operatorname{im}\lambda_B=\mathcal M_B.
    \end{equation}

    Conversely, suppose \(\ell(\overline E)\in\mathcal M_B\). Choose \(T\in Y\otimes B\) with
    \begin{equation}
        \lambda_B(T)=\ell(\overline E).
    \end{equation}
    The operator \(\overline E+\Stitch(T)\) is then an original stabiliser on every block, and hence belongs to \(G_0\). Thus
    \begin{equation}
        E
        =
        \Stitch(\Gamma(E))
        +
        \Stitch(T)
        +
        \bigl(\overline E+\Stitch(T)\bigr)
        \in\quilt G.
    \end{equation}
    This proves Eq.~\eqref{eq:gauge-label-criterion}.
\end{proof}

\begin{remark}
    The complement \(Z\) fixes concrete representatives for \(\Gamma(E)\) and \(\overline E\), but Eq.~\eqref{eq:gauge-label-criterion} is independent of this choice. Changing \(Z\) changes the completion only by an element of \(\Stitch(Y\otimes B)\), and hence changes \(\ell(\overline E)\) only by an element of \(\mathcal M_B\).
\end{remark}

\section{Distance theorems}
\label{sec:distancetheorems}
\subsection{Sub-distance errors and generating subquilts}

\begin{lemma}[Sub-distance operational errors]\label{lemma:subdistancegamma}
    Fix \(D\le\min_i d^i\). If
    \begin{equation}
        E\in\quilt C^\perp\setminus\quilt G,
        \qquad
        |E|<D,
    \end{equation}
    then \(\Gamma(E)\ne0\).
\end{lemma}

\begin{proof}
    If \(\Gamma(E)=0\), then each \(E^i\) commutes with \(H^i\). Since \(|E^i|\le|E|<D\le d^i\), every \(E^i\) is an original stabiliser. Hence \(E\in G_0\le\quilt G\), a contradiction.
\end{proof}

\begin{theorem}[Generator-subquilt distance theorem]\label{theorem:generatorsubquilt}
    Let $\quilt{\mathcal C}$ be a quilt with incidence code $B$. Suppose $\{\quilt{\mathcal C_q}\}_{q\in Q}$ is a generating set of subquilts in the sense of Definition~\ref{def:generatingsubquilts}. If every generator subquilt has operational distance at least $D$, then
    \begin{equation}
        \quilt d_{\mathrm{op}}\ge D.
    \end{equation}
\end{theorem}

\begin{proof}
    Suppose, for contradiction, that
    \begin{equation}
        E\in\quilt C^\perp\setminus\quilt G,
        \qquad
        |E|<D.
    \end{equation}
    By Lemma~\ref{lemma:stitchsyndrome},
    \begin{equation}
        \Gamma(E)\in Z\otimes B.
    \end{equation}
    For every generator subquilt, the restricted stitch syndrome satisfies
    \begin{equation}
        \Gamma(E)|_{I_q}\in Z\otimes B_q.
    \end{equation}
    The restriction \(E|_{I_q}\) commutes with the local \(\quilt H_c\), and the displayed incidence condition implies, through Eq.~\eqref{eq:stab-stitch-pairing}, that it commutes with the subquilt's \(\quilt H_{ac}\). It is therefore compatible with the subquilt PIC. Since
    \begin{equation}
        \bigl|E|_{I_q}\bigr|\le |E|<D,
    \end{equation}
    the assumed operational distance implies \(E|_{I_q}\) belongs to the subquilt gauge group. Eq.~\eqref{eq:gauge-label-criterion}, applied inside that subquilt, then gives
    \begin{equation}
        \ell(\overline E)|_{I_q}\in\mathcal M_q,
        \qquad\text{or equivalently}\qquad
        \ell(\overline E)\in\widetilde{\mathcal M}_q
        \qquad\text{for every }q\in Q.
    \end{equation}
    Therefore
    \begin{equation}
        \ell(\overline E)
        \in
        \bigcap_{q\in Q}\widetilde{\mathcal M}_q
        =
        \mathcal M_B.
    \end{equation}
    By Eq.~\eqref{eq:gauge-label-criterion}, this implies \(E\in\quilt G\), a contradiction.
\end{proof}

\subsection{Weight-$D$ quilts}

\begin{theorem}[Weight-$D$ quilt distance]\label{theorem:weightDquilt}
    A code quilt where
    \begin{enumerate}
        \item the constituent codes all have distance $D$;
        \item the local stitching operators have matching weights
        \begin{equation}
            |S^i_a|=|S^j_a|=w_a,
            \qquad
            1\le i,j\le N,
            \qquad
            1\le a\le W,
        \end{equation}
        with
        \begin{equation}
            \sum_{a=1}^W w_a=D;
        \end{equation}
    \end{enumerate}
    has operational distance $\quilt d_{\mathrm{op}}\ge D$.
\end{theorem}

\begin{proof}
    First note that $Y=\langle\mathbbm 1\rangle$. Indeed, if $0\ne y\ne\mathbbm 1$ lies in $Y$, then $\Stitch_i(y)$ is closed and has weight strictly less than $D$. If it is a nontrivial logical operator, this contradicts $d(\mathcal C^i)=D$. If it is a stabiliser, then $\Stitch_i(\mathbbm 1+y)$ is a nontrivial logical operator of weight strictly less than $D$, again a contradiction.

    Let $L_i=\Stitch_i(\mathbbm 1)$. Suppose, for contradiction, that there is an operational error
    \begin{equation}
        E=(E^1,\dots,E^N)
        \in
        \quilt C^\perp\setminus\quilt G,
        \qquad
        |E|<D.
    \end{equation}
    By Lemmas~\ref{lemma:stitchsyndrome} and~\ref{lemma:subdistancegamma}, its stitch syndrome satisfies
    \begin{equation}
        0\ne\Gamma(E)\in Z\otimes B.
    \end{equation}
    Write \(\Gamma=\Gamma(E)\).

    For each block $i$ and $t\in\ft$, define the local stitching completion
    \begin{equation}
        P_i(t)=\Stitch_i(\Gamma^i+t\mathbbm 1),
        \qquad
        |P_i(t)|=\sum_{a=1}^W(\Gamma_a^i+t)w_a.
    \end{equation}
    The two completed operators
    \begin{equation}
        F_i(0)=E^i+P_i(0),
        \qquad
        F_i(1)=E^i+P_i(1)
    \end{equation}
    are closed operators of $\mathcal C^i$, and they differ by $L_i$. Choose any value of $\tau_i\in\ft$ for which $F_i(\tau_i+1)$ is nontrivial; if both completions are nontrivial, either value of $\tau_i$ is allowed and the following inequality holds for both choices. The distance-$D$ condition gives
    \begin{align}
        D
        &\le |F_i(\tau_i+1)|\\
        &=|E^i+P_i(\tau_i+1)|\\
        &\le |E^i|+|P_i(\tau_i+1)|.
    \end{align}
    For each \(a\), exactly one of \(\Gamma_a^i+\tau_i\) and \(\Gamma_a^i+\tau_i+1\) is \(1\). Since the local stitching operators are disjoint, the two completions therefore select complementary sets of local stitching operators, and
    \begin{equation}
        |P_i(\tau_i)|+|P_i(\tau_i+1)|
        =
        \sum_{a=1}^W w_a
        =
        D.
    \end{equation}
    Hence
    \begin{equation}
        |E^i|\ge D-|P_i(\tau_i+1)|=|P_i(\tau_i)|=\sum_{a=1}^W (\Gamma^i_a+\tau_i)w_a.
    \end{equation}
    A vector $\tau=(\tau_1,\dots,\tau_N)$ satisfying these blockwise choices will be called valid. Summing over codeblocks gives
    \begin{equation}\label{eq:weightD-bound-v6}
        |E|\ge \sum_{a=1}^W w_a |\Gamma_a+\tau|.
    \end{equation}

    If a valid choice of $\tau\notin B$ exists, then each row $\Gamma_a+\tau$ is nonzero because $\Gamma_a\in B$. Eq.~\eqref{eq:weightD-bound-v6} then gives
    \begin{equation}
        |E|\ge \sum_{a=1}^W w_a=D.
    \end{equation}

    It remains to consider the case where no valid choice of $\tau\notin B$ exists. Choose a valid $\tau\in B$. For this $\tau$, the completed operators $F_i(\tau_i)$ have trivial logical effect on every codeblock. To see this, suppose instead that \(F_i(\tau_i)\) were nontrivial for some \(i\). The original choice of \(\tau\) is valid, so \(F_i(\tau_i+1)\) is nontrivial. Flipping only the \(i\)th coordinate gives a new vector \(\tau+e_i\), and the required nontrivial completion on block \(i\) is now \(F_i(\tau_i)\); all other blocks are unchanged. Thus \(\tau+e_i\) is valid. Since \(d_B\ge2\), no unit vector lies in \(B\), so \(\tau+e_i\notin B\), contradicting the assumption that no valid choice lies outside \(B\). Therefore each $F_i(\tau_i)$ is an original-code stabiliser, and
    \begin{equation}
        E=\Stitch(\Gamma+\mathbbm 1\otimes\tau)+\text{(local stabilisers)}.
    \end{equation}
    Since $\Gamma\in Z\otimes B\subseteq\ft^W\otimes B$ and $\mathbbm 1\otimes\tau\in Y\otimes B\subseteq\ft^W\otimes B$, the stitching term lies in the gauge group. Thus $E\in\quilt G$, contradicting the assumption that \(E\) is operationally nontrivial.
\end{proof}

\subsection{Homogeneous quilts and the perforated code}

\begin{definition}[Base quilt]
    Given a quilt $\quilt{\mathcal C}$, a \textit{base quilt} is any isomorphic quilt $\quilt{\mathcal C}_B\cong\quilt{\mathcal C}$ with the property that whenever two codeblocks, together with their local stitching operators, are locally Clifford equivalent, they are represented by the same pair $(H,S)$.
    This is only a notational device: by Corollary~\ref{corollary:isodistance}, passing to a base quilt does not change either distance or the number of logical qubits.
\end{definition}

\begin{definition}[Homogeneous quilt]\label{def:homogeneousquilt}
    A \textit{homogeneous quilt} is a quilt whose base quilt has identical local data $(H,S)$ on every codeblock.
\end{definition}

\begin{definition}[Perforated code]\label{def:perforatedcode}
    Given a codeblock \(\mathcal C^i\) inside a quilt, its \textit{perforated code} is the PIC
    \begin{equation}
        (\dotG^{\,i},H^i_c),
    \end{equation}
    where
    \begin{equation}
        \dotG^{\,i}=\left\langle H^i,\ \Stitch_i(x)\ (x\in\ft^W)\right\rangle.
    \end{equation}
    Its operational distance is denoted \(\dotd_i\). In a homogeneous quilt, after passing to the base quilt, these distances are all equal; we denote the common value by \(\dotd\).
\end{definition}

\begin{lemma}[Perforated operational quotient]\label{lemma:perforatedquotient}
    Let
    \begin{equation}
        U_i:=\operatorname{im}\lambda_i\le\mathcal L_i.
    \end{equation}
    Completing the local stitch syndrome induces an isomorphism
    \begin{equation}
        (H^i_c)^\perp/\dotG^{\,i}
        \cong
        \mathcal L_i/U_i.
    \end{equation}
\end{lemma}

\begin{proof}
    For \(e\in(H^i_c)^\perp\), there is a unique \(\gamma^i(e)\in Z\) such that
    \begin{equation}
        \overline e=e+\Stitch_i(\gamma^i(e))\in(H^i)^\perp.
    \end{equation}
    Since the added stitching operator belongs to \(\dotG^{\,i}\), the class of \(e\) is represented by the original-code logical label \([\overline e]\). Two such labels represent the same perforated operational class exactly when they differ by a closed local stitching product, whose logical image lies in \(U_i\).
\end{proof}

\begin{theorem}[Homogeneous quilt distance]\label{theorem:homogeneousquiltdistance}
    Let $\quilt{\mathcal C}$ be a homogeneous quilt whose original codeblocks have distance $D$. If the common perforated-code operational distance is $\dotd$ and
    \begin{equation}
        d_B\dotd\ge D,
    \end{equation}
    then
    \begin{equation}
        \quilt d_{\mathrm{op}}\ge D.
    \end{equation}
\end{theorem}

\begin{proof}
    By Corollary~\ref{corollary:isodistance}, it is sufficient to consider the base quilt in which all codeblocks support identical codes and local stitching operators. Write
    \begin{equation}
        \mathcal L:=H^\perp/H,
        \qquad
        U:=\operatorname{im}\lambda_i\le\mathcal L
    \end{equation}
    for the common original-code logical space and common measured local subspace.

    Suppose, for contradiction, that
    \begin{equation}
        E\in\quilt C^\perp\setminus\quilt G,
        \qquad
        |E|<D.
    \end{equation}
    By Lemma~\ref{lemma:stitchsyndrome},
    \begin{equation}
        \Gamma(E)\in Z\otimes B.
    \end{equation}

    For any $b^\perp\in B^\perp$, we have
    \begin{equation}
        \sum_{i=1}^N b^\perp_i\Gamma^i=0.
    \end{equation}
    Identify the physical Pauli spaces of the base-code blocks and form
    \begin{equation}
        P=\sum_{i=1}^N b^\perp_iE^i.
    \end{equation}
    This operator has zero stitch syndrome and
    \begin{equation}
        |P|
        \le
        \sum_{i:b^\perp_i=1}|E^i|
        \le
        |E|
        <D.
    \end{equation}
    It therefore commutes with \(H\), and the distance of the original code gives \(P\in H\). Since
    \begin{equation}
        \sum_{i=1}^N b^\perp_i\overline E^i
        =
        P+\Stitch_1\left(\sum_{i=1}^N b^\perp_i\Gamma^i\right)
        =
        P,
    \end{equation}
    we obtain
    \begin{equation}
        \sum_{i=1}^N b^\perp_i[\overline E^i]=0
        \qquad\text{in }\mathcal L.
    \end{equation}
    Since this holds for every \(b^\perp\in B^\perp\),
    \begin{equation}
        \ell(\overline E)\in\mathcal L\otimes B.
    \end{equation}

    Homogeneity gives
    \begin{equation}
        \mathcal M_B=U\otimes B.
    \end{equation}
    Since \(E\notin\quilt G\), Eq.~\eqref{eq:gauge-label-criterion} implies
    \begin{equation}
        \ell(\overline E)\notin U\otimes B.
    \end{equation}
    Its image in \((\mathcal L/U)\otimes B\) is therefore nonzero. In any basis of \(\mathcal L/U\), at least one coefficient is a nonzero codeword of \(B\), and hence is supported on at least \(d_B\) blocks. By Lemma~\ref{lemma:perforatedquotient}, \(E^i\) is a nontrivial operational error of the perforated PIC on each of those blocks. Each contributes at least \(\dotd\), so
    \begin{equation}
        |E|\ge d_B\dotd\ge D.
    \end{equation}
    This contradicts $|E|<D$ and proves the theorem.
\end{proof}

\subsection{Atomic and nuclear quilts}

\begin{definition}[Atomic and nuclear quilts]\label{def:atomicnuclear}
    An \textit{atomic quilt} $\atom{\mathcal C}$ is a quilt on $N$ codeblocks whose incidence code is the $N$-bit repetition code
    \begin{equation}
        B=\langle \mathbbm 1_N\rangle.
    \end{equation}

    A \textit{local code/stitch type} means one of the pairs $(H,S)$ appearing in the base quilt. When at least two local code/stitch types appear, the \textit{nuclear quilt} of an atomic quilt is the subquilt of the atomic quilt's base quilt obtained by taking one representative of each type. Its incidence code is the repetition code on those representatives. Thus a nuclear quilt always contains at least two codeblocks. If every type appears once, the atomic quilt is its own nuclear quilt. We do not define a nuclear quilt when only one type appears; in that case the atomic quilt is homogeneous.
\end{definition}

\begin{corollary}[Atomic quilt generators]\label{corollary:atomicgenerators}
    Let $\atom{\mathcal C}$ be an atomic quilt with at least two local code/stitch types and target distance $D$. Suppose the following quilts all have operational distance at least $D$:
    \begin{enumerate}
        \item the nuclear quilt;
        \item for each local code/stitch type appearing with multiplicity $m_t>1$, a connected set of homogeneous atomic subquilts joining the $m_t$ copies of that type.
    \end{enumerate}
    Then
    \begin{equation}
        \atom d_{\mathrm{op}}\ge D.
    \end{equation}
\end{corollary}

\begin{proof}
    The nuclear quilt enforces consistency between one representative of each local code/stitch type. The within-type homogeneous generators enforce consistency among all copies of the same type. Together these conditions make the generator hypergraph connected, hence
    \begin{equation}
        \bigcap_q\widetilde B_q=\langle\mathbbm 1_N\rangle.
    \end{equation}
    Now let \(\ell\) lie in every lifted measured subspace \(\widetilde{\mathcal M}_q\). The nuclear quilt supplies one \(y\in Y\) whose images \(\lambda_i(y)\) agree with \(\ell\) on the chosen representative of every type. Each connected within-type generator then propagates that same value from the representative to every other block of the type. Hence
    \begin{equation}
        \ell
        =
        \bigl(\lambda_i(y)\bigr)_{i=1}^N
        \in
        \mathcal M_B.
    \end{equation}
    Thus
    \begin{equation}
        \bigcap_q\widetilde{\mathcal M}_q=\mathcal M_B,
    \end{equation}
    so the listed subquilts form a generating set in the sense of Definition~\ref{def:generatingsubquilts}. The result follows from Theorem~\ref{theorem:generatorsubquilt}.
\end{proof}

\begin{corollary}[Atomic quilt perforated-distance certificate]\label{corollary:atomicperfcertificate}
    Let $\atom{\mathcal C}$ be an atomic quilt with at least two local code/stitch types and target distance $D$. If the nuclear quilt has operational distance at least $D$, and for every local code/stitch type $t$ with multiplicity $m_t$ the perforated operational distance $\dotd_t$ satisfies
    \begin{equation}
        m_t\dotd_t\ge D,
    \end{equation}
    then
    \begin{equation}
        \atom d_{\mathrm{op}}\ge D.
    \end{equation}
\end{corollary}

\begin{proof}
    For each repeated type $t$, consider the homogeneous atomic subquilt containing all $m_t$ copies of that type. Its incidence code is $\langle\mathbbm 1_{m_t}\rangle$, so its distance is $m_t$. The condition $m_t\dotd_t\ge D$ lets us apply Theorem~\ref{theorem:homogeneousquiltdistance} to each such within-type subquilt. Together with the assumed operational distance of the nuclear quilt, Corollary~\ref{corollary:atomicgenerators} gives $\atom d_{\mathrm{op}}\ge D$.
\end{proof}

\begin{remark}[Dual generation width]\label{rem:dualgenerationwidth}
    The incidence condition in a local certificate can be generated locally when $B^\perp$ is generated by low-weight checks. Define the dual generation width
    \begin{equation}
        g^\perp(B)=\min\{s:B^\perp\text{ is generated by codewords of weight at most }s\}.
    \end{equation}
    For atomic quilts, $g^\perp(B)=2$, and a connected graph of two-block generators suffices at the incidence level. This controls only the incidence-code condition in Definition~\ref{def:generatingsubquilts}; the selected restrictions must still define valid subquilts, and the measured-logical-subspace condition must be checked separately.
\end{remark}

\section{Operational distance under gauge fixing}\label{sec:gauge-fixing}

In this appendix we show that the stabiliser protocol is at least as well protected as the subsystem protocol at the code-capacity level. The stabiliser protocol is obtained from the code quilt by gauge fixing: the commuting stitching operators are promoted from gauge operators to stabilisers, and their values become available for error detection. Theorem~\ref{theorem:gauge-fixing-distance} states that fixing a maximal commuting set of gauge operators of a partially initialised code in this way cannot reduce its operational distance, provided that the fixed gauge operators are added to the trusted stabilisers whilst the operators we intend to measure remain untrusted. We then check that the stabiliser protocol is a gauge fixing of this kind, which proves Theorem~\ref{theorem:gauge-fixing-main} of the main text.

\begin{theorem}[Operational distance under complete gauge fixing]\label{theorem:gauge-fixing-distance}
    Let $(G,C)$ be a partially initialised code, and write $Z(G)=C\oplus M$. Let $A\le G$ be an isotropic subspace with $A\cap Z(G)=0$ such that
    \begin{equation}
        K=Z(G)\oplus A,\qquad G\cap K^\perp=K.
    \end{equation}
    Thus $K$ fixes a maximal commuting set of gauge degrees of freedom. With their fixed eigenvalues included among the trusted stabilisers, the resulting partially initialised stabiliser code is $(K,C_A)$, where $C_A=C\oplus A$, and
    \begin{equation}
        d_{\mathrm{op}}(K,C_A)\ge d_{\mathrm{op}}(G,C).
    \end{equation}
\end{theorem}

\begin{proof}
    Suppose $E\in C_A^\perp\setminus K$. Then $E\in C^\perp$. If $E$ belonged to $G$, it would commute with $Z(G)$ and with $A$, hence belong to $G\cap K^\perp=K$, a contradiction. Therefore
    \begin{equation}
        C_A^\perp\setminus K\ \subseteq\ C^\perp\setminus G.
    \end{equation}
    Minimising physical Pauli weight proves the claim. The choice of fixed eigenvalue signs does not affect these Pauli spaces or their distances.
\end{proof}

For a quilt, let $S=\Stitch(\ft^W\otimes B)$ be the commuting stitching space and $G_0$ the original stabiliser space. Then $\quilt G=G_0+S$. The stabiliser protocol fixes the stitching gauges, giving
\begin{equation}
    K=Z(\quilt G)+S=\quilt C+S.
\end{equation}
To check completeness, take $h+s\in\quilt G$ commuting with $S$, with $h\in G_0$ and $s\in S$. Since both $G_0$ and $S$ are isotropic, $h$ then commutes with all of $\quilt G$, and lies in $\quilt C$. Hence $\quilt G\cap S^\perp=\quilt C+S=K$, which gives the required maximality. A complement $A$ of $S\cap Z(\quilt G)$ in $S$ fixes the gauge degrees while the central measured products retain their role as unknown operational outputs.

\section{Analytical sparsity and congestion}\label{sec:sparsityanalytical}

This appendix establishes the analytical columns of the summary table of
Section~\ref{sec:sparsity}.  Throughout we work under the restrictions stated
there: the stitching operators are weight-\(N\), so that each local stitching
operator is weight-one on its block, and \(N\) is treated as a constant.

\subsection{Sparsity data of a generating set}

Both figures of merit are properties of a \emph{choice} of generating set, not
of the stabiliser subspace it spans, so we fix the terminology first.

\begin{definition}[Weight and congestion]\label{def:weightcongestion}
    Let \(\mathcal G\) be a finite set of Pauli operators on a register of
    qubits.  Its \emph{weight} and \emph{congestion} are
    \begin{equation}
        \wt(\mathcal G)=\max_{g\in\mathcal G}\wt(g),
        \qquad
        \gamma(\mathcal G)=\max_{q}\left|\left\{g\in\mathcal G: q\in\operatorname{supp}g\right\}\right|,
        \label{eq:weight-congestion}
    \end{equation}
    where \(q\) ranges over the physical qubits.  We call
    \(\left(\wt(\mathcal G),\gamma(\mathcal G)\right)\) the \emph{sparsity data}
    of \(\mathcal G\).
\end{definition}

The local data against which these are measured are those of the constituent
codes.  Fix on each block \(i\) a \emph{source-check presentation}
\(\mathcal R_i\subseteq H^i\): the specific family of local checks from which
the syndrome extraction circuit of \(\mathcal C^i\) is built.  Write
\begin{equation}
    w_i=\wt(\mathcal R_i),
    \qquad
    \kappa_i=\gamma(\mathcal R_i)
    \label{eq:local-sparsity-data}
\end{equation}
for its sparsity data, and set \(w=\max_i w_i\), \(\kappa=\max_i\kappa_i\).

\begin{definition}[Analytically bounded family]\label{def:analyticallybounded}
    A family of quilt operators is \emph{analytically bounded} if it admits a
    generating set whose sparsity data are bounded by a function of
    \(w\), \(\kappa\), \(N\) and the chosen basis of \(B^\perp\) alone.
\end{definition}

The content of Definition~\ref{def:analyticallybounded} is what the bound may
\emph{not} depend on: the interface width \(W\), and the weight of the chosen
representative of the measured logical product.  A family that is bounded in
this sense remains LDPC as the interface is widened and as the measured
representative is made heavier.

\subsection{Families with immediate bounds}

\begin{proposition}[Trivially bounded families]\label{prop:trivialsparsity}
    The following families are analytically bounded.
    \begin{enumerate}
        \item \emph{Physically measured operators, subsystem protocol.}  These
        are the source checks of the constituent codes together with the local
        stitching operators, with sparsity data at most
        \(\left(\max(w,N),\,\kappa+1\right)\).
        \item \emph{Untouched stabilisers.}  In either protocol these have
        sparsity data at most \((w,\kappa)\).
        \item \emph{Stitching consistency stabilisers, stabiliser protocol.}
        These have sparsity data at most \((N,1)\).
    \end{enumerate}
\end{proposition}

\begin{proof}
    For (1), each source check lies in some \(\mathcal R_i\) and so has weight
    at most \(w\); each weight-\(N\) stitching operator \(\Stitch(x\otimes b)\)
    with \(x\) a single coordinate is weight-one on each of the \(|b|\le N\)
    blocks it touches.  A physical qubit lies in at most \(\kappa\) source
    checks, and, because the local stitching operators on a block are disjoint,
    in at most one local stitching operator.

    For (2), the untouched stabilisers are by definition generated by those
    elements of \(\mathcal R_i\) lying in \(H^i_c=\ker\partial_i^*\).  They pass
    into the quilt unchanged, so both figures of merit are inherited from
    \(\mathcal R_i\).

    For (3), in the stabiliser protocol the stitching operators are themselves
    stabilisers.  Each is weight-\(N\), and disjointness of the local stitching
    operators means each qubit carries at most one.
\end{proof}

\begin{proposition}[Measurement-consistency detectors]\label{prop:measurementconsistency}
    Let \(\Stitch(T)\in\quilt H_m\) with \(T\in Y\otimes B\) be a closed
    stitching product realising a measured logical product
    \(\lambda_B(T)\in\mathcal M_B\).  Then
    \begin{equation}
        \wt\!\left(\Stitch(T)\right)=\sum_{i:\,T_{\cdot i}\neq 0}\left|T_{\cdot i}\right|,
        \label{eq:measurementconsistencyweight}
    \end{equation}
    which is the total weight of the chosen representatives of the measured
    logical product.  This family is therefore not analytically bounded.  Each
    such operator contributes at most one to the congestion of any qubit.
\end{proposition}

\begin{proof}
    The local stitching operators are weight-one and disjoint, so the weight of
    \(\Stitch_i(T_{\cdot i})\) is the number of selected coordinates, and
    weights add across blocks.  Since the product of all local stitching
    operators on block \(i\) is a representative of a nontrivial logical
    operator of \(\mathcal C^i\), the right-hand side of
    Eq.~\eqref{eq:measurementconsistencyweight} is the weight of the chosen
    representative, which is not bounded by \(w\), \(\kappa\) or \(N\).  The
    congestion claim is immediate from disjointness: a qubit carries at most one
    local stitching operator, hence lies in the support of \(\Stitch(T)\) with
    multiplicity at most one.
\end{proof}

Proposition~\ref{prop:measurementconsistency} is the single asymmetry between
the two figures of merit in the table of Section~\ref{sec:sparsity}: the
measurement-consistency detector is heavy, but there is only one, so it adds
only one to the degree of each qubit it touches.

\subsection{Inherited stabilisers}

By Lemma~\ref{lemma:quiltstabiliserstructure}, the cross-block part of the inherited centre is
\begin{equation}
    \quilt H_{ac}=\Stab\!\left(Y^\perp\otimes B^\perp\right).
    \label{eq:sparse-hac}
\end{equation}
This characterises the subspace algebraically but exhibits no generating set,
and the realisation maps of
Definition~\ref{def:stab-stitch-realisation} are defined through an arbitrary
section \(\sigma_i\).  The cost of a generator is controlled by that choice.

\begin{lemma}[Cost of a simple tensor]\label{lemma:simpletensorcost}
    For \(y\in Y^\perp\), the lifts of \(y\) on block \(i\) form the affine coset
    \begin{equation}
        (\partial_i^*)^{-1}(y)=\sigma_i(y)+H^i_c ,
        \label{eq:affinecoset}
    \end{equation}
    and if \(c_i(y)\) denotes the least physical weight attained on it, then
    \(y\otimes b\in Y^\perp\otimes B^\perp\) admits a realisation with
    \begin{equation}
        \wt\!\left(\Stab(y\otimes b)\right)\le\sum_{i:\,b_i=1}c_i(y).
        \label{eq:sparse-hac-bound}
    \end{equation}
\end{lemma}

\begin{proof}
    Two lifts of \(y\) differ by an element of \(\ker\partial_i^*=H^i_c\), which
    gives Eq.~\eqref{eq:affinecoset}.  Replacing \(\sigma_i(y)\) by a
    minimum-weight element of its coset is a permitted choice of section, and by
    Definition~\ref{def:stab-stitch-realisation} the global realisation is the
    block-disjoint product of the local ones, so weights add over the blocks
    selected by \(b\).
\end{proof}

The problem is therefore to choose a basis of \(Y^\perp\) whose elements admit
cheap lifts simultaneously on every block that must be combined.  We solve it
under a connectivity hypothesis on the source-check presentation.

\begin{definition}[Stitching graph and connected interface]\label{def:stitchinggraph}
    Given a source-check presentation \(\mathcal R_i\) on block \(i\), let
    \(\Lambda_i\) be the graph on the \(W\) stitching coordinates with an edge \(ab\)
    whenever a single row \(r\in\mathcal R_i\) satisfies
    \(\partial_i^* r=e_a+e_b\).  Products of two or more source checks do not
    contribute edges.  The interface is \emph{connected} if \(\Lambda_i\) is connected
    for every \(1\le i\le N\).
\end{definition}

\begin{lemma}[SkipTree traversal~\cite{swaroop2024adapters}]\label{lemma:skiptree}
    Let \(\mathcal T\) be a tree on \(W\) vertices.  There is an ordering
    \(v_1,\dots,v_W\) of its vertices such that
    \begin{enumerate}
        \item consecutive vertices satisfy \(\operatorname{dist}_{\mathcal T}(v_a,v_{a+1})\le 3\); and
        \item every edge of \(\mathcal T\) lies on the connecting path of at most two
        consecutive pairs.
    \end{enumerate}
\end{lemma}

\begin{proof}
    Root \(\mathcal T\) and define two mutually recursive traversals: \(F(v)\) emits
    \(v\) and then applies \(L\) to each child in turn, while \(L(v)\) applies
    \(F\) to each child in turn and then emits \(v\).  Run \(F\) at the root.

    For (1), argue by induction on subtree height.  Within a child subtree the
    claim holds inductively.  The first and last vertices emitted by a call on a
    subtree are each either its root or a child of that root, by inspection of
    the two cases.  A transition between consecutive child subtrees of a common
    parent \(p\) therefore travels from a vertex at depth at most one in the
    first subtree, up through \(p\), and down to a vertex at depth at most one in
    the second: at most three edges.  A transition between \(p\) itself and an
    adjacent child subtree uses at most two.

    For (2), consider an edge \(pc\) with \(c\) a child of \(p\).  Every
    connecting path using \(pc\) arises from a transition entering or leaving the
    subtree rooted at \(c\).  The traversal of that subtree is a single
    contiguous block of the emitted sequence, so there is exactly one transition
    into it and one out of it, whence \(pc\) is used at most twice.  Transitions
    interior to the subtree do not use \(pc\), and transitions between other
    children of \(p\) pass through \(p\) without descending into \(c\).
\end{proof}

\begin{theorem}[Sparse inherited basis]\label{theorem:sparseinheritedbasis}
    Suppose \(Y=\langle\mathbbm 1_W\rangle\), the interface is connected in the
    sense of Definition~\ref{def:stitchinggraph}, and the local stitching
    operators are paired across blocks in the order given by applying
    Lemma~\ref{lemma:skiptree} to a spanning tree of each \(\Lambda_i\).  Let
    \(\mathcal B\) be a basis of \(B^\perp\).  Then \(\quilt H_{ac}\) admits a generating set
    \(\mathcal G=\{\,\Stab(y_a\otimes b)\,\}\) indexed by \(1\le a<W\) and
    \(b\in\mathcal B\), in which every generator is a product of at most three
    source checks on each block it touches, with sparsity data
    \begin{equation}
        \wt(\mathcal G)\le 3\max_{b\in\mathcal B}\ \sum_{i:\,b_i=1}w_i,
        \qquad
        \gamma(\mathcal G)\le 2\max_{i}\ \kappa_i\,\beta_i,
        \label{eq:skiptree-finite-bound}
    \end{equation}
    where \(\beta_i=\left|\{b\in\mathcal B: b_i=1\}\right|\).
\end{theorem}

\begin{proof}
    Since \(Y=\langle\mathbbm 1_W\rangle\), \(Y^\perp\) is the even-weight subspace
    of \(\ft^W\).  Let \(\mathcal T_i\) be the chosen spanning tree of \(\Lambda_i\), which exists
    by connectivity, and let \(v^i_1,\dots,v^i_W\) be the ordering of the local
    stitching operators on block \(i\) given by Lemma~\ref{lemma:skiptree}.  By
    hypothesis \(v^i_a\) is paired with \(v^j_a\) for all \(i,j\), so we may
    label the stitching coordinates by \(a\) and take the path basis
    \begin{equation}
        y_a=e_a+e_{a+1},\qquad 1\le a<W,
        \label{eq:pathbasis}
    \end{equation}
    of \(Y^\perp\); these are independent and \(W-1=\dim Y^\perp\) in number.
    Since \(\mathcal B\) is a basis of \(B^\perp\), the simple tensors
    \(y_a\otimes b\) span \(Y^\perp\otimes B^\perp\), so by
    Eq.~\eqref{eq:sparse-hac} their realisations generate \(\quilt H_{ac}\).

    Fix \(a\) and a block \(i\).  By Lemma~\ref{lemma:skiptree}(1) the path in
    \(\mathcal T_i\) from \(v^i_a\) to \(v^i_{a+1}\) has at most three edges.  Each edge
    is labelled by a single source check whose dual anticommutation image is the
    corresponding coordinate pair, so the product of the source checks along the
    path telescopes to \(e_a+e_{a+1}=y_a\).  That product is therefore a lift of
    \(y_a\), of weight at most \(3w_i\), and the weight bound follows from
    Lemma~\ref{lemma:simpletensorcost} by summing over the blocks selected by
    \(b\).

    For the congestion bound, let \(q\) be a qubit on block \(i\).  A generator
    \(\Stab(y_a\otimes b)\) can contain \(q\) only if \(b_i=1\) and
    \(q\in\operatorname{supp}\sigma_i(y_a)\).  The latter requires \(q\) to lie in
    one of the source checks labelling the edges of the path for \(y_a\).  Qubit
    \(q\) lies in at most \(\kappa_i\) source checks of \(\mathcal R_i\), and by
    Lemma~\ref{lemma:skiptree}(2) each such check, being a single edge of
    \(\mathcal T_i\), serves at most two values of \(a\).  Hence at most \(2\kappa_i\)
    indices \(a\) can contribute, each for the \(\beta_i\) admissible choices of
    \(b\).
\end{proof}

\subsection{Homogeneous interfaces need no traversal}

Theorem~\ref{theorem:sparseinheritedbasis} requires a spanning tree on each block because it
assumes no relationship between the blocks: a pattern \(y\in Y^\perp\)
must be assembled on each block separately, out of whatever source checks that
block happens to offer.  When the blocks are copies of one another this is
unnecessary, and the resulting bound is better by a constant factor and free of
the connectivity hypothesis.

\begin{proposition}[Doubled source checks]\label{prop:homogeneousdoubling}
    Let the quilt be homogeneous in the sense of
    Definition~\ref{def:homogeneousquilt}, so that after passing to the base
    quilt every block carries the same local data \((H,S)\) and hence the same
    source-check presentation \(\mathcal R\) and the same dual map
    \(\partial^*\).  Choose \(\mathcal R_0\subseteq\mathcal R\) whose image
    \(\partial^*(\mathcal R_0)\) is a basis of \(Y^\perp\), and let
    \(\mathcal B\) be a basis of \(B^\perp\).  Then
    \begin{equation}
        \mathcal G=\left\{\ \prod_{i:\,b_i=1}h^{(i)}\ :\ h\in\mathcal R_0,\ b\in\mathcal B\ \right\}
        \label{eq:doubledbasis}
    \end{equation}
    is a basis of \(\quilt H_{ac}\), with sparsity data
    \begin{equation}
        \wt(\mathcal G)\le w\max_{b\in\mathcal B}|b|,
        \qquad
        \gamma(\mathcal G)\le \kappa\max_i\beta_i ,
        \label{eq:doubledbound}
    \end{equation}
    where \(\beta_i=\left|\{b\in\mathcal B:b_i=1\}\right|\) as before.  In
    particular, for \(B=\langle\mathbbm 1_N\rangle\) with the weight-two path
    basis of \(B^\perp\),
    \begin{equation}
        \wt(\mathcal G)\le 2w,
        \qquad
        \gamma(\mathcal G)\le 2\kappa .
        \label{eq:doubledbound-repetition}
    \end{equation}
\end{proposition}

\begin{proof}
    Homogeneity makes the common presentation \(\mathcal R\) and the common map
    \(\partial^*\) available on every block, so for \(h\in\mathcal R_0\) the
    element \(h^{(i)}\) has the \emph{same} anticommutation pattern
    \(y_h=\partial^*(h)\) on whichever block it is placed.  Taking
    \(\sigma_i=\sigma\) for all \(i\), where \(\sigma(y_h)=h\), is therefore a
    legitimate common choice of section, and under it the product in
    Eq.~\eqref{eq:doubledbasis} is exactly \(\Stab(y_h\otimes b)\).  Since
    \(b\in B^\perp\), Eq.~\eqref{eq:sparse-hac} places it in \(\quilt H_{ac}\).

    For spanning, \(\partial^*(\mathcal R_0)\) is a basis of \(Y^\perp\) by
    construction and \(\mathcal B\) is a basis of \(B^\perp\), so the simple
    tensors \(y_h\otimes b\) form a basis of \(Y^\perp\otimes B^\perp\); their
    realisations therefore form a basis of \(\quilt H_{ac}\).

    Each generator is a product of \emph{one} source check per block it touches,
    rather than up to three, giving the weight bound.  For congestion, a qubit
    \(q\) on block \(i\) lies in at most \(\kappa\) members of \(\mathcal R\),
    each contributing one generator for each of the \(\beta_i\) admissible
    \(b\).  The repetition case follows since the path basis of \(B^\perp\) has
    \(|b|=2\) and \(\beta_i\le 2\).
\end{proof}

It is worth being precise about when
Proposition~\ref{prop:homogeneousdoubling} applies, because the hypothesis is
stronger than it may appear.  Definition~\ref{def:homogeneousquilt} asks for
identical local data \((H,S)\) on every block, and \(S\) records the local
stitching operators, which are fixed by the chosen logical representative.  It
is therefore not enough for the blocks to carry the same code: the
representatives must agree as well, up to a quilt isomorphism in the sense of
Definition~\ref{def:quiltisomorphism}.  This is the setting of
Theorem~\ref{theorem:homogeneousquiltdistance}, and in particular of the
automorphism-generated quilts of Section~\ref{sec:distance}, where the
representative on each block is obtained from one representative by a
permutation and single-qubit gates.

The representatives that do satisfy it, for each of the constituent codes,
are provided as a separate bank (Appendix~\ref{sec:homogeneousbank}).
Like Theorem~\ref{theorem:sparseinheritedbasis}, this concerns
\(\quilt H_{ac}\) only; \(\quilt H_c\) is not bounded for homogeneous quilts
either.

Both bounds in Eq.~\eqref{eq:skiptree-finite-bound} are consequences of the
same structural statement, namely that every generator is a product of a
bounded number of source checks.  Weight and congestion are therefore
analytically bounded on the same families.

\begin{remark}[Sufficiency, not necessity]\label{rem:sufficiency}
    Connectivity of \(\Lambda_i\) is sufficient but not necessary for a sparse
    inherited basis.  A disconnected \(\Lambda_i\) may still be bridged by
    higher-weight syndrome rows, or by products of source checks.
\end{remark}

\begin{remark}[Relaxing the stitching graph]\label{rem:relaxedstitchinggraph}
    If an edge of \(\Lambda_i\) is permitted to be a product of at most \(c\) source
    checks, the argument of Theorem~\ref{theorem:sparseinheritedbasis} gives only
    a bound of \(3c\) source checks per block, and correspondingly
    \(\wt(\mathcal G)\le 3c\max_b\sum_{i:b_i=1}w_i\).  Such a relaxed graph
    therefore cannot certify the three-source-check claim for \(c>1\).
\end{remark}

\subsection{Families with no analytical bound}\label{sec:nobound}

Neither the sparsity nor the congestion of \(\quilt H_c\) is bounded by any of
the arguments above, and the same therefore applies to
\(\quilt H_c+\quilt H_{ac}\), which contains it.  We appeal instead to the
empirical results of Section~\ref{sec:sparsity} and
Appendix~\ref{sec:sparsityempirical}, which show that at practical code sizes
both quantities are well behaved.

\begin{remark}[The basis used in our circuits]\label{rem:notdeployed}
    The basis of Theorem~\ref{theorem:sparseinheritedbasis} is deterministic and
    comes with a proof, but it is not the one used in our circuits.  Because
    Section~\ref{sec:nobound} leaves \(\quilt H_c+\quilt H_{ac}\)
    uncontrolled, we instead optimise a basis of the combined
    subspace numerically; the procedure, and the pairing conventions it depends
    on, are given in Appendix~\ref{sec:sparsityempirical}.  Note that a different
    pairing defines a different quilt gauge group even when the local
    representatives and the measured logical product are unchanged, so
    operational distance must always be assessed for the exact pairing used.
\end{remark}

\section{Mixed targets and the overlap condition}\label{sec:overlap}

This appendix concerns a condition that affects the suitability of logical representatives when constructing code quilts to measure non-CSS operators. It is the condition imposed as item (3) of the feasibility requirements of Appendix~\ref{sec:repsearch}.

A \textit{mixed} target is one whose logical class on some block is neither pure $X$ nor pure $Z$. On such a block $i$, choose pure representatives $x^i$ and $z^i$ of the $X$ and $Z$ components of the class, so that the local representative is their product
\begin{equation}
    \ell^i=x^i+z^i .
\end{equation}
Its support divides into three disjoint parts,
\begin{equation}
    A=\operatorname{supp}(x^i)\setminus\operatorname{supp}(z^i),
    \qquad
    O=\operatorname{supp}(x^i)\cap\operatorname{supp}(z^i),
    \qquad
    A'=\operatorname{supp}(z^i)\setminus\operatorname{supp}(x^i),
\end{equation}
on which $\ell^i$ acts as $X$, $Y$ and $Z$ respectively. We call
\begin{equation}
    o_i=|O|
\end{equation}
the \textit{overlap} on block $i$. Under the strict stitching primitive of Appendix~\ref{sec:repsearch} every qubit of $\operatorname{supp}(\ell^i)$ carries its own local stitching operator, so the operators on $A$ are pure $X$, those on $O$ are $Y$, and those on $A'$ are pure $Z$.

The overlap controls a failure mode that has no analogue for CSS targets. Compare the quilt measuring the single mixed product with the \textit{parallel} quilt that measures the two pure products $\Stitch(\mathbbm 1\otimes b)$ built from $x$ and from $z$ separately. The parallel operation resolves strictly more information: its two outcomes determine the mixed outcome, but not conversely. Consequently the pure product built from $x$ alone is a stabiliser of the parallel merged code, while for the mixed quilt it is an operator that commutes with every inherited check and with the measured product, yet is not a gauge operator. It is therefore an operational error of the mixed quilt in the sense of Definition~\ref{def:operationalquilt}, and it is invisible to the mixed measurement. We refer to the classes of such errors as the \textit{relative sector}.

Two remarks fix the scope of the comparison. First, the two pure products must be simultaneously measurable, which requires
\begin{equation}\label{eq:overlapparity}
    \sum_{i=1}^N x^i\,\Omega\,(z^i)^T=0 .
\end{equation}
Since $x^i\Omega(z^i)^T=o_i\bmod 2$, this is a condition on the parities of the block overlaps, and for an atomic quilt in which every local pair anticommutes it forces $N$ to be even. Second, when Eq.~\eqref{eq:overlapparity} holds the relative-sector error commutes with the measured product and is thus a memory error of the merged code; when it fails, the same operator anticommutes with the measured product and instead corrupts the reported outcome.

A relative-sector error is constrained blockwise. Write
\begin{equation}
    \delta_i=\min_{h\in H^i,\;v\in\ft^W}\bigl|x^i+h+\Stitch_i(v)\bigr|
\end{equation}
for the least weight to which the local component on block $i$ can be reduced by stabilisers of that block and by local stitching operators. Every relative-sector representative restricts on each active block to an element of $x^i+H^i+\operatorname{im}S^i$, so its weight there is at least $\delta_i$, and the least weight $d_\Delta$ of a relative-sector error satisfies
\begin{equation}\label{eq:relativelower}
    d_\Delta\ \ge\ \min_{0\ne b\in B}\ \sum_{i\,:\,b_i=1}\delta_i ,
\end{equation}
which for an atomic quilt is simply $\sum_{i=1}^N\delta_i$. The bound is not tight: a cross-block stitching product is specified by a matrix whose rows lie in $B$, so the selection of local stitching operators is shared between blocks while only the stabiliser dressing is free on each, and the minimisers achieving the several $\delta_i$ need not be realisable simultaneously.

The overlap enters as a ceiling on each $\delta_i$.

\begin{lemma}[Overlap bounds the local relative sector]\label{lemma:overlapbound}
    $\delta_i\le o_i$.
\end{lemma}

\begin{proof}
    The local stitching operators supported on $A$ are pure $X$, and $x^i$ acts as $X$ on all of $A$. Let $v$ select exactly those operators. Because the operators are disjoint, multiplying by $\Stitch_i(v)$ cancels the support of $x^i$ on $A$ and leaves its support on $O$ untouched. The result is the pure-$X$ operator $X(O)$, which lies in the required coset and has weight $o_i$.
\end{proof}

Combining Lemma~\ref{lemma:overlapbound} with Eq.~\eqref{eq:relativelower}, the bound can certify $d_\Delta\ge D$ only if
\begin{equation}\label{eq:overlapsum}
    \min_{0\ne b\in B}\ \sum_{i\,:\,b_i=1}o_i\ \ge\ D ,
\end{equation}
and for an atomic quilt only if $\sum_i o_i\ge D$. This condition is imposed during the search of Appendix~\ref{sec:repsearch}. It is cheap to evaluate, being a property of the chosen representatives alone, and it discards at no cost those states from which no useful certificate could be obtained.

Two cautions on how Eq.~\eqref{eq:overlapsum} should be read. It is necessary for the \textit{certificate} of Eq.~\eqref{eq:relativelower} to reach $D$, not for the operation to have distance $D$; a target failing it may still be fault-tolerant, established by other means. Conversely it is not sufficient, since dressing by stabilisers of both sectors together with the mixed local stitching operators may reduce a local component below $o_i$. In the experiments reported here Eq.~\eqref{eq:overlapsum} is therefore used only to steer and prune the representative search, and every surviving candidate is assessed by the operational-distance search described in Appendix~\ref{sec:repsearch}.

\section{A charge-rank pairing heuristic}\label{sec:chargepairing}

This appendix describes the heuristic mentioned in Section~\ref{sec:repfinding} for choosing the pairing of the local stitching operators between two codeblocks. The aim is to choose a pairing under which no proper subset of the stitching operators is closed on both blocks at once, so that a pathological logical operator is forced to `use all of the stitching operators' between the codeblocks.

We use only the standard Abelian-charge description of stabiliser codes~\cite{haah2013commuting,liang2023extracting}.  For block $i$, let $\mathcal A_i$ be its binary charge space and define $Q_i:\ft^W\to\mathcal A_i$ by $Q_i e_a=q_a^i$, where $q_a^i$ is the charge at stitching endpoint $a$.  If a permutation $\pi$ pairs the endpoints of two blocks, with $P_\pi e_a=e_{\pi(a)}$, then a subset $u$ is charge neutral on both blocks exactly when
\begin{equation}
    u\in\ker\begin{bmatrix}Q_1\\Q_2P_\pi\end{bmatrix}.
\end{equation}
Because each complete representative is closed, $\mathbbm 1_W$ lies in this kernel.  Hence the maximal-rank condition
\begin{equation}
    \operatorname{rank}\begin{bmatrix}Q_1\\Q_2P_\pi\end{bmatrix}=W-1
\end{equation}
excludes every proper nonempty subset that is neutral on both blocks.  

In the BB examples tested, high charge rank did repair non-fault-tolerant pairings into fault-tolerant ones, but simply trying a random permutation was equally effective.

\section{Representative search}\label{sec:repsearch}

The constructions of Section~\ref{sec:quilts} take the local stitching operators $S^i$ as given. To gather the data presented in Sections~\ref{sec:distance} and~\ref{sec:sparsity}, we must find local operators that satisfy a \textit{target}: a tuple of logical classes, one per codeblock, whose product is to be measured.

Section~\ref{sec:repfinding} states which targets were chosen and the desiderata each accepted representative must meet. This appendix gives the procedure that turns such a target into concrete stitching data: the feasibility conditions in the notation of Appendix~\ref{sec:structuretheorems}, the annealer that searches for them, and the checks applied before a candidate enters the corpus.

\subsection{Feasibility conditions}
We work throughout with the \textit{strict} stitching primitive, in which each local stitching operator acts on one qubit of its block and corresponding operators are paired across blocks. Writing $\ell^i$ for the representative chosen on block $i$, this forces the common \textit{interface width}
\begin{equation}
    \wt(\ell^1)=\cdots=\wt(\ell^N)=W .
\end{equation}
A search state is a tuple of representatives, one per block, and is called \textit{feasible} when
\begin{enumerate}
    \item all $N$ representatives have equal weight;
    \item on each block the local kernel is rank one, $\ker\partial_i=\langle\mathbbm 1_W\rangle$;
    \item for mixed (non-CSS) targets, the sites at which a representative's $X$ and $Z$ supports coincide number at least $D$ in total across the blocks, as required by Eq.~\eqref{eq:overlapsum} of Appendix~\ref{sec:overlap}; and
    \item the stitching graph of each block is connected, in the sense made precise below.
\end{enumerate}
The first condition is by far the most restrictive in practice. The second is the case $\dim Y=1$ of Definition~\ref{def:codequilt}; imposing it blockwise is convenient because it also discharges the requirement that the kernels $\ker\partial_i$ agree across blocks, which would otherwise have to be checked jointly. The third prevents a mixed representative from degenerating towards a CSS operator on either block, and is necessary for the relative-sector certificate of Appendix~\ref{sec:overlap} to establish distance $D$; it is imposed here to prune and to steer the search, not as an obstruction.

The fourth condition is imposed as a terminal condition of the search.  The code definition specifies a source-check presentation, which need not equal the row-reduced basis used for rank calculations.  Form the graph whose vertices are the $W$ stitching coordinates and whose edges are the weight-two syndromes $e_a+e_b$ occurring among individual rows of this presentation under $\partial_i^*$.  No product of source checks is allowed to supply an edge.  This graph must be connected.  As Appendix~\ref{sec:sparsityanalytical} explains, this condition and a deterministic SkipTree pairing certify local lifts containing at most three source checks.  It is sufficient, but not necessary, for a sparse basis of $H_{ac}$.

Note that a feasible state need not satisfy the hypotheses of Theorem~\ref{theorem:weightDquilt}, which additionally demands $W=D$; representatives with $W>D$ are admitted here.

\subsection{Coset annealing}
Multiplying a representative by a stabiliser of its own block changes neither its logical class nor the measured product, so the search may move freely within a product of stabiliser cosets. We do so by simulated annealing. A move selects one block---and, for a mixed target, one of its $X$ or $Z$ components---and multiplies it by a uniformly chosen stabiliser generator of that block. Any move producing a state that violates the rank-one condition is discarded immediately; the remainder are accepted by the Metropolis rule at a temperature falling linearly over the run. The objective penalises, in decreasing order of weight, the deficit in condition (3), the spread of the interface widths, and the total weight of the representatives, so that feasibility is secured first and low weight pursued only afterwards.  In the connected-output search, connectivity is an additional required terminal condition of this same annealer: intermediate states may remain disconnected, but a disconnected endpoint is not emitted as a corpus representative.

For each cohort we precompute a library of low-weight representatives of the logical classes of its constituent codes. Simply drawing from that library, without any annealing, suffices for only a minority of targets: it rarely offers two representatives of equal weight for a given pair of classes. Coset annealing reaches equal width in the remaining cases, and in our experiments it does so within a few tens of moves.
Several candidates are kept as output samples of the annealer, ranked based on their stitching weight $W$.

\subsection{Verification}
Each selected candidate is assembled into the full quilt gauge group $\quilt G$ of Definition~\ref{def:codequilt}. Equal width, the rank-one kernel, source-check connectivity, the three-source-check SkipTree bound, and the resulting logical accounting are recomputed as deterministic consistency checks; in this setting the kernel condition already supplies the no-overmeasurement condition. When constructing a canonical sparse pairing, the unpaired coordinates are first ordered by increasing physical qubit index and then reordered by the deterministic per-block SkipTree construction of Appendix~\ref{sec:sparsityanalytical}, with positions of equal new coordinate paired across blocks. Operational distance is assessed for the exact pairing used. Where Theorem~\ref{theorem:weightDquilt} applies it gives the analytic certificate; otherwise we search $\quilt C^\perp\setminus\quilt G$ for sub-$D$ witnesses using a Rust distance finder adapted to non-CSS subsystem codes, as detailed in Appendix~\ref{sec:tweedietest}. Finding such a witness rejects a candidate; failure to find one is reported as randomised evidence.

\section{The homogeneous representative bank}\label{sec:homogeneousbank}

A \emph{homogeneous} code quilt is one in which every codeblock carries the same
logical representative up to a quilt isomorphism in the sense of
Definition~\ref{def:quiltisomorphism}. These are the quilts to which
Theorem~\ref{theorem:homogeneousquiltdistance} and
Proposition~\ref{prop:homogeneousdoubling} apply. Accompanying this work is a
bank of representatives suitable for such quilts, drawn from the bivariate
bicycle, generalised bicycle, rotated surface and colour
codes~\cite{quantumquilts}.

Each entry is a $Z$-type logical representative of one constituent code whose
perforated operational distance satisfies $\dotd\ge d/2$, recorded with the
evidence for that bound. A major advantage of homogeneous quilts is that
certifying a distance half that of the target is computationally far less
demanding than certifying the target itself. Indeed, for the 72-qubit code,
certifying that the perforated distance is at least $3$ is brought within range
of exact SAT-solver-based proofs. Two kinds of evidence therefore appear, and an
entry may carry both: an exact certificate, which proves that no operator of
lower weight exists, and a count of randomised
\texttt{QDistRnd}~\cite{pryadko2022qdistrnd} trials that found none.

The bank holds one entry per automorphism orbit. If a representative of the
coset $a$ satisfies the bound and $L=\varphi(a)$ for a code automorphism
$\varphi$, then applying $\varphi$ to every block yields a distance-preserving
quilt measuring $L$ on each, so a single entry covers its whole orbit.
Table~\ref{tab:homogeneous-coverage} gives, for each code, the orbits of the
translation group acting on the $Z$-logical cosets, the number carrying an
accepted representative, and the number of cosets those orbits contain.

\begin{table}[htbp]
\centering
\begin{tabular}{@{}lrrrrr@{}}
\toprule
Code & $k$ & Orbits & Covered & \multicolumn{2}{c}{CSS-type cosets covered}\\
\cmidrule(l){5-6}
\midrule
72-qubit bivariate bicycle & 12 & 155 & \textcolor{nored}{98} & \textcolor{nored}{$5{,}388$} & of $8{,}190$\\
144-qubit bivariate bicycle & 12 & 155 & \textcolor{partorange}{147} & \textcolor{partorange}{$8{,}106$} & of $8{,}190$\\
288-qubit bivariate bicycle & 12 & 155 & \textcolor{yesgreen}{155} & \textcolor{yesgreen}{$8{,}190$} & of $8{,}190$\\
162-qubit symmetric bivariate bicycle & 8 & 35 & \textcolor{yesgreen}{35} & \textcolor{yesgreen}{$510$} & of $510$\\
\midrule
30-qubit generalised bicycle & 8 & 17 & \textcolor{yesgreen}{17} & \textcolor{yesgreen}{$510$} & of $510$\\
62-qubit generalised bicycle & 10 & 33 & \textcolor{yesgreen}{33} & \textcolor{yesgreen}{$2{,}046$} & of $2{,}046$\\
126-qubit generalised bicycle & 12 & 65 & \textcolor{yesgreen}{65} & \textcolor{yesgreen}{$8{,}190$} & of $8{,}190$\\
254-qubit generalised bicycle & 14 & 129 & \textcolor{yesgreen}{129} & \textcolor{yesgreen}{$32{,}766$} & of $32{,}766$\\
510-qubit generalised bicycle & 16 & 257 & \textcolor{yesgreen}{257} & \textcolor{yesgreen}{$131{,}070$} & of $131{,}070$\\
\midrule
distance-12 rotated surface & 1 & 1 & \textcolor{yesgreen}{1} & \textcolor{yesgreen}{$2$} & of $2$\\
distance-18 rotated surface & 1 & 1 & \textcolor{yesgreen}{1} & \textcolor{yesgreen}{$2$} & of $2$\\
distance-13 square colour & 1 & 1 & \textcolor{yesgreen}{1} & \textcolor{yesgreen}{$2$} & of $2$\\
distance-19 square colour & 1 & 1 & \textcolor{yesgreen}{1} & \textcolor{yesgreen}{$2$} & of $2$\\
\bottomrule
\end{tabular}
\caption{\textit{Coverage of the homogeneous bank.} Orbits are those of the
translation group acting on the $Z$-logical cosets of each code: the two torus
shifts for the bivariate bicycle codes, the cyclic shift for the generalised
bicycle codes. The symmetric 162-qubit code is the $[[162,8,12]]$ code
of~\cite{eberhardt2024logical}. ``Covered'' counts orbits carrying at least one accepted
representative, and the last column the number of CSS-type logical cosets those
orbits contain, each of which admits a homogeneous measurement; the $ZX$
duality doubles the $Z$-type count, as described below. The three
bivariate bicycle codes have the same orbit structure, $155$ orbits of sizes
$\{36^{100},12^{20},9^{25},3^{10}\}$, independently of the order of the
translation group. The surface and colour codes encode a single logical qubit. Colour marks complete coverage (green), near-complete (orange) and partial (red).}
\label{tab:homogeneous-coverage}
\end{table}

Every entry is $Z$-type. The $ZX$ duality of each code carries the bank onto the
$X$-type cosets; for the bivariate bicycle codes that duality swaps the two
qubit halves and negates the torus coordinates. Mixed targets are not reached,
since the translations and the duality both preserve Pauli type.

\section{Empirical sparsity and congestion}\label{sec:sparsityempirical}

Appendix~\ref{sec:sparsityanalytical} certifies a basis for the cross-block
part \(\quilt H_{ac}\) of the inherited centre alone, and
Section~\ref{sec:nobound} records that neither \(\quilt H_c\) nor the
combined subspace \(\quilt H_c+\quilt H_{ac}\) is controlled by that argument.
The subsystem protocol forms its detectors from the combined subspace, so the
sparsity data of that subspace must be established by measurement.  This
appendix describes the basis optimisation used to do so, and reports the
results over the full corpus of Section~\ref{sec:repfinding}.

A stabiliser subspace does not determine the sparsity data of
Definition~\ref{def:weightcongestion}: the same subspace admits many generating
sets, and weight and congestion are properties of the choice.  Weight is a row property of the chosen basis and
congestion a column property, and a basis of uniformly low weight can still
produce a poor detector graph if a few data qubits are reused by many rows.
Choosing the generating set is therefore an optimisation problem, and
the reported statistics are statistics of its solutions.

\subsection{The glued interface graph}

Throughout this appendix the quilts are atomic
(Definition~\ref{def:atomicnuclear}), as are all those in the corpus.
Each source check \(r\in\mathcal R_i\)
carries a \emph{stitch syndrome} \(\partial_i^* r\in Y^\perp\), recording which
interface coordinates it anticommutes with.  The interface coordinates are
shared by every block, so a check of stitch-syndrome weight two is an edge of a
single graph spanning the whole quilt, not of one block's graph.  Any cycle of
that glued graph is therefore a product of source checks with vanishing total
stitch syndrome, that is, an element of the inherited centre, and a
cross-block generator is produced directly rather than by pairing sections
chosen separately on each block.

The corpus of Section~\ref{sec:repfinding} enforces the connectivity
hypothesis of Definition~\ref{def:stitchinggraph} as a terminal condition
of the representative search, so every accepted quilt has a connected glued
interface graph.

\subsection{The candidate pool}

Source checks are partitioned by stitch-syndrome weight into three families,
each contributing candidates differently.

\begin{enumerate}
    \item \textit{Weight zero: free checks.} These lie in \(H^i_c=\ker\partial_i^*\)
    and are already central, so they are admitted to the pool directly.
    \item \textit{Weight two: edges.} These are the edges of the glued graph.  Their
    cycle space is generated from the fundamental cycles of a breadth-first
    tree rooted at every interface coordinate in turn, together with the
    two-edge cycles formed by parallel edges, that is, by pairs of checks with
    the same syndrome pair.  This is Horton's pool~\cite{horton1987cyclebasis}, which is guaranteed to
    contain a minimum cycle basis of the weight-two subgraph when the length
    of a cycle is counted in source checks.  We expect short cycles to be a
    good heuristic for low physical weight, although cancellation between
    checks means that the pool is not guaranteed to contain the basis of least
    physical weight.
    \item \textit{Weight four or more: hyperedges.} A single such check is not
    closed.  It is closed by pairing up its syndrome support along paths of the
    glued graph at least cost, a minimum \(T\)-join, and the resulting generator
    is the check together with that join.  We compute it as a recursive
    cheapest pairing over the terminals using precomputed all-roots
    breadth-first distances.
\end{enumerate}

Measured over the corpus, the fraction of
source checks carrying a stitch syndrome of weight four or more ranges from
\(0.0\%\) for \texttt{SC\_BB\_ZZ} at \(n=288\) to \(10.5\%\) for
\texttt{BB2\_GENERIC\_MATCHED} at \(n=72\).  The general non-CSS targets are
therefore the demanding case: for them a construction restricted to the cycle
space of the weight-two subgraph would discard a substantial part of the
available structure, whereas for the \(ZZ\)-type targets the weight-two
subgraph very often suffices---in the \texttt{BB2\_ZZ}, \texttt{SC\_BB\_ZZ} and \texttt{SC\_BB2\_ZZZ}
cohorts a majority of blocks carry no hyperedge at all.  The \(T\)-join family
lets the same construction serve both.

Each candidate is stored as the set of source checks used
together with their Pauli product, so that the source-check count
can be reported alongside the physical weight.

\subsection{Selection}

The number of generators required is fixed in advance by a rank computation,
\begin{equation}
    \texttt{required}=\sum_{i=1}^{N}\operatorname{rank}\left(H^i\right)-\operatorname{rank}\left(\left\{\partial_i^* r: r\in\mathcal R_i,\ 1\le i\le N\right\}\right);
    \label{eq:requiredrank}
\end{equation}
selection stops there, and the pool is verified to span.  Two
selections are run on the same pool.

\begin{enumerate}
\item \textit{Weight-first selection.}
Sort the candidates by physical Pauli weight and admit each in turn if it is
independent of those already taken.  Because independence of Pauli operators is
a matroid, this greedy is min--max optimal over the pool: among all bases
drawn from the pool, the one it returns minimises its own maximum generator
weight.  Candidates are ordered by weight, then by number of source checks used, then
by the row itself, so the output is deterministic.

\item \textit{Congestion-aware selection.}
Bucket the candidates identically by weight and source-check count, but order
each bucket, when it is reached, by the current qubit load of its candidates'
supports.  This breaks weight ties towards spreading the
basis across data qubits.  Because it only reorders within buckets of equal
weight, it cannot increase the maximum weight: the column statistics improve at
no cost in the row statistics.
\end{enumerate}

Both selections are computed for every quilt, so the sparsity--congestion
trade-off can be measured rather than asserted.  The figures
below report the weight-first selection, so that the quoted congestion is the
value obtained \emph{without} optimising for it.

\subsection{Reported statistics}

For each quilt we report the interface width \(W\), the block count and the
pairing strategy, so that the statistics can be conditioned on how the quilt
was made, together with
\begin{itemize}
    \item \emph{row statistics}: the maximum source-check count per generator,
    and the physical Pauli weight
    per generator---mean, median, ninetieth percentile and maximum---split
    by free against non-free generators, since the free checks are inherited
    unchanged and would otherwise flatter the distribution; and
    \item \emph{column statistics}: the qubit degree, that is, the number of
    chosen generators touching each data qubit.
\end{itemize}
The quantities plotted are the two maxima, \(\wt(\mathcal G)\) and
\(\gamma(\mathcal G)\) of Definition~\ref{def:weightcongestion}, since the detector graph is governed
by the worst generator and the worst qubit.  Note that these figures do not
report the measurement-consistency operators, whose weight is \(NW\).

\subsection{Results over the corpus}

\begin{figure}[p]
    \centering
    \includegraphics[width=0.8\linewidth]{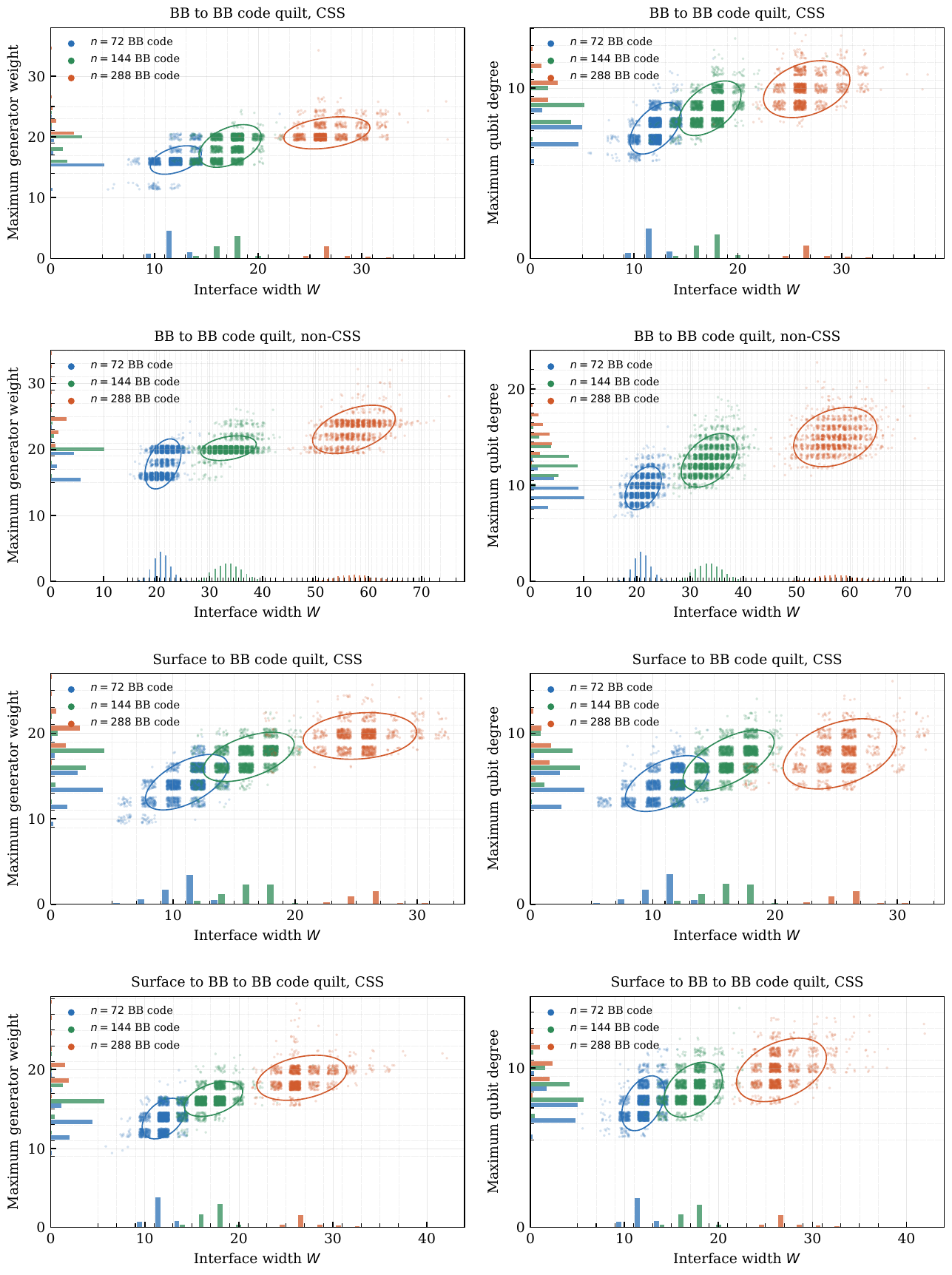}
    \caption{\textit{Sparsity and congestion of the optimised
    \(\quilt H_c+\quilt H_{ac}\) basis, over the full corpus.}  Left column,
    maximum generator weight; right column, maximum qubit degree; both against
    the interface width \(W\).  Rows are the four classes of target of
    Section~\ref{sec:repfinding}: \texttt{BB2\_ZZ}, \texttt{BB2\_GENERIC\_MATCHED},
    \texttt{SC\_BB\_ZZ} and \texttt{SC\_BB2\_ZZZ}.  Each point is one accepted
    representative, coloured by the size of the bivariate bicycle code involved,
    with a two-sigma covariance ellipse per size.  Population histograms are
    drawn inside each panel against both edges, with the code sizes interleaved
    rather than overlaid; their heights are a fixed fraction of the axis span and
    carry no units.  The last row is the only class with \(N=3\), so it is the
    one that exercises the \(N\)-block form of the glued construction.  The
    quantities plotted are those of the weight-first selection, so the congestion
    shown is the value obtained \emph{without} optimising for it.}
    \label{fig:sparsityallcohorts}
\end{figure}

The optimisation was run on every quilt of the corpus of
Section~\ref{sec:repfinding}: four classes of target, each at the three code
sizes, giving 25,000 quilts in total, all of which were measured.
Figure~\ref{fig:sparsitycorpus} of Section~\ref{sec:sparsity} shows the two
classes of bivariate-bicycle-to-bivariate-bicycle quilt;
Figure~\ref{fig:sparsityallcohorts} collects all four classes on both axes.

Each point is one accepted representative, plotted against its interface width
\(W\); the three code sizes are shown separately so that it can be read off
whether sparsity and congestion are a single stochastic function of \(W\), with
the width distribution depending on the code, or different functions of \(W\)
for each code size.  Two clusters can always be joined by a line, so only the
intermediate size can settle that question.

Two features are common to every class.  Both quantities grow slowly and
sublinearly in \(W\), so neither diverges over the range of interface widths
the corpus reaches.  The clusters for the three code sizes are offset along the
\(W\) axis but lie close to a common trend, indicating that the sparsity data
is set by the interface width rather than by the code size.  In the three CSS
classes the maximum qubit degree also stays close to twice the degree of the
underlying bivariate bicycle codes, never exceeding 14 against a code degree of
6.  The non-CSS class is the exception: its median maximum qubit degree
reaches 15 at \(d=18\), and its largest value over the corpus is 23.

\subsection{Comparison with the SkipTree basis}

The basis used in our circuits is the output of this optimisation, not the
deterministic SkipTree basis of
Theorem~\ref{theorem:sparseinheritedbasis}.  The two differ in scope and in
guarantee.  The SkipTree basis spans \(\quilt H_{ac}\) only and comes with a
proof; the optimised basis spans the larger subspace
\(\quilt H_c+\quilt H_{ac}\) that the protocol requires.  In our experience, optimising a basis of the combined subspace in this way also
produces sparser and less congested bases overall than using the deterministic
SkipTree basis for \(\quilt H_{ac}\) and optimising \(\quilt H_c\) separately, because the glued
construction is free to use cycles crossing several blocks where the
deterministic construction is committed to a per-block spanning tree.

Two caveats apply.  First, the optimisation ranges over a pool,
so its guarantees are relative to that pool: weight-first selection is optimal
over the Horton and \(T\)-join candidates, not over all bases of the subspace.
Second, a different pairing of the interface coordinates defines a different quilt
gauge group even when the local representatives and the measured logical
product are unchanged, so the operational distance of a quilt is only ever
asserted for the exact pairing used to produce it.

\section{Circuit-distance reduction and temporal protection}\label{sec:circuit-distance-methods}

\subsection{Summary of techniques}

In this appendix we explain the techniques used to estimate the circuit distance of the various circuits described in Section \ref{sec:distance}.
Rather than searching the detector error model generated from the circuit directly, we decompose the problem of estimating circuit distance into spacelike and timelike sectors, using a mapping technique this appendix makes precise. The essential flavour of the method is due to Shaw and Terhal~\cite{shaw2026optimising}, but we perform the analysis computationally rather than purely algebraically in order to avoid a complex case-by-case analysis of all kinds of failure mode in a logical measurement circuit with both spacelike and timelike observables, and with split and merged intervals that break the perfect symmetry the proof in~\cite{shaw2026optimising} relies on.

As described in Section \ref{sec:circuits}, we designed the syndrome extraction circuits to be alternating, in the sense of~\cite{shaw2026optimising}. As explained in detail below, we also ensure that two important observables of the circuit $\mu$ and $\nu$ are specified with supports that are separated by $d-1$ rounds during which neither observable has any support (for a preview, see Figure~\ref{fig:H}).
This means that every other round of measurement (including the `split' idle rounds before and after the merged portion of the circuit) is run in reverse. Since the syndrome extraction rounds are self-inverse, this does not affect the basic function of the circuit, but does affect error propagation in the following way: any spatial string of faults that passes through a boundary between rounds $t$ and $t+1$ can be `folded' onto just one of the rounds by projecting each fault in round $t+1$  onto its corresponding time-reversed location in round $t$.
Similarly, we can fold faults `forwards in time', from round $t$ onto faults in round $t+1$. Both directions are illustrated in Figure~\ref{fig:fold}.
If the detector structure is symmetric around the time boundary $t | t+1$, these folds map undetectable fault histories onto undetectable faults. They also map fault histories that flip a timelike observable (a `vertical sheet' in Figure~\ref{fig:H}) to fault histories that still flip a timelike observable, although they may map fault histories that flip only a spacelike observable (a `flat sheet' in Figure~\ref{fig:H}) to trivial errors. 

The circuit distance analysis then proceeds in three stages:
\begin{enumerate}
    \item We fold every round of the $d+2$ round circuit (2 idle, $d$ merged) onto a single prototype round, resulting in a much smaller detector error model which retains all the observables except for a single spacelike observable. We certify that every fault maps to at most one target fault in the smaller DEM.
    \item We run the distance finder of Appendix~\ref{sec:circuitfinder} to estimate the circuit distance of the smaller DEM. This gives an estimated lower bound on the weight of any fault history that flips a timelike observable in the full DEM.
    \item We use the structure of the circuit and a simple analytical argument to show that any undetectable fault history that flips only the remaining spacelike observable must have weight at least $d$, completing the analysis. 
\end{enumerate}

\subsection{Fault complexes and circuit distance}

The detector response of a Clifford circuit defines a binary linear map $D:F\to\mathcal D$ from fault configurations to detector outcomes. For a specified set of operational outputs, let $G_f$ be the subspace of silent faults with trivial action on those outputs. With $\iota$ the inclusion, the circuit has a fault complex
\begin{equation}
    G_f\xrightarrow{\iota}F\xrightarrow{D}\mathcal D,
    \qquad H_f=\ker D/\operatorname{im}\iota.
\end{equation}
Note that $F$ is not a space of Pauli strings; it is a vector space whose elements are `fault histories'---formal sums of basis elements corresponding to each elementary fault in the circuit. We assign unit cost to each allowed elementary fault. For example, $F$ contains basis elements corresponding to all 15 possible two-qubit faults that follow a CNOT. Dependencies among them, such as the equivalence $ZI + IX = ZX$, are encoded via $G_f$, as is the fact that an $X$ error on the control qubit immediately before the CNOT is equivalent to an $XX$ error immediately afterwards. The nonidentity homology classes in $H_f$ are the undetected operational errors. Circuit distance is the least physical fault cost of a representative of any such class. Dually, cocycles $\ker \iota^*$ correspond to linear functionals on fault histories whose value is unchanged by trivial faults in $G_f$, including the detectors $\operatorname{im} D^*$. Nontrivial cohomology classes in $H^f = \ker \iota^* / \operatorname{im} D^* $, which are dual to the error classes $H_f$ (note superscript vs subscript $f$) correspond to functionals that determine which class, if any, of operational error occurred during the circuit. These are also known as `correlators', `logical sheets' (particularly in simple topological codes), or `circuit observables' (in Stim).

In the usual description of a detector error model, we use the matrix $D$ verbatim to define detectors, and choose an explicit basis of representatives $L$ of the cohomology classes $H^f$. The advantage of abstracting to a three-term fault complex is that we can directly consider the fault classes $H_f$, which transform covariantly with the fault histories $F$. We thereby sidestep some basis-dependent subtleties that result from the fact that observables transform contravariantly with fault histories.

\subsection{Detector-error-model folding}
We now interpret a class of chain maps from one fault complex (the `source complex', $s$) to another (the `target complex', $t$ ).

\begin{equation}
\begin{tikzpicture}[baseline=(fs.base),x=3.1cm,y=1.25cm,>=Stealth]
    \node (gs) at (0,0) {$G_{f,s}$};
    \node (fs) at (1,0) {$F_s$};
    \node (ds) at (2,0) {$\mathcal D_s$};
    \node (gt) at (0,-1) {$G_{f,t}$};
    \node (ft) at (1,-1) {$F_t$};
    \node (dt) at (2,-1) {$\mathcal D_t$};
    \draw[->] (gs) -- node[above] {$\iota_s$} (fs);
    \draw[->] (fs) -- node[above] {$D_s$} (ds);
    \draw[->] (gt) -- node[below] {$\iota_t$} (ft);
    \draw[->] (ft) -- node[below] {$D_t$} (dt);
    \draw[->] (gs) -- node[left] {$P_G$} (gt);
    \draw[->] (fs) -- node[right] {$P$} (ft);
    \draw[->] (ds) -- node[right] {$Q$} (dt);
\end{tikzpicture}
\label{eq:fault-chain-map}
\end{equation}

In particular, we are interested in `weight non-increasing' chain maps whose middle homomorphism $P$ maps each elementary fault in $F_s$ to at most one elementary fault in $F_t$, so we have $|P(f)| \le |f|$.

The chain map descends to homology 
\begin{equation}
    P_*: H_{f,s} \rightarrow H_{f,t}.
\end{equation}
We say that the fault classes that are not mapped to the identity by $P_*$ `survive the fold'.
Dually, the observables (cohomology classes) in $H^{f,t}$ that detect these surviving fault classes map contravariantly via a pullback,
\begin{equation}
    P^*: H^{f,t}\rightarrow H^{f,s},\qquad \text{with} \qquad P^* h^{f,t} (f) = h^{f,t}\left ( P(f)\right ).
\end{equation}
This allows us to define observables that `survive the fold' as those cohomology classes that contain a pullback in the image of $P^*$. Observables of $s$ that do not contain any such image are `killed' by the fold.

The source and target fault complexes now have an important relationship: any undetectable, nontrivial fault $f$ in any fault class of $s$ that survives the fold must map to an undetectable, nontrivial fault $P(f)$ of the target complex with equal or lesser weight.
Equivalently, if $f$ flips the circuit observable $h^{f,s}$ of $s$ that detects some fault class that survives the fold, $P(f)$ flips some observable $h^{f,t}$ of $t$.
We therefore have
\begin{equation}
    d_{f,s}^{\mathrm{sur}}\ge d_{f,t},
    \label{eq:fold-distance-bound}
\end{equation}
where $d_{f,s}^{\mathrm{sur}}$ is the minimum weight of any error of $s$ that flips any of the surviving observables.
If we can obtain a confident lower bound on this quantity, empirically or otherwise, we reduce the problem of estimating the circuit distance of $s$ to that of lower-bounding the weight of undetectable faults that are trivial on the surviving observables, essentially converting those observables into detectors.

In detector-error-model coordinates, the certificates take the form
\begin{equation}
    D_tP=QD_s,\qquad
    L_tP=L_s^{\mathrm{sur}}+RD_s.
    \label{eq:fold-certificates}
\end{equation}
The second identity expresses preservation of the operational class on silent faults. It also ensures that $P$ maps harmless cycles to harmless cycles, defining $P_G$ in Eq.~\eqref{eq:fault-chain-map}. The target retains all detector constraints compatible with the fault map, so that $\ker D_t=P(\ker D_s)$. The supplied matrices and verification scripts specify the map for each circuit.

The fold we use retains the logical observables of the unmeasured logical qubits, as well as the observable that compares the eigenvalue of the measured logical operator at the output of the circuit with its eigenvalue at the input. Whether the recorded measurement outcome is consistent with these eigenvalues is not retained by the fold, and is treated separately below. A lower bound on the distance of the full circuit requires both parts, and applies only to the elementary faults included in $F_s$. The certificates provided with this paper~\cite{quantumquilts} list, for each circuit, the elementary faults considered and the observables that survive the fold.

\subsection{Macroscopic picture: logical sheets}

\begin{figure}
    \centering
    \includegraphics[width=\linewidth]{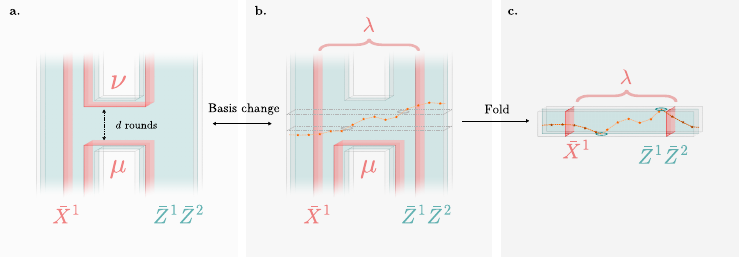}
    \caption{\textit{Surface code analogy for logical sheets under the fold map.}
    Shown is the standard spacetime depiction of a `merge and split' circuit for the surface code. Fault histories that terminate on the boundaries and therefore trigger no detectors correspond to cycles of the fault complex, and therefore each belong to a homology class of the fault complex, with nontrivial errors in a nonidentity class. Detectors and observables are both cocycles of the complex, with observables the nonidentity cohomology classes.
    \textbf{a.} and \textbf{b.} Two bases for the logical observables (nontrivial cohomology classes) of a surface code lattice surgery experiment. 
    We transform between the two via the relation $\lambda = \mu + \nu$. Both contain the ambient logical observables $\bar{X}^1$ and $\bar{Z}^1 \bar{Z}^2$.
    \textbf{c.} The target complex for the fold map. Fault locations in the alternating syndrome extraction circuit rounds are folded onto their corresponding faults in the prototype round in such a way that the image of fault strings in the target complex `reflect off' the boundaries of the target complex and remain continuous. This allows more detectors to be preserved under the fold map. The ambient logical observables and $\lambda$ survive the fold, whereas $\mu$ and $\nu$ are individually killed.}
    \label{fig:H}
\end{figure}

We now use the definitions of surviving and killed circuit observables to describe at a high level the properties of the map we implement.
We use a standard surface code `merge and split' circuit, in which observables can literally be drawn as 2D sheets in spacetime, for intuition.
For the example, illustrated in Figure~\ref{fig:H}, we consider a logical $\bar{X}^1 \bar{X}^2$ measurement between two surface code patches.

More generally, in a logical measurement operation between $N$ codes in which a single irreducible logical operator is measured, there are $2k = 2\sum_{i=1}^N k^i$ circuit observables that must be preserved for the operation to complete successfully:
\begin{enumerate}
    \item \textit{Ambient quantum logicals.} These are the $2k - 2$ logical operators that commute with the measured logical operator, modulo the measured operator itself. In the surface code example, these are generated by $\bar{X}^1$ and $\bar{Z}^1 \bar{Z}^2$.
    \item \textit{Target logical memory.} This observable detects whether the eigenvalue of the measured logical operator is consistent at the output of the circuit with its eigenvalue at the input of the circuit. In the example, this is the dual sheet $\bar{X}^1\bar{X}^2$. We refer to this observable as $\lambda$.
    \item \textit{Target logical measurement.} This observable detects whether the measured classical output bit of the circuit is consistent with the eigenvalue of the target logical operator at the input of the circuit. We refer to this value as $\mu$, or the `cap' observable. The name is justified by its shape as illustrated in Figure~\ref{fig:H}.
\end{enumerate}

We may also choose a different basis for the observables of the circuit by defining $\nu = \mu + \lambda$ (recall all three objects are cohomology classes), so that an alternative basis consists of the ambient logicals, $\mu$, and $\nu$. We refer to $\nu$ as the `cup'; it detects whether the quantum output of the circuit is consistent with the classical recorded outcome. 
Crucially, we can design our circuit to use the cap and cup observables, and we can choose representatives of them whose support is separated in time by the number of stitching rounds $R$, so that any connected error that flips both has weight at least $\min(d,R)$. The observables in the circuits provided as supplementary data with this paper are built on this basis~\cite{quantumquilts}. The two bases are illustrated in Figure~\ref{fig:H}.

Our fold map, detailed below, maps every round of syndrome extraction onto a single prototype stitching round.
This fold kills the observables $\mu$ and $\nu$ individually (the cap and cup), but their sum $\lambda$ survives the fold. (This does not violate linearity; the surviving observables $\operatorname{im} P^*$ are a linear subspace.) The ambient logical observables all survive the fold, giving a target complex with $2k - 1$ observables, that is reduced in volume by an order $d$.

\subsection{Microscopic picture: the alternating fold}
\begin{figure}
    \centering
    \includegraphics[width=\linewidth]{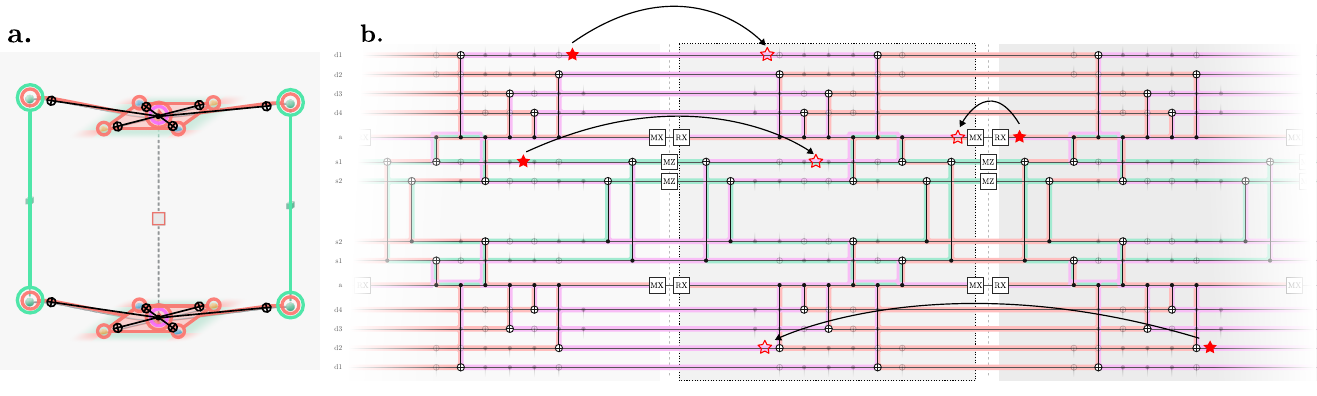}
    \caption{\textit{Folding an alternating stitching circuit onto a prototype round.}
    Depicted are three alternating rounds of syndrome extraction for one product stabiliser generator and two $ZZ$ stitching operators between two bivariate bicycle codes.
    \textbf{a.} A schematic showing the 14 qubits and 3 detector species involved in the circuit. There are two ancilla qubits, and 12 data qubits of which four support stitching operators. There are two species of $X$-type detectors corresponding to the product of two original-code stabilisers, and a single high-weight $Z$-type detector that corresponds to the measured observable.
    \textbf{b.} The full circuit. Faults in rounds before and after the prototype round (shown outlined) are mapped onto spacetime locations in the prototype round. Pure measurement readout faults are mapped to the identity fault.}
    \label{fig:fold}
\end{figure}

Following~\cite{shaw2026optimising}, we define $P: F_s \rightarrow F_t$ by mapping each gate error in rounds other than the prototype round to its corresponding (time-reversed, as the case may be) location in the prototype round.
The map is illustrated for the 14 qubits associated with a product stabiliser across 2 bivariate bicycle codes in Figure~\ref{fig:fold}. Note that both the gate structure and the detector structure are symmetric across each time boundary.  
Unlike in~\cite{shaw2026optimising}, pure measurement errors---where a wrong physical measurement outcome is recorded but the quantum state is unchanged---are mapped to the identity, so they do not exist in the smaller complex.
The idle split rounds at the start of the circuit contain only a subset of the gates of the merged circuit, as they do not contain the morphing CNOTs or stitching measurements that occur in the merged stitching rounds.
The map of faults from these rounds onto the prototype round is therefore not surjective.
The exact maps are available here \cite{quantumquilts}.
\subsection{Protection of the recorded outcome}

Recall from the previous subsection that $\lambda$ is the target logical memory
observable, which survives the fold, whereas the cap $\mu$ and the cup
$\nu=\lambda+\mu$ are individually killed by it. We use detector combinations to
represent $\mu$ and $\nu$ at opposite temporal boundaries; these representatives
are the rows
\begin{equation}
    c=\mu+aD_s,\qquad u=\nu+bD_s,
    \qquad c+u=\lambda+(a+b)D_s.
    \label{eq:cap-cup-identities}
\end{equation}
They evaluate the same operational flags on silent faults.

Choose a detector matrix $B$ with the same row space as $D_s$, and assign a temporal height to every row of $B$ and to the cap and cup. An elementary fault defines a hyperedge containing every detector and boundary row that it flips. Suppose each such hyperedge has height span at most one, and the cap and cup heights differ by $r$. These conditions concern complete physical faults, including correlated two-qubit errors.

\begin{theorem}[Temporal protection]\label{theorem:cap-cup}
    Under the preceding conditions, if every silent error flipping $\lambda$ has cost at least $d$, then every silent error flipping $\mu$ has cost at least $\min(d,r)$.
\end{theorem}
\begin{proof}
    Suppose a silent error flipping $\mu$ has cost less than $\min(d,r)$. Choose a minimum-cost representation by elementary faults and partition these faults into components connected through shared detectors in $B$. Each component is separately silent, since all selected faults incident on a detector belong to the same component. Also, each component has weight less than $d$, so its $\lambda$ value is zero.

    At least one component flips $c$. Eq.~\eqref{eq:cap-cup-identities} implies that this component also flips $u$. Its detector connectivity supplies a path of distinct fault hyperedges joining the two boundaries. Each hyperedge changes height by at most one, so the path contains at least $r$ elementary faults. This contradicts the assumed cost.
\end{proof}

Connectivity in this proof is defined through trusted detectors; the cap and cup serve as endpoints. The argument allows faults of arbitrary detector degree.

If the folded model has certified distance at least $d$, it supplies the required bound on $\lambda$ as well as on the other surviving outputs. Applying Theorem~\ref{theorem:cap-cup} to every omitted recorded-outcome direction gives
\begin{equation}
    d_{\mathrm{full}}\ge \min\{d,r_{\min}\},
    \label{eq:complete-circuit-distance}
\end{equation}
where $r_{\min}$ is the smallest certified cap--cup separation. With enough measurement rounds, the temporal bound reaches the folded-distance bound. A physical witness attaining that bound establishes equality.

\subsection{Certificates and empirical searches}

The circuits published with the present work do not themselves carry detector coordinates~\cite{quantumquilts}: the temporal height of each fault location is recorded in the fold description that accompanies each circuit, and is recomputed from the circuit when the temporal certificate is built. The circuits are defined so that the `cap' and `cup' observables $\mu$ and $\nu$ are supported on the opening and closing rounds of the stitching operation respectively, with $R$ rounds of separation. We choose $R$ to be at least the circuit-distance target of the corresponding code size, taking $R=6$, $12$ and $18$ for the quilts built from the 72-, 144- and 288-qubit bivariate bicycle codes, whose targets are $6$, $10$ and $18$. The separation is then verified case by case from the circuits and fold descriptions, which give a cap window of $[0,1]$, a cup window of $[R,R+2]$ and a height gap of exactly $R$ in every case, so that $\min(d,R)=d$ throughout.
In the structure of the circuit, each fault flips detectors belonging to at most two consecutive rounds. 
We also provide both the full and folded detector error models and explicit linear maps between them.
Together with the cost-nonincreasing fold, these checks establish the hypotheses under which Eq.~\eqref{eq:complete-circuit-distance} holds.
What remains is the search of the folded model itself. The finder we use for it, and how far its failure to find a fault can be trusted, are the subject of Appendix~\ref{sec:tweedietest}.

\section{Reliably estimating distances: the Tweedie test}\label{sec:tweedietest}
The problem with estimating the distance of a quantum code or circuit is that, whilst various ingenious routines exist to efficiently find low-weight logical operators and faults, proving that no lower-weight fault exists is, in general, an NP-hard problem~\cite{kapshikar2023hardness}.
How do we know we have run the distance finder long enough? And if we keep trying new codes and eventually find one for which we cannot find a sub-$d$ logical operator, is that more likely because the final code is distance preserving, or because the distance finder was bound to fail to find an operator eventually?
We cannot solve these problems completely.
However, the large volume of codes and circuits considered in this work allows us to perform numerical analysis to estimate the sensitivity of the distance finding algorithm, given only certain explicit qualitative assumptions.
What is more, for small codes we can test the validity of these assumptions exactly, suggesting they are sensible at larger distances.

\subsection{Statistically modelling distance estimation}\label{sec:distancemodel}

We provide a routine that takes in a set of codeblocks and a target logical coset $L = (L^1, \dots, L^N)$, with $L^i \in {G^i}^\perp / Z(G^i)$, and stochastically outputs a representative $l = (l^1,\dots,l^N)$, with $l^i \in \ft^{2n^i}$. In our case this is based on simulated annealing and described in Appendix~\ref{sec:repsearch}, but the details are irrelevant here.
Throughout this section, we assume that the configuration of codeblocks is fixed, but the target coset may vary, and will ultimately be sampled uniformly as a bit string over some subspace of $\ft^{\sum_i 2k^i}$.
From a representative of the target coset, we may deterministically construct a quilt (partially initialised) code $\quilt{C}$, which has some operational distance $\quilt{d} \le \min_i{d^i}$ (for simplicity, we assume all the codeblocks have the same distance $d$).

We also have access to a stochastic distance finding algorithm. This happens to be a modified version of \texttt{QDistRnd}~\cite{pryadko2022qdistrnd} optimised for non-CSS subsystem codes, which is once again irrelevant. It takes in a quilt, and attempts to find an operationally nontrivial logical operator of weight less than $d$.
If it succeeds in doing so, an event we refer to as a \textit{concession}, we know the quilt is no good, and need to find another representative or give up on that logical target.
If the distance finder fails, this in principle tells us very little.
Every trial of this stochastic test is independent.

Let the probability that a single trial of the finder on the quilt fails to find a low-weight logical be $e^{-\lambda}$, so that $\lambda \equiv \lambda(\quilt{C}) \ge 0$ is the rate at which the quilt concedes. Clearly, if $\quilt{d} = d$, then $\lambda = 0$. In general, it is subject to some distribution $\lambda \sim \tilde{G}(L)$ on $[0,\infty)$, where the distribution $\tilde{G}(L)$ is a function of the target coset (as well as the codeblocks, representative generation algorithm, and distance finding algorithm, which we hold fixed).

Define a run to be the process:
\begin{enumerate}
    \item Uniformly sample a coset $L$.
    \item Generate one representative $l$ of $L$.
    \item Run up to $T$ trials of the distance finder, stopping early if the quilt concedes a low-weight operator.
\end{enumerate}

If we perform this experiment with $S$ target cosets, a quilt survives $T$ trials with probability $e^{-\lambda T}$, and we can calculate the expected fraction of generated quilts $p_\text{surv}$ that have not conceded after $T$ trials as
\begin{align}
    \mathbbm{E}[p_\text{surv}(T)] &= \sum_L P(L) \int_{0}^{\infty}e^{-\lambda T}\,\tilde{G}(\lambda,L)\,d\lambda\\
    &= \int_{0}^{\infty}e^{-\lambda T}\,G(\lambda)\,d\lambda ,
\end{align}
where $G(\lambda) = \sum_L P(L)\,\tilde{G}(\lambda, L)$ is the distribution of rates over runs.
The distance-preserving quilts contribute an atom to $G$ at $\lambda = 0$ (an atom in the sense of measure theory: a single point that carries non-zero probability).
Writing $p_\text{good}$ for the weight of that atom, which is the fraction of sampled quilts that are distance preserving, and $G_{+}$ for the distribution of the non-zero rates,
\begin{equation}
    G(\lambda) = p_\text{good}\,\delta(\lambda) + (1 - p_\text{good})\,G_{+}(\lambda),
\end{equation}
so that
\begin{equation}
    \mathbbm{E}[p_\text{surv}(T)] = p_\text{good} + (1 - p_\text{good})\,\mathcal{L}(T),
    \qquad \mathcal{L}(T) = \int_0^\infty e^{-\lambda T}\, G_{+}(\lambda)\,d\lambda .
    \label{eq:plateau}
\end{equation}
This much is exact: the survival curve is a plateau plus the Laplace transform of $G_{+}$, the distribution of rates of the defective quilts.
The quantity we care about is the probability that a quilt which has survived $T$ trials is nevertheless defective, the false negative rate
\begin{equation}
    \mathrm{FNR}(T) = \frac{(1 - p_\text{good})\,\mathcal{L}(T)}{p_\text{good} + (1 - p_\text{good})\,\mathcal{L}(T)} .
    \label{eq:fnr}
\end{equation}

Our one modelling assumption is the shape of $G_{+}$. We take the Tweedie (exponentially tilted stable) family~\cite{jorgensen1997dispersion}, a standard choice for unobserved heterogeneity in survival analysis~\cite{hougaard1986survival}, defined by its Laplace transform
\begin{equation}
    \mathcal{L}(T) = \exp\!\left[-\frac{\delta\gamma^\theta}{\theta}\left(\left(1 + \frac{T}{\gamma}\right)^{\theta} - 1\right)\right],
    \qquad \delta, \gamma > 0,\quad 0 \le \theta < 1 .
    \label{eq:tweedie}
\end{equation}
The shape parameter $\theta$ interpolates between qualitatively different tails.
As $\theta \to 0$ the rates are gamma distributed and $\mathcal{L}(T) = (1 + T/\gamma)^{-\delta}$ decays as a power law: defective quilts with arbitrarily small rates exist, with density $\propto \lambda^{\delta - 1}$.
For any $\theta > 0$ the density of rates vanishes faster than any power as $\lambda \to 0$ ($\theta = 1/2$ is the inverse Gaussian), and $\mathcal{L}$ decays as a stretched exponential.
As $\theta \to 1$ every defective quilt has the same rate.
The pessimistic end of the family is therefore $\theta = 0$.

We fit $(p_\text{good}, \delta, \gamma, \theta)$ by maximum likelihood.
A quilt that concedes at trial $t$ contributes $(1 - p_\text{good})[\mathcal{L}(t-1) - \mathcal{L}(t)]$, and one that survives all $T$ trials contributes the right-hand side of Eq.~\eqref{eq:plateau}.
All intervals quoted are $95\%$ profile-likelihood intervals over all four parameters; they are conditional on the family~\eqref{eq:tweedie}, and on a single distribution describing every quilt in the cohort.
The Tweedie test then consists of deciding whether a distance finder, calibrated on a sufficient number of randomly sampled codes, is found to be described well by the distribution in such a way that we can state a tight bound on its rate of failure after a given number of trials.

\subsection{Calibrating the code-capacity distance finder}\label{sec:codecapcalibration}

Some of the quilts in the data provided with this paper were found before we settled on the method of analysis described in this appendix, under earlier search and screening procedures.
We therefore do not fit the model to them.
Instead, we calibrate the sensitivity of the finder only on freshly generated representatives, in the calibration experiments listed in Table~\ref{tab:codecap}.
In each experiment we take a set of logical targets, generate one representative for each (the first that the search produces, so that the sample is not selected by the finder), and run up to $T = 2\times 10^5$ trials on the resulting quilt.

\begin{figure}[!htbp]
\centering
\includegraphics[width=0.67\linewidth]{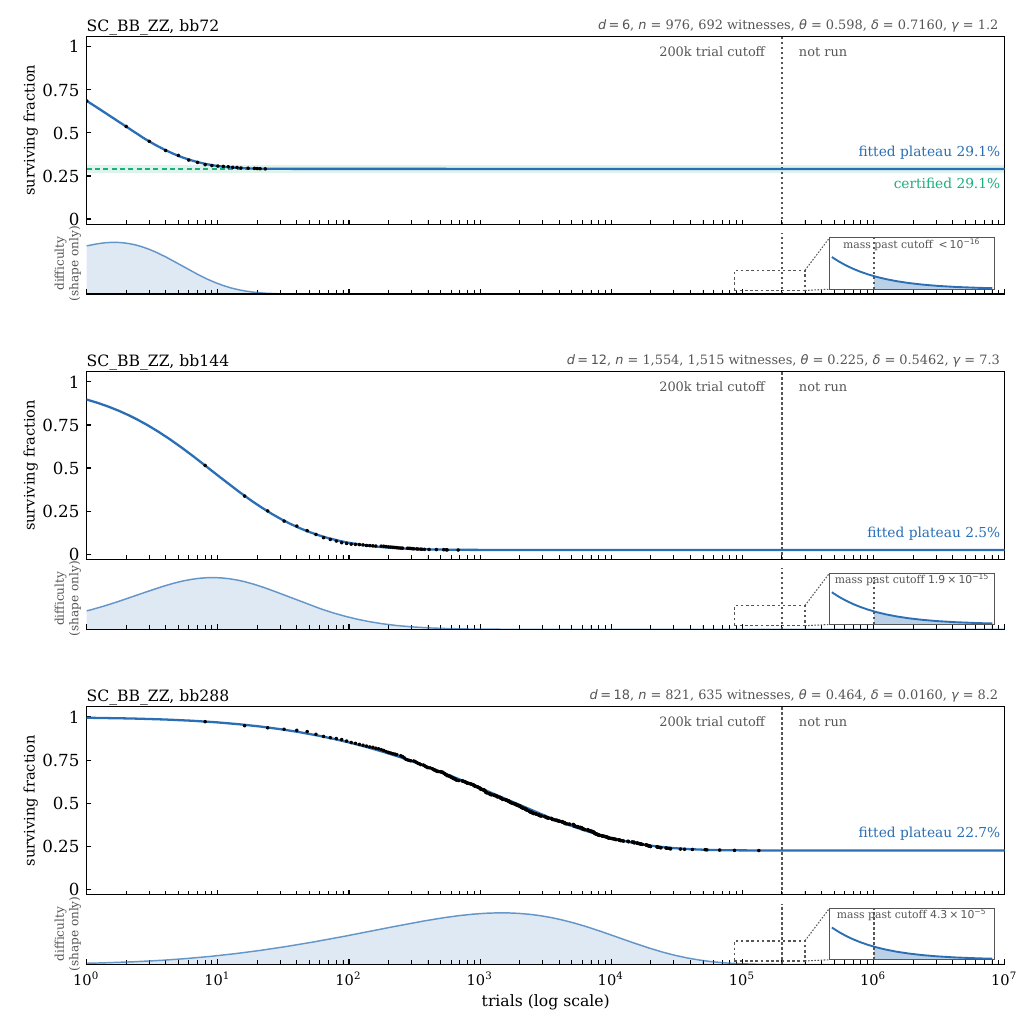}
\caption{\textit{Calibration of the code-capacity finder.} Fraction of freshly generated quilts that have not conceded, against the number of trials, for the \texttt{SC\_BB\_ZZ} class at $d = 6$, $12$ and $18$ (top to bottom; the panel titles give the fitted parameters). Points are the data, one quilt per logical target; the curve is the Tweedie fit of Eq.~\eqref{eq:plateau}, and the dotted line marks the budget of $2\times10^5$ trials. In the top panel the dashed green line is the exact fraction of distance-preserving quilts, from the exhaustive search. The strip under each panel shows the shape of the fitted rate of concessions per decade, a decade being a factor of ten in the number of trials. The inset gives the fitted fraction of all quilts that are defective and yet survive the whole budget. At the two larger sizes the finder checks for a concession once every eight trials, and the likelihood uses the corresponding interval.}
\label{fig:codecapcalibration}
\end{figure}

Figure~\ref{fig:codecapcalibration} shows three of these experiments, one at each code size, for the \texttt{SC\_BB\_ZZ} class, which has the lowest plateau of the four at every size.

At $d = 6$ the model can be checked against the truth. A meet-in-the-middle search over all operators of weight at most five decides exactly whether a quilt is distance preserving.
We drew 976 fresh quilts and ran both the finder and the exhaustive search on each.
The exhaustive search finds 692 of them defective; the finder conceded exactly those 692, all within 23 trials, with no false negatives and no false positives.
The fitted plateau, $29.1\%$, equals the certified fraction of distance-preserving quilts, $29.1\%$ (top panel of Figure~\ref{fig:codecapcalibration}).
This validates the likelihood and the plateau estimate. It does not test the extrapolation, since at this size nothing remains to extrapolate into.

Table~\ref{tab:codecap} gives the result of every calibration experiment.
At $d = 6$ every quilt is also decided exactly by the meet-in-the-middle search, so the fitted plateau can be compared with the truth.

\begin{table}[!htbp]
\centering
\footnotesize
\setlength{\tabcolsep}{3pt}
\begin{tabular}{@{}llccccccccc@{}}
\toprule
Codes & Target logical class
      & Quilts
      & Conceded
      & $p_\text{good}$
      & $\theta$
      & $\delta$
      & $\gamma$
      & Trials $T$
      & $\mathrm{FNR}(T)$
      & \begin{tabular}[b]{@{}c@{}}Certified\\ exactly\end{tabular} \\
\midrule
\midrule
\multirow{4}{*}{\shortstack[l]{$n=72$ BB code\\ $+\ d=6$ RS code}}
  & \texttt{SC\_BB\_ZZ}            & 976 & 692 & 29.1\% & 0.60 & 0.72 & 1.2 & $2\times10^5$ & $<2\times10^{-14}$ & \cYes \\
  & \texttt{BB2\_ZZ}               & 998 & 151 & 84.9\% & 1.00 & 0.27 & -- & $10^3$ & $<10^{-72}$ & \cYes \\
  & \texttt{BB2\_GENERIC\_MATCHED} & 1{,}000 & 33 & 96.7\% & 0.71 & 0.16 & 17 & $10^3$ & $<3\times10^{-7}$ & \cYes \\
  & \texttt{SC\_BB2\_ZZZ}          & 998 & 26 & 97.4\% & 0.88 & 0.13 & 0.49 & $10^3$ & $<5\times10^{-12}$ & \cYes \\
\midrule
\multirow{4}{*}{\shortstack[l]{$n=144$ BB code\\ $+\ d=12$ RS code}}
  & \texttt{SC\_BB\_ZZ}            & 1{,}554 & 1{,}515 & 2.5\% & 0.22 & 0.55 & 7.3 & $2\times10^5$ & $<1.1\times10^{-5}$ & \cNo \\
  & \texttt{BB2\_ZZ}               & 422 & 211 & 50.0\% & 0.49 & 0.077 & 23 & $2\times10^5$ & $<2.2\times10^{-6}$ & \cNo \\
  & \texttt{BB2\_GENERIC\_MATCHED} & 404 & 91 & 77.5\% & 0.22 & 0.36 & 330 & $2\times10^5$ & $<1.3\times10^{-5}$ & \cNo \\
  & \texttt{SC\_BB2\_ZZZ}          & 25 & 2 & \multicolumn{4}{c}{too few concessions to fit} & $2\times10^5$ & -- & \cNo \\
\midrule
\multirow{4}{*}{\shortstack[l]{$n=288$ BB code\\ $+\ d=18$ RS code}}
  & \texttt{SC\_BB\_ZZ}            & 821 & 635 & 22.7\% & 0.46 & 0.016 & 8.2 & $2\times10^5$ & $2\times10^{-4}$ ($<10^{-3}$) & \cNo \\
  & \texttt{BB2\_ZZ}               & 120 & 30 & 74.7\% & 0.11 & 0.30 & 1{,}600 & $2\times10^5$ & $4\times10^{-3}$ ($<0.04$) & \cNo \\
  & \texttt{BB2\_GENERIC\_MATCHED} & 156 & 16 & \multicolumn{4}{c}{too few concessions to fit} & $2\times10^5$ & -- & \cNo \\
  & \texttt{SC\_BB2\_ZZZ}          & 13 & 0 & \multicolumn{4}{c}{no concessions} & $2\times10^5$ & -- & \cNo \\
\bottomrule
\end{tabular}
\caption{\textit{Calibration of the code-capacity finder on freshly generated quilts.} Shown are the best-fit Tweedie parameters and the false negative rate at the budget of $T = 2\times10^5$ trials ($10^3$ for the last three classes at $n = 72$): the best estimate with its $95\%$ upper limit in brackets, or the upper limit alone where the best estimate is negligible. `Certified exactly' marks the experiments in which every quilt was also decided by exhaustive search. In those experiments the fitted $p_\text{good}$ equals the true fraction to the precision shown, and no surviving quilt was found to be defective. Three classes concede too rarely to be fitted, and are listed without parameters.}
\label{tab:codecap}
\end{table}

At $d = 12$ fewer than one surviving quilt in $8\times10^{4}$ is defective in any class, even at the upper end of the interval.
At $d = 18$ the search is harder: the false negative rate is $2\times10^{-4}$ for \texttt{SC\_BB\_ZZ} (below $10^{-3}$) and $4\times10^{-3}$ for \texttt{BB2\_ZZ} (below $4\times10^{-2}$), the latter from only 30 concessions.
In the two classes with the most concessions, $\theta$ is well determined ($0.22$ at $d = 12$ and $0.46$ at $d = 18$) and the power-law member $\theta = 0$ is excluded.

\subsection{The circuit distance finder}\label{sec:circuitfinder}

\begin{figure}[!htbp]
\centering
\includegraphics[width=0.45\linewidth]{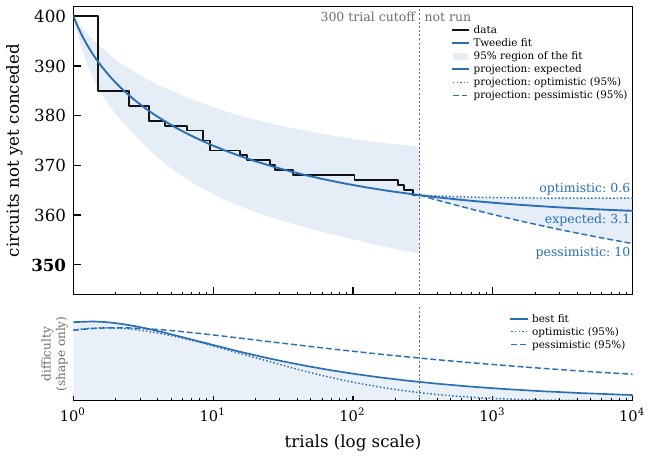}
\caption{\textit{Calibration of the circuit distance finder at $n = 144$.} Circuits not yet conceded, against the number of trials, for the three pooled classes. Note that the vertical axis starts at 350, not zero. The data are in black and everything fitted is in blue. The solid curve is the Tweedie fit, with its $95\%$ region shaded. No circuit was run past trial 300 (vertical dashed line); from there the lines project how many of the 364 surviving circuits would remain under the best fit (solid) and under the most optimistic (dotted) and most pessimistic (dashed) parameters within the $95\%$ region, labelled with the number of further concessions each projects by trial $10^4$. The strip underneath shows, for the same three sets of parameters, the shape of the fitted rate of concessions per decade, as in Figure~\ref{fig:codecapcalibration}; the three curves share one vertical scale.}
\label{fig:findercalibration}
\end{figure}

We developed a novel circuit distance finding tool for this task, whose code is given in \cite{quantumquilts}.
Each circuit is first folded onto a single prototype round (Appendix~\ref{sec:circuit-distance-methods}), which reduces the size of the problem by a factor $O(d)$ and gives a detector matrix $D$ and an observable matrix $L$ over a set of unit-weight faults.
A \textit{witness} is a set of faults $x$ with $Dx = 0$, $Lx \ne 0$ and weight below the target circuit distance.

Our finder searches level by level. A state is a set of faults, identified by its residual syndrome and logical label; level $w$ holds states of exactly $w$ faults.
Each state selects one of its violated detectors and branches over the faults that touch it. Any completion of the state must use one of those faults, so branching loses nothing.
States are ranked by the number of violated detectors and only the best are kept at each level, which is the one step that can lose a witness.
The choice of detector to branch on is randomised, so that repeated searches of one circuit are close to independent: one search is a trial in the sense of Section~\ref{sec:distancemodel}, and a circuit that yields a witness has conceded.
We found this finder far more effective per unit of compute time than BP-OSD~\cite{roffe2020decoding}, \texttt{QDistRnd}~\cite{pryadko2022qdistrnd} and \texttt{qdistevol}~\cite{webster2024qdistevol}: it finds weight-10 witnesses for the syndrome extraction circuits of the $[[144,12,12]]$ code within a minute on a few cores, where the others took hours or found none at all.

\subsection{Calibrating the circuit distance finder}\label{sec:circuitcalibration}
We use the same statistical model as in the code-capacity case.
However, at the circuit level, running trials is far more computationally demanding, so we make the less-than-ideal assumption that the shape of the difficulty curve $G_+(\lambda)$ is common across the classes at each distance, and that only the fraction of distance-preserving circuits $p_\text{good}$ varies between classes. Table~\ref{tab:findercalibration} gives the results.

At $d = 6$, an exhaustive search using meet-in-the-middle can rule out every witness of weight at most five, so we can verify exactly whether a circuit is distance preserving.
Of 200 circuits, 50 in each class, the finder conceded 80 (34 in \texttt{SC\_BB\_ZZ}, 25 in \texttt{BB2\_ZZ}, 21 in \texttt{BB2\_GENERIC\_MATCHED} and none in \texttt{SC\_BB2\_ZZZ}), each at its first trial, and the exhaustive search certifies that the other 120 have no witness.
The finder made no error, and there is no distribution of rates to fit.

\begin{table}[!htbp]
\centering
\footnotesize
\setlength{\tabcolsep}{3pt}
\begin{tabular}{@{}ll@{\hspace{2pt}}lccccccccc@{}}
\toprule
Codes & \multicolumn{2}{l}{Target logical class}
      & Circuits
      & Conceded
      & $p_\text{good}$
      & $\theta$
      & $\delta$
      & $\gamma$
      & Trials $T$
      & $\mathrm{FNR}(T)$
      & \begin{tabular}[b]{@{}c@{}}Certified\\ exactly\end{tabular} \\
\midrule
\midrule
\multirow{4}{*}{\shortstack[l]{$n=144$ BB code\\ $+\ d=12$ RS code}}
  & \texttt{SC\_BB\_ZZ}            & \multirow{3}{*}{$\left.\vphantom{\begin{tabular}{@{}l@{}}a\\a\\a\end{tabular}}\right\}$\,pooled} & 150 & 5 & \multirow{3}{*}{89.9\%} & \multirow{3}{*}{0} & \multirow{3}{*}{0.34} & \multirow{3}{*}{0.44} & \multirow{3}{*}{300} & \multirow{3}{*}{\shortstack[c]{$1.3\%$\\ ($<7.8\%$)}} & \multirow{3}{*}{\cNo} \\
  & \texttt{BB2\_ZZ}               & & 50 & 15 & & & & & & & \\
  & \texttt{BB2\_GENERIC\_MATCHED} & & 200 & 16 & & & & & & & \\
  & \texttt{SC\_BB2\_ZZZ}          & & 400 & 0 & \multicolumn{4}{c}{no concessions} & 300 & -- & \cNo \\
\bottomrule
\end{tabular}
\caption{\textit{Calibration of the circuit distance finder on the $n = 144$ code, in the format of Table~\ref{tab:codecap}.} The three classes joined by a brace are analysed together and share one row of results. The false negative rate is at $T = 300$ trials, with its $95\%$ upper limit in brackets.}
\label{tab:findercalibration}
\end{table}

At $d = 12$, where the target circuit distance is 10, no exhaustive search is available.
Of 800 circuits, each run for up to 300 trials, 36 conceded, four of them after more than 100 trials.
First concessions alone cannot separate a few slowly conceding circuits from many that almost never concede, so we ran each circuit that conceded again, without stopping at concessions, to measure its rate directly.
A circuit that first conceded at trial $t$, and then conceded in $k$ of $m$ further trials, contributes
\begin{equation}
    (1 - p_\text{good}) \int_0^\infty G_{+}(\lambda)\, e^{-\lambda (t - 1 + m - k)} \left(1 - e^{-\lambda}\right)^{k+1} d\lambda
\end{equation}
to the likelihood, with $G_{+}$ given by Eq.~\eqref{eq:tweedie}.

Pooling the three classes that conceded (Figure~\ref{fig:findercalibration}), the fit gives $p_\text{good} = 89.9\%$ ($83.3$--$93.0\%$), $\theta = 0$ ($0$--$0.26$) and
\begin{equation}
    \mathrm{FNR}(300) = 1.3\% \quad (0.15\text{--}7.8\%).
\end{equation}
That is, a circuit that survives 300 trials is distance preserving with probability $98.7\%$, and with probability at least $92\%$ at $95\%$ confidence.
The fourth class, \texttt{SC\_BB2\_ZZZ}, never conceded and so cannot be calibrated.
Each of the four $d = 12$ circuits provided with this paper survived 300 trials.

At $d = 18$ a trial is too expensive to assemble a calibration population, so we cannot fit the model or quote a false negative rate.
We ran the finder only on the circuits considered for this paper, for up to 300 trials each.
In \texttt{BB2\_ZZ} and \texttt{SC\_BB2\_ZZZ}, the one circuit searched showed no witness. In \texttt{BB2\_GENERIC\_MATCHED}, the first circuit conceded and a second, built from a redrawn representative, showed no witness.
In \texttt{SC\_BB\_ZZ}, six of the seven candidate circuits for the logical target conceded a witness of weight 17 or, in one case, 16, although all seven had passed the code-capacity screen.
None of the six was a marginal case: each conceded at its very first trial, and every one of the 47 trials we ran on them returned a witness.
The seventh showed no witness in 300 trials.
As a control, we also searched the idle memory circuit of a single $n = 288$ block with the same syndrome extraction schedule: 300 trials found no witness of weight below 18, while a search allowed to reach weight 18 finds one readily.
The finder is therefore demonstrably sensitive to sub-distance faults at this size, which lends some confidence to the circuits that survive, but we cannot put a number on that confidence as we can at the smaller sizes.

\section{Circuit decoding experiment details}\label{sec:simulationdetails}

This appendix provides additional details about the circuit-level decoding experiments whose results are reported in Section~\ref{sec:circuitsims}. The circuits themselves, the decoder configuration, and the shot and failure counts behind every point plotted are provided alongside this work~\cite{quantumquilts}.

\subsection{Circuit and observable structure}

The circuits each consist of the following stages:
\begin{enumerate}
    \item Noiseless Pauli product measurements of the stabilisers of each codeblock that establish the codespace: these simulate the interface with a past idle memory interval. 
    \item Noiseless Pauli product measurements of the joint operators $\bar X_j X^{(r_j)}$ and $\bar Z_j Z^{(r_j)}$, one reference qubit $r_j$ per logical qubit, and of the measured logical product against its own reference qubit. Each such pair of measurements projects the logical qubit and its reference into a logical Bell state, which allows the $X$-type and $Z$-type observables of the circuit to be simulated and decoded simultaneously. The reference qubits take part in no gate and carry no noise at any point.
    \item A single noisy syndrome extraction round of the idle memory circuit of each codeblock.
    \item $r$ noisy rounds of alternating code quilt syndrome extraction rounds (or `stitching rounds') according to the description of the subsystem protocol given in Section \ref{sec:circuits}.
    \item A final noisy `split' round where the idle memory circuits of each block are performed.
    \item The same noiseless Pauli product measurements are repeated, first for the protected logical Bell operators and then for the measured logical products against their reference qubits. Comparing each against its counterpart at the opening boundary gives the observables of the circuit.
    \item Each reference qubit of a measured logical product is then measured on its own, noiselessly. The recorded classical outcome of the measurement is thereby available as a circuit observable in its own right.
    \item A closing round of noiseless Pauli product measurements of the stabilisers of the split codes, simulating the interface with future idle memory rounds.
\end{enumerate}

As described in Appendix~\ref{sec:circuit-distance-methods}, each circuit has $2k = 2\sum_{i=1}^N k^i$ observables, of two kinds. For each of the $k-|\mathcal M|$ logical qubits that are not measured, a pair of observables compares the $X$-type and the $Z$-type half of its Bell pair between the opening and closing boundaries; these are the ambient logical observables, and there are $2(k-|\mathcal M|)$ of them. For each measured logical product there are two further observables, the `cap' $\mu$ and the `cup' $\nu$ of Appendix~\ref{sec:circuit-distance-methods}: the cap compares the recorded classical outcome against the eigenvalue at the input, and the cup compares it against the eigenvalue at the output. Their sum $\lambda=\mu+\nu$ is the memory direction of the measured logical, so an error flipping both leaves the quantum information intact and is not a logical failure of the memory.

A decoding failure is counted whenever the decoder fails to return a solution (e.g. because Relay-BP fails to converge after the maximum allowed attempts) or whenever the solution returned differs in any of the $2k$ bits from the correct solution, which is provided by the Stim sampler.
In our simulations, we track counts of which observables were incorrectly decoded in each failure. In particular, we track the number of occurrences where the cap observable $\mu$ was decoded incorrectly (any number of the other observables may also have been decoded incorrectly).
As described in Appendix \ref{sec:circuit-distance-methods}, this corresponds to a classical measurement error: the recorded outcome is inconsistent with the eigenvalue of the measured logical operator at the input of the circuit. The breakdown into classical measurement errors and all other errors is shown in Table \ref{tab:readout-share}.

\subsection{Noise model}

The noisy parts of the circuit are simulated with standard uniform circuit-level depolarising noise, in which each of the following channels acts with probability $p$. Two-qubit depolarising noise is applied to both qubits after every CNOT; single-qubit depolarising noise is applied at the remaining gate locations; and a Pauli error in the complementary basis is applied after every reset and before every measurement. In some of the simulations, qubits sitting idle also experience single-qubit depolarising noise with probability $p$. The boundary Pauli product measurements of the previous subsection, and the reference qubits, carry no noise.

\subsection{The `kitchen-sink' decoder}

Decoding uses Relay-BP~\cite{relaybp2025} with $\alpha=0.8$, $\gamma_0=0.15$, disordered memory strengths drawn from $[-0.05,0.66]$, $80$ pre-iterations, and up to $60$ relay sets of $60$ iterations each.

We decode each shot against a sequence of equivalent presentations of the same decoding problem, related by invertible transformations of the detectors, with the syndrome transformed to match; the freedom exploited is that of the basis optimisation of Appendix~\ref{sec:sparsityempirical}, and each presentation is additionally passed through a GARI-style~\cite{maan2026gari} transformation.
We refer to the scripts in the accompanying data \cite{wolanski2026quiltsdata} for details. A prediction is accepted once two presentations converge to solutions that agree in logical class. Over the data reported here, $99.1\%$ of shots were resolved by the first two presentations and the mean number of attempts per shot was $2.03$, although the deepest shot required $90$.

Two residual outcomes are counted as failures rather than discarded. A shot is \textit{uncorroborated} when the sequence of presentations is exhausted with at least one converged solution but no two in agreement, in which case the first converged prediction is scored; and \textit{exhausted} when no presentation converges at all. Across $10.6$ million shots these account for $159$ and $229$ shots respectively.

We remark that this decoder is very slow. It takes multiple seconds per presentation, per shot, and some shots must be decoded using several presentations before corroborated convergence. Efficient decoding of non-CSS logical operations remains an open problem.

\vspace{0.5cm}
\href{https://youtu.be/kK-fF0d-Slw?si=_Cf6XDmgowY2M8tY&t=179}{Cheers!}
\section{Symbol glossary}\label{sec:symbolglossary}

\footnotesize
\rowcolors{2}{gray!10}{white}
\begin{longtable}{@{}p{0.18\linewidth}p{0.24\linewidth}p{0.38\linewidth}p{0.12\linewidth}@{}}
\textbf{Symbol} & \textbf{Type} & \textbf{Meaning} & \textbf{First use}\\
\hline
\(\ft\) & field & The binary field \(\mathbb F_2\). & Preamble\\
\(\Pauli_n\) & vector space & Binary phaseless Pauli space on \(n\) qubits. & Def.~\ref{def:codequilt}\\
\(\Omega\) & bilinear form & The binary symplectic form used to compute Pauli commutation. & Def.~\ref{def:codequilt}\\
\(E^*\) & vector space & \(\operatorname{Hom}_{\ft}(E,\ft)\). & App.~\ref{sec:structuretheorems}\\
\(X^\perp\) & subspace & Annihilator of \(X\) in the dual/opposite space. & App.~\ref{sec:structuretheorems}\\
\((G,C)\) & PIC & Gauge subspace \(G\) together with its trusted central subspace \(C\le Z(G)\). & App.~\ref{sec:operationaldistance}\\
\(d_{\mathrm q},d_{\mathrm c},d_{\mathrm{op}}\) & integers & Quantum, classical, and operational distances of a PIC. & App.~\ref{sec:operationaldistance}\\
\(N\) & integer & Number of codeblocks in a quilt. & Def.~\ref{def:codequilt}\\
\(\mathcal C^i\) & stabiliser code & The \(i\)th constituent codeblock. & Def.~\ref{def:codequilt}\\
\(n^i,k^i,d^i\) & integers & Parameters of \(\mathcal C^i\). & Def.~\ref{def:codequilt}\\
\(H^i\) & subspace of \(\Pauli_{n^i}\) & Stabiliser subspace of \(\mathcal C^i\). & Def.~\ref{def:codequilt}\\
\(W\) & integer & Number of local stitching operators on each block. & Def.~\ref{def:codequilt}\\
\(S^i\) & matrix/subspace data & Rows are the local stitching operators on block \(i\). & Def.~\ref{def:codequilt}\\
\(B\) & subspace of \(\ft^N\) & Incidence code specifying which codeblocks each stitching product touches. & Def.~\ref{def:codequilt}\\
\(d_B\) & integer & Minimum Hamming distance of \(B\). & Def.~\ref{def:codequilt}\\
\(p_\uparrow^{(i)}\) & Pauli on full quilt & Operator \(p\) placed on block \(i\), with zero support elsewhere. & Def.~\ref{def:codequilt}\\
\(\Stitch_i\) & linear map & Local realisation map \(x\mapsto xS^i\). & Def.~\ref{def:codequilt}\\
\(\Stitch(T)\) & Pauli product & Cross-block stitching product specified by \(T\in\ft^{W\times N}\). & Def.~\ref{def:codequilt}\\
\(\partial_i\) & linear map & Anticommutation map from local stitching selections to functionals on \(H^i\). & Def.~\ref{def:codequilt}\\
\(Y\) & subspace of \(\ft^W\) & Common kernel of the local anticommutation maps \(\partial_i\). & Def.~\ref{def:codequilt}\\
\(\quilt{\mathcal C}\) & PIC & The quilt PIC, whose underlying gauge group is \(\quilt G\). & Defs.~\ref{def:codequilt} and~\ref{def:operationalquilt}\\
\(\quilt G\) & gauge subspace & Gauge group/subspace of the quilt. & Def.~\ref{def:codequilt}\\
\(G_0\) & stabiliser subspace & Direct sum of the unstitched codeblock stabilisers. & Def.~\ref{def:operationalquilt}\\
\(\quilt C\) & stabiliser subspace & Inherited/trusted centre \(G_0\cap\quilt H\) of the quilt PIC. & Def.~\ref{def:operationalquilt}\\
\(\quilt d_{\mathrm{op}}\) & integer & Operational distance of \((\quilt G,\quilt C)\). & Def.~\ref{def:operationalquilt}\\
\(U^i,M_{U^i}\) & Clifford/symplectic matrix & Local Clifford and its induced binary symplectic action. & Def.~\ref{def:quiltisomorphism}\\
\(\partial_i^*\) & linear map & Dual anticommutation map from \(H^i\) to \((\ft^W)^*\). & Def.~\ref{def:stab-stitch-realisation}\\
\(\sigma_i\) & section & Chosen section \(Y^\perp\to H^i\) of \(\partial_i^*\). & Def.~\ref{def:stab-stitch-realisation}\\
\(\Stab_i,\Stab(T^*)\) & linear map/Pauli product & Local and global stabiliser realisation maps. & Def.~\ref{def:stab-stitch-realisation}\\
\(H^i_c\) & subspace of \(H^i\) & Original stabilisers commuting with all local stitching operators. & Lem.~\ref{lemma:quiltgaugestructure}\\
\(\quilt H_c\) & stabiliser subspace & Lifted commuting original stabilisers. & Lem.~\ref{lemma:quiltgaugestructure}\\
\(G_{ac}\) & gauge subspace & Chosen anticommuting original-stabiliser representatives. & Lem.~\ref{lemma:quiltgaugestructure}\\
\(G_s\) & gauge subspace & Stitching part of the gauge group. & Lem.~\ref{lemma:quiltgaugestructure}\\
\(\quilt H\) & stabiliser subspace & Centre of the quilt gauge group. & Lem.~\ref{lemma:quiltstabiliserstructure}\\
\(\quilt H_{ac}\) & stabiliser subspace & Central products of anticommuting stabiliser representatives. & Lem.~\ref{lemma:quiltstabiliserstructure}\\
\(\quilt H_m\) & stabiliser subspace & Measured closed stitching products. & Lem.~\ref{lemma:quiltstabiliserstructure}\\
\(\mathcal L_i\) & quotient space & Logical quotient \((H^i)^\perp/H^i\). & Def.~\ref{def:measuredlogicalproducts}\\
\(\lambda_i\) & linear map & Local logical image of closed stitching products. & Def.~\ref{def:measuredlogicalproducts}\\
\(\lambda_B\) & linear map & Logical image map for closed stitching products in \(Y\otimes B\). & Def.~\ref{def:measuredlogicalproducts}\\
\(\mathcal M_B\) & subspace & Logical products measured by the quilt. & Def.~\ref{def:measuredlogicalproducts}\\
\(B|_I\) & subspace of \(\ft^I\) & Coordinate restriction \(\{b|_I:b\in B\}\). & Def.~\ref{def:subquilt}\\
\(\quilt{\mathcal C}|_I\) & subsystem code & Subquilt induced by the block subset \(I\). & Def.~\ref{def:subquilt}\\
\(I_q,B_q\) & set/code & Codeblock subset and restricted incidence code of generator subquilt \(q\). & Def.~\ref{def:generatingsubquilts}\\
\(\widetilde B_q\) & subspace of \(\ft^N\) & Lifted consistency code from generator subquilt \(q\). & Def.~\ref{def:generatingsubquilts}\\
\(\mathcal M_q,\widetilde{\mathcal M}_q\) & subspaces & Measured logical subspace of \(q\), and its lift to all blocks. & Def.~\ref{def:generatingsubquilts}\\
\(Z\) & subspace of \(\ft^W\) & Complement/section for \(\ft^W/Y\). & Def.~\ref{def:stitchsyndrome}\\
\(\Gamma\) & linear map \(\quilt H_c^\perp\to Z\otimes\ft^N\) & Stitch-syndrome map. & Def.~\ref{def:stitchsyndrome}\\
\(\overline E\) & Pauli operator & Completion \(E+\Stitch(\Gamma(E))\), which is closed on every original codeblock. & Def.~\ref{def:stitchsyndrome}\\
\(D\) & integer & Target distance used in distance statements. & Lem.~\ref{lemma:subdistancegamma}\\
\(\ell(\overline E)\) & element of \(\bigoplus_i\mathcal L_i\) & Vector of original-code logical labels of the completed error. & Lem.~\ref{lemma:stitchsyndrome}\\
\(P_i(t),F_i(t)\) & local Paulis & Stitching completion and completed local operator in the weight-\(D\) proof. & Thm.~\ref{theorem:weightDquilt}\\
\(\tau\) & vector in \(\ft^N\) & Choice of local completion branches in the weight-\(D\) proof. & Thm.~\ref{theorem:weightDquilt}\\
\(\dotG^{\,i},\dotd_i\) & gauge subspace/integer & Gauge group and operational distance of the perforated PIC on block \(i\). & Def.~\ref{def:perforatedcode}\\
\(U_i\) & subspace of \(\mathcal L_i\) & Local logical image \(\operatorname{im}\lambda_i\) gauged by the perforated code. & Lem.~\ref{lemma:perforatedquotient}\\
\(\dotd\) & integer & Common perforated operational distance in a homogeneous quilt. & Def.~\ref{def:perforatedcode}\\
\(\atom{\mathcal C}\) & PIC & Atomic quilt, with repetition incidence code. & Def.~\ref{def:atomicnuclear}\\
\(\nucleus{\mathcal C}\) & PIC & Nuclear quilt: the atomic quilt on one representative block of each local code/stitch type. & Def.~\ref{def:atomicnuclear}\\
\(m_t\) & integer & Multiplicity of local code/stitch type \(t\). & Cor.~\ref{corollary:atomicgenerators}\\
\(g^\perp(B)\) & integer & Smallest width needed to generate \(B^\perp\). & Rem.~\ref{rem:dualgenerationwidth}\\
\(\mathcal R_i\) & set of Paulis & Source-check presentation on block \(i\): the local checks from which syndrome extraction is built, not an arbitrary basis of \(H^i\). & App.~\ref{sec:sparsityanalytical}\\
\(w_i,\kappa_i\) & integers & Maximum weight and maximum qubit degree of \(\mathcal R_i\); \(w=\max_i w_i\) and \(\kappa=\max_i\kappa_i\). & Eq.~\eqref{eq:local-sparsity-data}\\
\(\wt(\mathcal G),\gamma(\mathcal G)\) & integers & Sparsity data of a generating set \(\mathcal G\): its largest generator weight, and the largest number of its generators meeting one qubit. & Def.~\ref{def:weightcongestion}\\
\(c_i(y)\) & integer & Least physical weight of a lift of \(y\in Y^\perp\) on block \(i\). & Lem.~\ref{lemma:simpletensorcost}\\
\(\Lambda_i,\mathcal T_i\) & graphs & Stitching graph on the \(W\) interface coordinates, whose edges come from single source checks of stitch-syndrome weight two, and a chosen spanning tree of it. & Def.~\ref{def:stitchinggraph}\\
\(\mathcal B,\beta_i\) & basis/integer & Chosen basis of \(B^\perp\), and the number of its elements supported on block \(i\). & Thm.~\ref{theorem:sparseinheritedbasis}\\
\(\mathcal R_0\) & subset of \(\mathcal R\) & Source checks whose stitch syndromes form a basis of \(Y^\perp\), doubled across blocks in the homogeneous case. & Prop.~\ref{prop:homogeneousdoubling}\\
\(x^i,z^i\) & Paulis on block \(i\) & Pure \(X\) and pure \(Z\) representatives whose product is the mixed local representative \(\ell^i\). & App.~\ref{sec:overlap}\\
\(A,O,A'\) & qubit sets & The \(X\)-only, overlap and \(Z\)-only parts of the support of a mixed representative. & App.~\ref{sec:overlap}\\
\(o_i\) & integer & Overlap on block \(i\): the number of sites carrying \(Y\), \(|O|\). & App.~\ref{sec:overlap}\\
\(\delta_i\) & integer & Least weight of a relative-sector component on block \(i\) after dressing by local stabilisers and local stitching operators. & App.~\ref{sec:overlap}\\
\(d_\Delta\) & integer & Distance of the relative sector of a mixed measurement. & App.~\ref{sec:overlap}\\
\(r\) & integer & Number of stitching rounds in a circuit. & Sec.~\ref{sec:circuitsims}\\
\(L,\,l\) & coset/representative & A target logical coset and one representative of it, from which a quilt is built. & App.~\ref{sec:tweedietest}\\
\(T\) & integer & Number of trials of a distance finder run on one quilt or circuit. & App.~\ref{sec:tweedietest}\\
\(\lambda\) & non-negative real & Rate at which a quilt or circuit concedes: one trial fails to find a low-weight operator with probability \(e^{-\lambda}\). Distinct from the maps \(\lambda_i,\lambda_B\) and from the observable \(\lambda\) of App.~\ref{sec:circuit-distance-methods}. & App.~\ref{sec:tweedietest}\\
\(G(\lambda),G_+(\lambda)\) & distributions & Distribution of rates over freshly generated quilts, and its part at non-zero rate. & App.~\ref{sec:tweedietest}\\
\(p_\text{good}\) & probability & Fraction of freshly generated quilts or circuits that are distance preserving; the weight of \(G\) at \(\lambda=0\). & App.~\ref{sec:tweedietest}\\
\(p_\text{surv}(T)\) & fraction & Fraction of quilts or circuits that have not conceded after \(T\) trials. & Eq.~\eqref{eq:plateau}\\
\(\mathcal L(T)\) & function & Laplace transform of \(G_+\): the probability that a defective quilt survives \(T\) trials. & Eq.~\eqref{eq:plateau}\\
\(\mathrm{FNR}(T)\) & probability & False negative rate: the probability that a quilt or circuit which has survived \(T\) trials is nevertheless defective. & Eq.~\eqref{eq:fnr}\\
\(\theta,\delta,\gamma\) & real parameters & Shape, scale and tilt of the Tweedie family assumed for \(G_+\). Distinct from the congestion \(\gamma(\mathcal G)\) and from \(\delta_i\). & Eq.~\eqref{eq:tweedie}\\
\(F,\mathcal D\) & vector spaces & Fault histories of a circuit, with unit cost per elementary fault, and its detector outcomes. & App.~\ref{sec:circuit-distance-methods}\\
\(D:F\to\mathcal D\) & linear map & Detector response of a circuit. Distinct from the target distance \(D\). & App.~\ref{sec:circuit-distance-methods}\\
\(G_f,H_f\) & subspace/quotient & Silent faults with trivial action on the outputs, and the fault classes \(\ker D/\operatorname{im}\iota\). & App.~\ref{sec:circuit-distance-methods}\\
\(P,P_G\) & linear maps & The fold: a cost-nonincreasing chain map from the fault complex of the full circuit to that of one prototype round, and its restriction to silent faults. & Eq.~\eqref{eq:fault-chain-map}\\
\(\mu,\nu,\lambda\) & observables & Cap and cup observables, supported on the opening and closing rounds of the stitching operation, and their sum \(\lambda=\mu+\nu\), which survives the fold. & App.~\ref{sec:circuit-distance-methods}\\
\(R,r_{\min}\) & integers & Number of stitching rounds separating cap from cup, and the smallest certified cap--cup separation. & Thm.~\ref{theorem:cap-cup}\\
\end{longtable}
\rowcolors{1}{}{}
\normalsize

\pagebreak
\printbibliography[heading=bibintoc]
\end{document}